# FRONT PAGE

# Satellite assisted disrupted communications in OMNeT++: Experiments and IoT Case Study

Georgios Koukis,  georkouk14@ee.duth.gr
Department of Electrical & Computer Engineering, Democritus University of Thrace (DUTH),
Xanthi, Greece, 04-07-2022

## Abstract

Space industry is trending anew with satellites operating in lower and lower altitudes promising global coverage and terrestrial type of speeds. In fact, communications through satellite infrastructures have become cost effective solutions not only for isolated areas in which terrestrial network development is infeasible, but also as a viable alternative in cases of emergency. Moreover, Wireless Sensor Networks (WSN) can benefit from the wide coverage of space infrastructure due to their high population, their disrupted communication nature and the possible lack of ground support.

In this work we discuss the utilization of micro-satellite constellations as effective infrastructures for the communication among ground stations or even among "smart" devices in IoT scenarios. We design and implement a series of experiments in OMNeT++ (with the OS3 framework) and evaluate their results in different scenarios. Initially, we establish the necessary theoretical background for space communications, including satellite and constellation design features, with existing and novel satellite services in various areas of interest. Furthermore, we detail the OMNeT++ and OS3 frameworks and introduce the significant variables/parameters for our experiments. Our scenarios are presented in three groups, departing from the straightforward one sender - one receiver communication and proceeding with a topology of multiple neighboring ground stations transmitting pings. We conclude with an IoT Case Study in which we import real measurements from sensors of the SmartSantander test-bed. Through our experiments we observe the effect of simulation parameters such as the constellation design (e.g., the number of satellites and planes) and the intersatelliteLinks regarding the produced RTT and ping loss, while we also highlight their potential contribution on networking communications when disruptions dominate.

# Table of Contents

# List of Figures

# List of Tables

# List of Abbreviations

ABFN: Adaptive Beamforming Networks

ANC: Analog Network Coding

AODV: Ad Hoc On-Demand Distance Vector

BER: Bit Error Rate

BRDF: Leonardo-Bidirectional Reflectance Distribution Function

CCN: Content-Centric Networking

CES: Coast Earth Stations

CNF: Cache-And-Forward

CPU: Control Process Unit

CRD: Communication Relay Drones

CSMA/CA: Carrier-sense multiple access with collision avoidance

CYGNSS: Cyclone Global Navigation Satellite System

DMC: Disaster Monitoring Constellation

DNC: Digital Network Coding

DONA: Data-Oriented Network Architecture

DRS: Data Relay Satellites

DSA: Delay Sensitive Application

DTN: Delay Tolerant Networking

D2D: Device-to-Device

ECI: Earth-Centered Inertial

EDRS: European Data Relay System

EPIRB: Emergency Position Indicating Radio Beacon

ESA: European Space Agency

GEO: Geostationary Earth Orbit

GLONASS: Global Navigation Satellite System

GPS: Global Positioning System

GS: Ground Station

G2G: Ground to Ground

G2S: Ground to Satellite

HAB: High-Altitude Balloons

HTS: High Throughput Satellite

ICN: Information Centric Networking

IFCPD: Interference-Free Contact Plan Design

IoT: Internet of Things

IPN: Inter-Planetary Network

IPv4/IPv6: Internet Protocol version 4/6

ISL: Inter-Satellite Links

ITU: International Telecommunication Union

LEO: Low Earth Orbit

MANET: Mobile Ad hoc Networks

MCC: Mission Control Center

MEO: Medium Earth Orbit

MMSC: Maritime Mobile Satellite Communications

MSSN: Maritime Surveillance Sensor Network

MTC: Machine-Type Communication

NDN: Named Data Networking

NDP: Neighbor Discovery Protocol

NORAD: North American Aerospace Defense Command

NS2: Network Simulator 2

OBP: On-Board Processing

OBR: On-Board Routing

OMNeT++: Objective Modular Network Testbed in C++

OPNET: Optimized Network Engineering Tools

OSM: OpenStreetMaps

OSPFv4: Open Shortest Path First version 4

OS3: Open Source Satellite Simulator

PACECR: Prototype Airborne Contested Environment Communication Relay

PER: Packet Error Rate

PLC: power line communication

PSIRP: Publish–Subscribe Internet Routing Paradigm

QoE: Quality of Experience

RAAN: Right Ascension of the Ascending Node

RRN: Ring Road Network

RTT: Round Trip Times

SAR: Search and Rescue

SDG: Sustainable Development Goals

SDN: Software Defined Networking

SDP: Simplified Deep space Perturbations

SER: Symbol Error Rate

SES: Ship Earth Stations

SGP: Simplified General Perturbation

S2S: Satellite to Satellite

S2G: Satellite to Ground

SUMO: Simulation of Urban Mobility

SVS: Scalable Video Streaming

TCP: Transmission Control Protocol

TDRSS: Tracking and Data Relay Satellite System

TLE: Two-Line Element

TSN: Time-Sensitive Networking

UAV: Unmanned Aerial Vehicle

UN: United Nations

VANET: Vehicular Ad hoc Network

VEINS: Vehicles in Network Simulation

VSL: Variable Speed Limit

V2X: Vehicle-to-everything

WMN: Wireless Mesh Networking

WSN: Wireless Sensor Networks

3GPP: Third Generation Partnership Project

# **Acknowledgements**

At first I would like to express my deepest appreciation to my supervisor Prof. Vassilis Tsaoussidis for his consistent support and guidance during the writing of this thesis. His door was always open for discussion, providing me with invaluable feedback.

Furthermore, I am also grateful to the rest of the research team, Ioanna, Vassilis and Giorgos for their collaborative efforts during the current work. They were eager to discuss my concerns about issues that arose and advise me whenever I had questions.

Finally, I must express my very profound gratitude to my family and friends for providing me with unfailing support and continuous encouragement throughout my years of study and through the process of writing this thesis. This work, like most of my accomplishments, would not have been possible without them.

# A. Introduction

This work aims to evaluate the reliability and effectiveness of satellite constellations in various circumstances. We discuss the potential use of space infrastructure and more specific Low Earth Orbit (LEO) satellite constellations at 600km altitude as substitute for communications among ground stations or in general among any "smart" device that can produce and communicate data. Assuming that communication is feasible, meaning that these devices can connect to a satellite infrastructure, satellites can become useful in various cases such as an environmental emergency situation or an occasion demanding of additional security. Specifically, LEO satellite constellations can play a significant role in smart-city and IoT scenarios where disruptive communications are usual, relaying data through a space-terrestrial infrastructure, either as the primarily communicational channel in isolated areas (where terrestrial facilities are impractical), or acting as backup links in cases of compromised or damaged equipment. Beyond that, the integration of novel technologies (e.g., ICN and DTN) and the use of alternative means of relay, such as drones or ships, could expand the range of available options that could be integrated in the communicational infrastructure and also provide on-demand medium recruitment, according to the characteristics of each mission.

The main tool for the simulation of the experiments is the Objective Modular Network Testbed in C++ (OMNeT++). Specifically we use an updated OS3 framework [4] which simulates and visualize satellite communications. Among the simulation parameters we highlight the: i) *communication range/footprint*, which depends on satellite's altitude and elevation angle, ii) *updateInterval*, representing the granularity of mobility changes, iii) *sendInterval*, indicating the waiting time period between pings' transmissions and iv) constellation characteristics such as the *number of satellites, number of planes, satellites per plane* and *intersatelliteLinks*, which enables the communication among satellites of the same plane.

We have conducted three groups of experiments evaluating the effect of some previously explained parameters (number of satellites, number of planes, satellites per plane and the existence of intersatelliteLinks) on the mean RTT, the range RTT and the ping loss, testing different constellations and topologies. We make a "logical" hypothesis that denser constellations with the intersatelliteLinks parameter enabled would produce improved results. Starting with the *"Simple"* group case, we evaluate the communication of two nodes (one Sender and one Receiver) with certain parameters and then we alternate these parameters, attempting to improve the observed results such as the lack of communication or the high RTTs. After that, in the "*More sophisticated*" scenario, a denser topology with more ground stations (representing sensors) is being examined. As these ground stations are adjacent, they utilize the same satellites, resulting in overlapping signals, collisions and consequently, lack of communication or extremely high RTTs. We change some parameters (such as the startTime and the number of satellites) to alleviate these occurred problems. Finally, we conclude with an IoT Case Study (*"ReWire Case Study"*) where the real data acquired within the on-going project ReWire [1][2][3] are used. Particularly we import some of these data taken from the Smart Santander test-bed (from sensors scattered in the city) in the OMNeT++ as input, to observe RTT and ping loss changes, in a 24 hour experiment where pings are transmitted to different test-beds in EU and USA, according to the real sensors' timestamps derived from a .csv file.

Through most of the experiments our aforementioned hypothesis is confirmed, observing a reduction of the mean and range RTT with lower ping loss, as we increase the satellites' number and enable intersatelliteLinks. On the contrary, in the IoT Case Study with the actual measurements, there are exceptions of sensors resulting in better performance with fewer satellites in the constellation and without the intersatelliteLinks parameter. Observing the outcome of the former example, we conclude that in some cases the position of the ground nodes with respect to the satellites' orbit prevails as a factor, influencing the results greatly in comparison to the multitude of satellites.

We came across a few difficulties we believe should be mentioned, including the installation of the framework and the use of some of its parameters. Firstly, despite of the instructions provided in [4+], the framework installation presents problems in the Windows environment, which forced us to work and run our simulations on an Intel Core i7 CPU laptop with Ubuntu preinstalled. Furthermore, we were not able to use some visualization parameters (such as the *displayRoutingTables*, which could help in understanding the alternative routes) or simulate denser and more populated constellations (with more satellites and planes, such as the complete Starlink's LEO constellation) due to hardware limitations.

The present work could be used for the evaluation of the expected RTT and ping loss, regarding the communication among ground stations through LEO space constellations. Our main contribution includes the implementation of a Case Study in which we import real measurements from IoT devices into the OMNeT++ and utilize them for space assisted communication experiments. Moreover, we observe and determine the impact of specific parameters such as the number of ground stations, satellites, planes & satellites per plane, and intersatelliteLinks on the reliability of the communications. Although modifications are necessary to enhance the credibility of our work as simulation examples of actual space-terrestrial communications, our view could be a motive for further research. The implementation of space communication protocols, mechanisms for collision detection and avoidance, routing optimization algorithms or buffer requirements of satellites are a few significant topics for discussion. Lastly, the utilization of ships or drones as relays should be considered from a technical perspective.

The rest of the work is outlined as follows. In the Background & Related work Section, we provide some characteristics of the satellites (*Section B.I*) such as the constellation, their communication, the coverage calculation and the effect of the altitude and the elevation angle on the satellite's footprint. Afterwards, the types of satellites according to their altitude are presented (*Section B.II*). In *Section B.III* we investigate the services provided by satellites, presenting existing and in development ones. Lastly, in *Section B.IV* we mention part of the related work in respect to the applications of OMNeT++, the space implementations and frameworks, and some research of alternative relays such as drones or ships. In *Section C* the methodology of the work is explained. We discuss the used tools/frameworks in addition to some important simulation variables/parameters (*Section C.I*) and describe our experiment scenarios (*Section C.II*). Also, the ReWire project within which we acquired the real data is briefly discussed in *Section C.III*. In *Section D* we provide our experimentation analysis and finally in *Section E* we discuss our future works, suggesting applications in smart-cities or IoT cases and possibly integrate space segments in a future alternative version of the ReWire project.

# B. Background & Related work

## I. Satellite characteristics

A multitude of satellites can be deployed either in the form of a constellation or in the form a cluster [6]. While constellations consist of multiple replicas of the same satellite, clusters contain non-identical cooperative satellites. Highlighting the first case, according to the National Research Council (U.S.) Committee on Earth Studies, constellations refer to *"collection of satellites whose relative positions are controlled in each of multiple orbits".* [7]

Some major constellation characteristics include the *pattern* (e.g., polar, rosette, flower, Walker-Delta or hybrid), the *type*, based on their altitude (e.g., LEO, MEO, GEO), the *number* of orbits and satellites in each orbit. A satellite constellation can be organized in several orbital planes with the altitude and inclination affecting the latency, the data rate of the communications, and the power consumption [6]. The period in which satellites are available for communication with fixed ground stations may differ from a few minutes for LEO satellites, to a complete day for GEOs. There are three elements present in every space mission: 1) the space segment (e.g., the satellite constellation), 2) the ground segment - including a set of GSs responsible for control and management tasks and 3) the user segment - referring to the rest of ground communication devices such as IoT devices, and smart-phones. [8]

The coverage area or footprint of a satellite is presented as a circular area on the Earth's surface indicating the zone within which satellites can communicate with GSs. The primary variables affecting the footprint are presented in *Figure 1*: $\varepsilon_0$ the elevation angle, $\alpha_0$ the nadir angle, $\beta_0$ the central angle, $d$ the slant range (the distance from a satellite to a GS) and $H$ the satellite attitude above the Earth's surface, while $R_e$=6378x10$^3$m is the constant distance from a GS to the Earth's center.

The following equations describe the process of calculating the above variables (see *Figure 1*) where $r = H + Re$.

$$\varepsilon_0 + \alpha_0 + \beta_0 = 90^o \quad (1)$$

$$d * cos\,\varepsilon_0 = r * sin\,\beta_0 \quad (2)$$

$$d * sin\,a_0 = R_e * sin\,\beta_0 \quad (3)$$

Applying the cosines law, we get:

$$r^2 = R_e^2 + d^2 - 2 * R_e * d * cos(90 + \varepsilon_0) \quad (4)$$

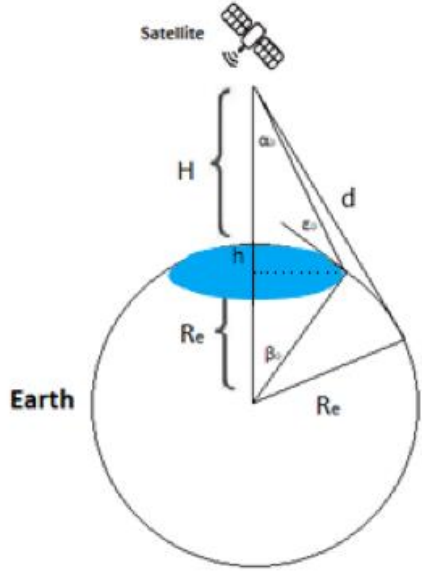

*Figure 1: Primary variables in GS − satellite communication*

Observing the largest triangle we calculate the *sinα₀* which depends on the elevation angle $\varepsilon_0$ according to Eq.(5). Consequently, the maximum coverage can be achieved for $\varepsilon_0$=0→cos$\varepsilon_0$=1. As mentioned in [9], to avoid obstacles caused by natural barriers at too low elevation angles, usually the minimum angle is determined in the range of 2°-10°.

$$sin\,\alpha_0 = \frac{R_e}{R_e+H} * cos\,\varepsilon_0 \quad (5)$$

The surface of the coverage area is calculated by Eq.(6), where *h* is the height of the produced spherical cap:

$$S_{Coverage} = 2 * \pi * R_e * h \qquad (6)$$

From the smallest triangle built:

$$cos\,\beta_0 = \frac{R_e-h}{R_e} = 1 - \frac{h}{R_e} => \frac{h}{R_e} = 1 - cos\,\beta_0 \quad (7)$$

Finally, from Eq. (6) & (7) we can express the satellite's footprint as a percentage of the Earth's surface area:

$$Coverage\,(\%) = \frac{S_{Coverage}}{S_{Earth}} = \frac{2*\pi*R_e*h}{4*\pi*R_e^2} = \frac{1}{2} * (1 - cos\,\beta_0) \quad (8)$$

*Table 1* describes the coverage area at different satellite altitudes and elevation angles applying the above equations with $R_e$, *H* and $e_0$ known.

*Table 1: LEO satellite coverage area per altitudes and per different elevation*

| Elevation (ε₀) | Satellite Altitude | | | | | | | |
|---|---|---|---|---|---|---|---|---|
| | 160km | 500km | 600km | 700km | 800km | 900km | 1000km | 1500km |
| | Coverage (%) | | | | | | | |
| 0° | 1,22 | 3,63 | 4,30 | 4,94 | 5,57 | 6,18 | 6,78 | 9,52 |
| 2° | 0,89 | 3,04 | 3,64 | 4,24 | 4,82 | 5,39 | 5,95 | 8,54 |
| 4° | 0,66 | 2,54 | 3,09 | 3,64 | 4,17 | 4,70 | 5,22 | 7,66 |
| 6° | 0,49 | 2,12 | 2,62 | 3,12 | 3,61 | 4,10 | 4,58 | 6,86 |
| 8° | 0,37 | 1,78 | 2,23 | 2,68 | 3,13 | 3,57 | 4,02 | 6,15 |
| 10° | 0,28 | 1,50 | 1,90 | 2,30 | 2,71 | 3,12 | 3,53 | 5,50 |
| 25° | 0,06 | 0,46 | 0,62 | 0,80 | 0,98 | 1,17 | 1,37 | 2,39 |
| 40° | 0,02 | 0,17 | 0,24 | 0,31 | 0,38 | 0,47 | 0,55 | 1,03 |

Communication among satellites is achieved via inter-satellite links (ISLs). They improve data routing, maximize throughput and minimize latency, enhancing the network connectivity and autonomy while satellites' complexity and cost are increased [6]. ISLs can also be further divided into intra-plane (for satellites in the same orbital plane) and inter-plane ISLs (for satellites of different planes). The connection with the GSs is achieved through Ground-to-Satellite Links (GSLs), communicating: i) *Directly with the destination*: the simplest architecture without ISLs or intermittent GSs, requiring the ground node to be located in the satellite's coverage area, ii) *With Ground infrastructure*: when no ISLs exist and the source transfer data to the nearest GS from where they are transmitted to the destination using terrestrial communication networks, iii) *With Space infrastructure*: the source sends data to the satellite which in turn delivers them to the destination or iv) *With Ground and Space infrastructure*: the source node utilize both satellites and ground links. To summarize, there are four logical links related to the communication sector, the: ground to ground (G2G), ground to satellite (G2S), satellite to satellite (S2S) and satellite to ground (S2G) links. [8]

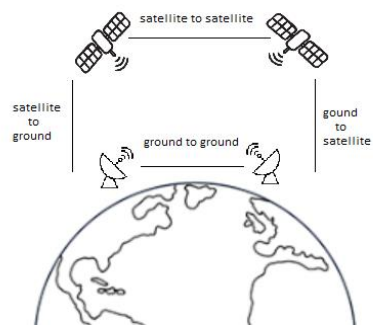

*Figure 2: Communication links*

## II.   Types of Satellites / Types of orbits

According to the distance between the satellites and the surface of Earth, satellites can be categorized into geostationary Earth orbit (GEO), medium Earth orbit (MEO), and low Earth orbit (LEO) satellites. Their basic properties can be observed in *Table 2.*

*Table 2: GEO, MEO and LEO satellite general characteristics*

|  | LEO | MEO | GEO |
|---|---|---|---|
| **Satellite Height** | 500-1500km | 8000-18000km | 35784km |
| **Line of sight time** | 0.5h | 2-4h | 24h |
| **Rotation Period** | 1.5h | 5-12h | 24h |
| **Travelling speed** | 28.163km/h | 16.093km/h | 11.070km/h |
| **Satellites required** | 30-60 | 10-20 | 3 |
| **Satellites Life** | Short/5yrs | Long/10-15yrs | Longest/15+yrs |
| **Number of Handoffs** | High | Low | Least |
| **Gateways Cost** | Very Expensive | Expensive | Cheap |
| **Gateways required** | Local-Numerous | Regional-Flexible | Fixed-Few |
| **Propagation Delay/Latency** | Least | High | Highest |

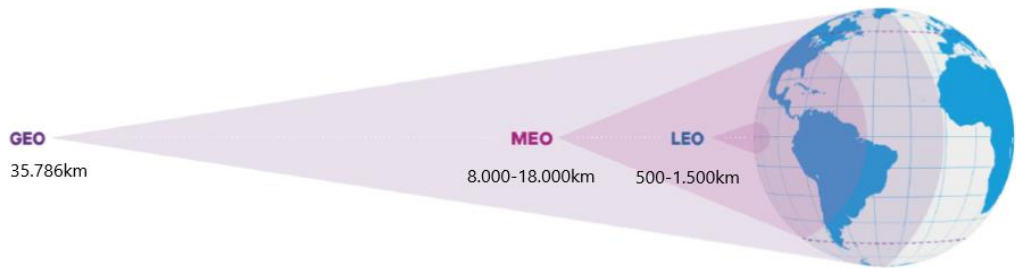

*Figure 3: GEO, MEO and LEO satellite distance*

1. **LEO:** LEO satellites operate relatively close to Earth's surface. According to the European Space Agency (ESA), they are located at an altitude of less than 1500km and could be as low as 160km, traveling at speeds of 28.163km/h [10]. The orbit is commonly used for satellite imaging, as the low altitude allows them to take higher resolution images. In recent years there has been a significant trend towards LEOs [11] with an increasing interest in low-cost small satellites ((<50kg), such as nanosatellites or CubeSats) for the provision of broadband internet and low-bandwidth communications. Because of their smaller coverage, compared to MEO and GEO, numerous satellites are necessary to cover a desirable area, as well as more complex ground systems in order to keep close-to-constant contact with the space infrastructure. Additionally, the radiation and atmospheric drag existing in lower altitudes result in a much shorter lifetime (an average of five years) with the constant maintenance or satellite replacement to be mandatory. Some examples of LEO satellites come from companies such as Iridium, Globalstar, Orbcomm, OneWeb, Telesat and SpaceX.

To conclude, we review the characteristics and tradeoffs of LEO satellites in the following: 1) less complicated and lower cost satellites, 2) low latency/propagation delay, 3) demand for more satellites due to their smaller footprint, 4) more speed/energy to overcome forces, 5) less life span and need for more frequent maintenance, 6) high image resolution due to the closer distance to the Earth, 7) revisit time limitations with frequent handovers and 8) high levels of interference.

2. **MEO:** Medium Earth orbit comprises a range of orbits between LEO and GEO. MEO was historically used for navigation applications like the Global Positioning System (GPS), GLONASS and Galileo but more recently, high throughput satellite (HTS) MEO constellations have been deployed to deliver low-latency and high-bandwidth data connectivity to service providers, government agencies, and commercial enterprises [11][12]. They are also used for tele/communication purposes (e.g., Sirius Satellite Radio, Odyssey, Inmarsat-P and O3b) and data backhaul.

3. **GEO:** Lastly, there are satellites placed in the GEO. They follow Earth's rotation, taking 23 hours 56 minutes and 4 seconds for a full rotation, appearing "stationary" over a fixed position [10]. In order to track Earth's rotation, they move at speeds of 3km/s in 35.786km altitude. GEO is used by satellites that need to stay constant above one particular place over Earth, such as tele/communication satellites (e.g., transmitting television broadcast signals). Thus, an antenna on Earth can be fixed to always stay pointed towards that satellite without moving, providing uninterrupted connection. They can also be used for weather monitoring and imaging purposes, as their bigger telescopes can provide persistent monitoring and data collection with adequate resolution. Satellites located GEO can cover a large area of the Earth, with three equally-spaced satellites to be able to provide near global coverage. Several GEO HTS communication systems have been deployed, such as the Inmarsat Global Xpress, Viasat, IPStar, and Chinasat-16. However, latency becomes noticeable as significant time is required for an RTT (up to 600ms [13]), presenting problems for latency-sensitive applications such as voice or video communication.

### III.    Satellite services

There is a plethora of services provided by satellites. We have focused on a few, categorizing them as: 1.Communication – Internet, 2.Military, 3.Remote Sensing – Earth Observation, 4.Maritime & Agriculture, 5.Power energy networks - Smart Grid, 6.Relay Communication Systems and 7.IOT/M2M constellations, 5G and DTN.

#### 1.  Communication – Internet

The demand for faster broadband Internet and high speed communications is growing nowadays [8] recognizing the potential advantages of a global network coverage from the space infrastructure. Specifically, the utilization of LEO satellite constellations could become beneficial due to their comprehensive coverage and large capacity [14]. Such constellations can provide access to information in areas where terrestrial networks are difficult to deploy or cost-prohibitive and contribute with flexible resource allocation [15], provisioning long-distance backhaul and directly serving users. Also, considering the complementary usage of high-altitude balloons (HABs) and even unmanned aerial vehicles (UAVs) [16][45], end-to-end Internet connectivity can be supported even in areas with no capability for immediate conversational data interchange. In addition, utilizing approaches such as the Delay Tolerant Networking (DTN) that could be integrated into the fabric of the Internet [13][17][18], with Internet access extension technologies (e.g., the exploration of unused bandwidth), we could alleviate issues that until recently were difficult to handle with the classic form of the terrestrial internet infrastructure.

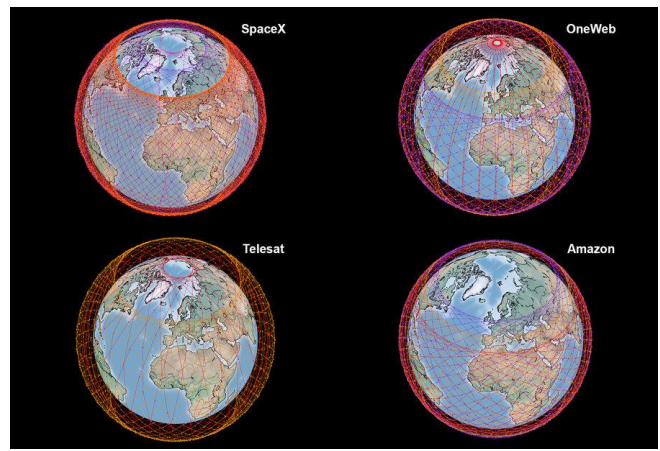

*Figure 4:  The four largest internet mega-networks [https://techxplore.com/news/2021-06-largest-internet-meganetworks.html]*

## 2. Military

Satellites can provide high visibility against tactical enemy's operations supplying with the essential information required for a strategic strike assessment. Their rapid access to near real time information and imagery (mainly the nano and microsatellites) is significant for the surveillance and aerial reconnaissance augmentation process [19]. In addition, their readily deployment in times of crisis may fill the missing gap for military operations, while also utilize the intelligent "things" (such as UAVs) which are involved in the battlefield sensing, communicating and collaborating [20]. Finally, due to the constant motion of the LEO satellites, non-approved users cannot easily intercept, corrupt or jam them.

## 3. Remote Sensing – Earth Observation

Often identified with remote sensing, the importance of Earth observation (EO) can be enhanced by its presence in the United Nations' (UN) 2030 agenda for the Sustainable Development Goals (SDG), [21] a set of 17 goals aimed at the improvement of people's living conditions. Numerous sectors have been benefited including: i) environmental/climate monitoring (e.g., the GR01-DUTHSat - investigating the upper atmosphere and measuring electron frequency [22], or the Leonardo-Bidirectional Reflectance Distribution Function (BRDF) constellation - studying the reflection of solar flux and relation to climate and its effects [23]), ii) meteorology phenomena (e.g., tracking volcanic ash, measuring water vapor or atmospheric tracking (e.g., Cyclone Global Navigation Satellite System (CYGNSS) predicting cyclone path [24]), iii) maritime identification, iv) pollution monitoring (e.g., remote oil and gas extraction or spill detection), v) vegetation monitoring and precision agriculture [25-28], vi) real time detection in disaster monitoring cases (e.g., Disaster Monitoring Constellation (DMC) – providing emergency Earth imaging for disaster cases) and vii) national security for border areas. Also, new applications are emerging in areas traditionally not related to space technology, such as viii) archeology and cultural heritage, ix) real estate, x) migrant and refugee fluxes. [29]

## 4. Agriculture

In the field of agriculture, novel technologies are emerging trying to deal with the ever-increasing needs of humanity for higher quality products and larger quantities. Several concepts are included such as the precision agriculture monitoring [25], the satellite data and remote sensing data [26] used for land cover classification and agriculture monitoring with UAV/drone usage [27][28]. Services like crop health monitoring, crop classification, soil moisture estimation and weather prediction are included benefiting the produced products.

### 5. Power energy networks - Smart Grid

The underlying communication infrastructure plays a significant role in the management and distribution of energy networks. Satellites due to their unique characteristics can contribute in various ways such as to: 1) Provide services in wider range of areas, 2) Reduce the impact of frequent transitions between different media that could degrade the availability and reliability, 3) Provide high flexibility and efficiency in bandwidth utilization, while reducing the installation and maintenance costs of newly added power network devices. [30]

### 6. Relay Communication Systems

According to the ITU [31], satellites relay signals from one point on or near the Earth's surface to another. Therefore, data relay satellites (DRS) are located to relay information to and from non-geostationary satellites, spacecraft, vehicles/vessels and fixed Earth stations. Satellites receive information from a GS and forward them over a radio frequency cross-link to a relay satellite, which consequently transmits the data back to the GS. There are several DRS used or planned to be used in the near future utilizing techniques like ISLs or the "store and forward" that will be mentioned. [17][18]

An active ESA project is the European Data Relay System (EDRS) [32][33], a constellation of GEO DRS, aiming to provide links among LEO satellites and other spacecrafts. EDRS can assist in cases such as security forces or rescuing workers who demand near-real-time satellite data of a disaster area for their time-critical services and require communication support in cut-off areas affected by natural or man-made actions. Likewise, the U.S. Tracking and Data Relay Satellite System (TDRSS) [34] is a network of American communication satellites and GSs used by NASA attempting to increase the amount of data transferred and the duration of space-terrestrial communications.

The authors in [35] mention the advantages of an ultra-dense LEO satellite constellation over a high-frequency band as potential solution for terrestrial data offloading due to the wide coverage, high-capacity backhaul and flexible network access services of the space infrastructure. Data from the offloaded terrestrial users are initially transmitted to the LEO-based small cells which then upload the received data to the core network via satellite backhaul links. Various research areas can be discussed for data offloading, especially in heterogeneous network cases (e.g., terrestrial-satellite network or cellular-mobile-WiFi network), such as resource allocation, energy efficiency, interference management and pricing mechanisms.

### 7. IoT/M2M constellations, 5G and DTN

LEO satellite constellations have become an interesting research area for IoT and 5G machine-type communication (MTC) applications. In IoT, where "smart" objects are often remote or disperse and unable to directly access terrestrial networks, satellite infrastructures could provide a more cost-effective solution compared to terrestrial technologies, enhancing the interoperability among the numerous applications and services [11]. They could also exploit their ability to broadcast, multicast or geocast in cases of multiple IoT devices [29]. Therefore, satellites collect data from various sensor nodes and transmit to GSs for data management. Subsequently, the processed data follow the reverse path from GSs to the satellites and then to the actuator nodes. Furthermore, IoT applications are divided into two groups according to their delay tolerance [36]: 1) Delay Tolerant Applications (DTAs) and 2) Delay-Sensitive Applications (DSAs). DTAs are named after the Delay Tolerant Networking (DTN), an approach to computer network architecture with the primarily characteristics of frequent and prolonged disconnections and long propagation delays. On the contrary, DSAs' requirements are more stringent, demanding lower latencies and higher reliability. Comparing to DTAs, DSAs entail a different space infrastructure design with ISLs, on-board processing (OBP) and on-board routing (OBR) [16]. Consequently, constellations without ISLs are suitable for DTAs while the existence of ISLs is necessary for real-time application.

Continuing with DTN, it has been originally developed (defined in RFC4838 [37]) to ensure the reliable message delivery in the *"challenging Interplanetary Network (IPN)"*. As mentioned in [38], DTN is defined as *"a digital communication networking technology that enables data to be covered reliably among communicating entities when round-trip times may be highly variable and/or very long"*. The challenging space communications are characterized by data losses, low throughput and error prone, along with intermittent connectivity, asymmetric link data rates and link disruptions [39]. Therefore, a design that tolerates large periods of delay and transmits at high bitrates when links are available seems obligatory. Therefore, transmissions are paused when the communication space link is unavailable, and resumes when line-of-sight is established. The DTN architecture introduces an overlay bundle protocol (BP) (defined in RFC5050 [40]), with each node of the DTN architecture able to implement the *"store and forward"* mechanism [13]. Essentially, delays and disruptions are handled within each DTN "hop" as nodes provide the necessary storage for the application data before they forward them to the next node on the path.

Additionally, the evolving 5G/6G networks assisted by satellite communication infrastructures have been a trending research topic. According to [41], realizing the increasing interest, standard development organizations such as the Third Generation Partnership Project (3GPP) has published specific requirements for an integrated satellite-terrestrial network infrastructure in the context of 5G [42]. Particularly, LEO satellites could play a key part in extending cellular 5G networks to air, sea and other remote areas enabling IoT sensors and M2M communication [43]. Integrating satellites with 5G infrastructure also improves the Quality of Experience (QoE) of high capacity applications while intelligently using routing and offloading traffic, valuable spectrum could be saved and the resilience of the network could be improved. With the help of 5G-enabled satellites, users' needs for low latency on-demand streaming is satisfied [41], while according to the ESA [44] integrated 5G networks support a plethora of applications in medicine, emergency services, media, autonomous drones/cars, ships, space exploration and remote education.

It is also worth mentioning a novel networking paradigm, alternative to the IP-based internetworking, the information-centric networking (ICN) [45][46] and its applications in the satellite infrastructure [47][48]. Departing from the traditional host-centric access paradigm, where access to a desired content is mapped to its location, an information-centric model associates the access with the content itself, irrespective of its location. ICN is based on accessing named-content in the network, instead of host-to-host communication, with some novel research paradigms including the UMOBILE model (*"a universal, mobile-centric, and opportunistic communications architecture integrating connectivity approaches into a single architecture"*) [46] and MINOS platform ("*a Multi-protocol Software-Defined Networking Solution for the IoT"*) [49] worth underlined from the literature. There are several ICN approaches which are under investigation, such as the publish–subscribe Internet routing paradigm (PSIRP), content-centric networking (CCN), named data networking (NDN), network of information (NetInf), data-oriented network architecture (DONA), cache-and-forward (CNF) and software defined networking (SDN).

Finally, in [48] the key features of ICN and their impact in the integration of satellite-terrestrial networks are highlighted, exploring the benefits of satellite networks such as the wide-area coverage and the ability to broadcast/multicast. As mentioned, satellites could be used to i) update multiple caches at the same time with low delay and cost, ii) provide content-aware traffic management and prioritization utilizing satellite's capacity, iii) simplify the topology management and forwarding as mechanisms already existed in satellite networks, iv) reduce the overhead and delay for routing updates or name resolution tables, and v) better address the long propagation delays of satellite links, providing a unifying framework for DTN exploiting hop-by-hop or segment-by-segment congestion control. [51]

# IV.    Related work

There are multiple examples of work that utilize OMNeT++ in various network domains such as 4G, 5G, Manet, Vanet, IoT, AdHoc, Wireless Sensor Network (WSN), peer-to-peer networks (P2P), storage area networks (SANs) and implement network protocols primarily using the INET framework [50]. INET has evolved from the IPSuite originally developed at the University of Karlsruhe, providing detailed protocol models (e.g., TCP, IPv4, IPv6, Ethernet, Ieee802.11, OSPFv4), while novel paradigms such as SDN adaptations and ICN are being developed by the academic community.

Regarding the performance in WSN simulations, [52] compares OMNeT++ with NS2 and OPNET, indicating that the former performs better in terms of execution time and memory usage, especially in large-scale scenarios. In [53] a distributed, scalable algorithm which estimates the location of nodes in a WSN is described, while [54] introduces a collection of tools in OMNeT++ adding smart antenna capability to central WSN nodes. Furthermore, [55] presents a new CO-simulation framework based on MATLAB and OMNeT++ (called COSMO) aiming to build credible simulations for indoor wireless networks. The paper in [56] describes an energy efficient MAC protocol design, developed for reliable communications in wireless mobile OMNeT++ scenarios. A detailed analysis of a mobile WSN (MWSN) design using OMNet++ and MiXiM framework is presented [57], evaluating its performance and the effect of some parameters such as the mobile node's architecture, movement and the different time intervals. In [58], the design and simulation of a WSN for Rural Exploitation Zones is described in OMNet++ and [59] examines a multi-hop wireless network with realistic behavior using OMNet++ and INET framework. Authors in [60] explore an OMNeT++ simulation for reliable point-to-point wireless transmission and realistic Segmentation & Reassembly (SAR) with error control mechanisms in a distributed WSN. Finally, the work in [61] proposes and evaluates a cognitive WMN that utilizes multiple gateways to serve a client through multi-path routing and takes appropriate decisions using game theory techniques.

Furthermore, the approach of a rapid Vehicular ad hoc network (VANET) is presented [62], incorporating OMNeT++ with Vehicles in Network Simulation (VEINS) and Simulation of Urban Mobility (SUMO) frameworks and allowing for an easy traffic simulation using maps imported from OpenStreetMaps (OSM). In [63] the implementation of two VANET broadcasting algorithms (N-Hop Broadcast and Street Broadcast Reduction) and the potential problems in actual devices are addressed, using OMNeT++, SUMO and Google Earth. A parallel version of the INET Framework for Vehicle-to-everything (V2X) communications is considered in [64], with a multi-thread code instead of the original single-thread, resulting in significantly reduced computational time in city-scale scenarios. The paper in [65], aims to improve the performance of AODV and OLSR routing protocols in high-density urban areas, by proposing a vehicle-node density parameter. In [66] authors discuss ways to enhance Road Safety and Traffic Management using Variable Speed Limit (VSL) through VANETs to overcome issues due to traditional systems, using OMNet++, SUMO and VEINS. Moreover, an improvement of the packet delivery in the routing protocol compared to standard AODV is discussed in [67], while [68] analyzes a Reactive AODV Routing Protocol for Mobile Ad hoc Networks (MANETs) using the OMNET++ simulator.

We proceed with the cases of communication relays such as drones and ships which have become significant research topics. UAVs and drones could be employed as mobile relays or micro-base stations/access points to collect and transmit information, becoming an efficient, low cost and energy-saving alternative [69]. Military sources have experimented in the use of UAVs as relays (wars in Iraq and Afghanistan) to extend the range of terrestrial communications. As mentioned in [70], communication relay drones (CRDs) could provide dedicated data paths passing information on more reliable connections. In addition, CRDs could re-establish links in cases of compromised communications due to jamming, environmental interference, data corruption or equipment malfunction. Similar to drones and UAVs, aircrafts could also act as relays. For instance, the U.S. Air Force attempt to transform some fighters and bombers, into flying wireless routers enhancing ground troops' communications [72]. Also in [71], the Prototype Airborne Contested Environment Communication Relay (PACECR) is discussed as a backup route option, with aircrafts replacing or augmenting a satellite infrastructure, forming a local-area network and providing high-capacity backbone link relays. Furthermore, the history of Maritime Mobile Satellite Communications (MMSC) systems since the 1970s is reported in [73] for commercial and military purposes. It is described as a two-way satellite communication between Ship Earth Stations (SES) and Coast Earth Stations (CES) via GEO and Non-GEO satellite constellations. Examples such as Survival Craft Stations, Search and Rescue (SAR) vessels and Emergency Position Indicating Radio Beacon (EPIRB) alert stations operate with the services of MMSC. Moreover, a ship-to-ship communication system is considered in [74], where ships exchange information with the assist of multiple relay ships, selecting a method between analog network coding (ANC) and digital network coding (DNC), adaptively, based on channel state information.

Considering the OMNeT++ implementations for relays, in [75] the BEE-DRONES framework is demonstrated - a novel framework for data collection in UAV-aided WSNs - able to enhance the WSN lifetime and optimize the quality of gathered data in terms of minimal spatial correlation. Also, security issues are investigated such as in [76] where a jamming-Resilient Multipath Routing Protocol for Flying Ad

Hoc Networks is considered, comparing its performance with other routing protocols. Also, paper [77] proposes a sky caching-aided spatial querying scheme with the support of a flying drone which collaboratively generates a set of dummy locations to hide the user's current location from the server. Authors in [78] provide a lightweight distributed detection scheme, referred to as Lids, to defend against flooding attacks in the "Internet of Drones" environment, with each drone storing its received packets, sharing them with other drones and sending them to nearby GSs for consistency check and flooding attacks detection. Regarding the implementation of ships, a particular example of SAR with error control was implemented in [79], presenting promising results in respect to BER, PER, window size, and system efficiency. Additionally, [80] proposes a simulation framework concerning the energy consumption and communication protocols based on OMNeT++, simulating Maritime Surveillance Sensor Networks (MSSNs) for information acquisition. In [81] the evaluation of a relay-aided device-to-device (D2D) alternative method is described, to support short-distance direct communications in cases of poor channel links and limited coverage within the OMNeT++. Finally, OMNeT++ simulations were conducted to verify the rationality and feasibility of a multipath relay transport control method for real-time video services, supporting end-to-end media multipath transmissions through relay controllers and servers. [82]

Although space works with the OMNeT++ are more limited, we present a number of interesting endeavors. A novel satellite network simulation platform (SNSP) based on OMNeT++ and MiXiM framework in presented in [83], providing a satellite environment to test communication protocols combined with Delay Tolerant Network (DTN) main ideas. The authors in [84] propose SLP-GPSR, a Satellite Lifetime Predicted Greedy Perimeter Stateless Routing based on Greedy Perimeter Stateless Routing for LEO satellite networks, aiming to solve the unstable connection problem caused by high-speed satellite movements. Furthermore in [85], a comparison and optimization of several CSMA/CA back-off distribution functions for LEO satellite links is conducted, with the results of the proposed Boolean-type distribution shown to be optimal. In [86] an interference-free contact plan design (IFCPD) scheme is presented with a DTN routing protocol for space-terrestrial networks, evaluating performance parameters in OMNeT++. A CubeSat network is addressed in [87], exploring the trade-offs of satellite-to-ground (S2G) and satellite-to-satellite (S2S) communications and proving a decrease in the energy consumption with modifications to a specialized MAC and MANET routing protocol. In [88] and [89] the applicability of the IEEE 802.15.4 standard for WSN devices, regarding the planetary exploration context, is evaluated using Simulink and OMNET++ simulation models. A different work presents a tool designed to simulate broadband satellite communication systems and techniques based on the advanced DVB-RCS and DVB-S2 standards implemented in Matlab and OMNeT++ [90]. Moreover, to reduce the overall computational burden of complex network scenarios such as the Adaptive Beamforming Networks (ABFN), authors in [91] propose an efficient co-simulation model with the integration of OMNeT++ and Matlab. In [92] a satellite network jamming and defense simulation in OMNET++ and MATLAB is presented, concluding that network jamming can be efficiently prevented with the proposed prevention algorithms.

Afterwards, we investigate various works on the effects of some space design parameters (e.g., the constellation density and the inter satellite links) on metrics such as the resulted delay. [94] presents the system architecture of three constellations, highlighting their similarities and differences, estimating the total system throughput and aiming to minimize the total number of stations required to support the system throughput. The authors in [95] design a LEO/MEO double-layer network structure that can meet the service quality and reduce the complexity of the network system based on service quality indicators such as the coverage, bandwidth, delay, packet loss rate and throughput. In addition, the hop-count of existing studies in complex and computationally expensive network simulations is evaluated in [93], especially in dense constellations with large number of satellites. In [96], authors aim to minimize the number of satellites in an ultra-dense LEO satellite constellation through an optimization algorithm, while satisfying the backhaul requirement of each terrestrial-satellite terminal. Paper [8] provides a comprehensive overview of the physical and logical links, along with the essential architectural and technological components that enable the integration of LEO constellations into 5G and B5G systems and explore novel techniques to maximize the achievable data rates. In work [97] the use of ground-based relays as a substitute for ISLs is investigated to provide low-latency wide area networking and showing that even without ISLs, such networks can still beat optical fiber networks for latency. Finally, various novel work and propositions have been made as space agencies are exploring new alternatives to cope with the increasing amounts of scientific data collected in a more suitable way regarding the characteristics of the trending low-cost, high-population satellite networks. One of them is detailed in [98] proposing a broadcast-based, peer-to-peer (P2P) return model for satellite data.

In [99] authors summarize various conducted research in the area of inter-satellite communications for small satellites and provide a complete architecture based on the Open System Interconnection (OSI) model. They also present a comprehensive list of design parameters useful for achieving inter-satellite communications for multiple small satellite missions while they propose solutions for the usual presented challenges. Paper [100] provides a research of the topology and routing design in Mega-constellation networks (MCNs) through the analysis of ISL paths. They investigate the effects of constellation parameters (e.g. orbit inclination, number of planes & satellites per plane and phasing factor) on the ISL hops and propose a theoretical model to estimate the ISL hop/relay count between ground users. They experiment on a Starlink constellation in which each satellite establishes four permanent ISLs with its neighboring satellites: two intra-plane and two inter-plane ISLs. Moreover in [101], the effects of constellation size and orbit inclination on the hop-count metric are analyzed. Authors in [102] investigated the hop-count and latency between two locations and compare their performance results in different constellations with ground fiber networks. In [103] the routing issues in a Starlink constellation are studied, evaluating the end-to-end latency in relation to multipath options. Finally, [104] presents the trends and future prospects of space laser communications.

Concluding, in cases of challenging networking scenarios such as disruptive communications and delay, DTN can be exploited [45][46][38]. A detailed description of the space inter-networking issues that DTN protocols overcome is investigated in [99]. One of the scenarios involving DTN is the Ring Road Network (RRN) approach, a world-wide message-ferry network built upon LEO satellites used to transfer data [105]. Each satellite acts as a "data mule", receiving, carrying, and delivering data from places that lack Internet connectivity. Hence, transmissions are completed during an "overflight" and the additional data remain in local storage for future flights. Moreover, there is a multi-hop DTN relay system, which transmits messages from Earth to Moon via LEO satellites, reducing the end-to-end latency of the delivered message [18]. The SPICE test-bed is presented in [106], a state-of-the-art Delay Tolerant Networking test-bed for satellite and space communications deployed at the Space Internetworking Center (Greece). The core of the test-bed relies on the Bundle Protocol and its architecture has been designed to support multiple DTN implementations and a variety of underlying and overlying protocols. In [110], authors propose the Connectivity plAn Routing PrOtOcOL (CARPOOL), a reference routing protocol, exploring the deployment of DTN-capable nodes to extend free Internet coverage in metropolitan areas and in highly-dense environments. In [107], a DTN-based architecture for dense urban environments is proposed, exploiting the existing infrastructure of public transport networks to provide Internet services through delay tolerant applications without significantly affecting user experience. Paper [108] is focused on routing in space DTNs, in particular on contact graph routing (CGR) and its most representative enhancements, while the applicability and the obtained performance of the DTN protocol stack and of the CGR have been evaluated by presenting results from real experimental experiences. Finally, in [109] authors employ DTN to form an internetworking overlay that exploits the surplus capacity of last hop wireless channels in order to prolong battery life for mobile networking devices, while showing experimentally that the DTN overlay can shape traffic, allowing the wireless interface of the mobile device to switch to the sleep state during idle intervals without degrading performance.

Most of the aforementioned works do not highlight the use of measurements from real-world test-beds as indicators of signal transmissions in the OMNeT++. In addition, prior works regarding the implementation of IoT Case Study scenarios in space assisted communications are limited. Finally, the evaluation of the different parameters that affect the reliability of space-terrestrial communications - which we investigated - is significant for the design of the appropriate satellite constellation.

# C. Methodology

## I. Tools and simulation parameters

As mentioned in Section A, the main tool for the simulation of our experiments is OMNeT++. According to [111a], OMNeT++ is *"an extensible, modular, component-based C++ simulation library and framework, primarily for building network simulators"*. It has been publicly available since 1997, designed to be a general and powerful open-source discrete event simulation tool, used by academic, educational and research-oriented commercial institutions for the simulation of computer networks and distributed/parallel systems [112]. OMNeT++ is an intermittent tool between open-source, research-oriented simulation software such as Network Simulator 2 (NS2) and commercial alternatives like Optimized Network Engineering Tools (OPNET), supporting layered modules and simulating complicated objects while offering the tools to write such simulations: a graphic editor, graphic analysis tool, and parallel simulation scheme. Specific application areas are supported by simulation models and frameworks like INET, SUMO and VEINS that can be found in [111b] or in their respective GitHub repositories.

A popular framework regarding space implementations is the Open Source Satellite Simulator (OS3) [111c]. Developed by the Communication Networks Institute, TU Dortmund in 2012, it is a framework for simulating various kinds of satellite-based communications based on OMNeT++, able to *"automatically import real satellite tracks and weather data to simulate conditions at a certain point in the past or in the future, and offer powerful visualization"*. The problem with the original OS3 is the outdated documentation and compatibility with the newer versions of OMNeT++ and INET. To overcome these issues we built our model around an updated framework of OS3 [5][4]. Further details can be provided in [113] in which authors describe the design and implementation of the framework within OMNeT++ and INET. They also describe some simulation examples using LEO satellite constellations and compare their results with existing works.

The OS3 framework includes and uses the NORAD SGP4/SDP4 algorithms. These simplified perturbation models are sets of mathematical models (SGP, SGP4, SDP4, SGP8 and SDP8) for the calculation of orbital state vectors of satellites and space debris relative to the Earth-centered inertial (ECI) coordinate system. They are often referred to collectively as SGP4, due to the frequent use of that particular model. SGP4 is the most widely used analytical propagation algorithm, its source code and algorithms has been publicly available since 1980 (in the Space Track Report #3 [114]) and it offers a good trade-off regarding the running time and accuracy [115]. These models predict the effect of perturbations caused by the Earth's shape, drag, radiation, and gravitation effects from other bodies such as the sun and moon. Simplified General Perturbation (SGP) models perturbation models apply to near earth objects with an orbital period of less than 225 minutes, while Simplified Deep Space Perturbations (SDP) models refer to objects which need more than 225 minutes to complete one orbit, corresponding to an altitude of 5,877.5 km. Some important parameters for the calculations are the: *-BStar* (the drag coefficient, representing the susceptibility of an object to drag force), *-Eccentricity* (determining the orbit's shape – 0 is perfectly circular and 1 is parabolic), *-Argument of Perigee/Periapsis* (the angle between the perigee/periapsis, i.e. the point in the orbit of an object orbiting the Earth that is nearest to its center, and the ascending node, i.e. the angular position at which a celestial body passes from the southern side of a reference plane to the northern side), *-Mean Anomaly* (the fraction of an elliptical orbit's period that has elapsed since the orbiting body passed periapsis), *-Mean motion* (measured in revolutions per day), *-Revolutions at Epochs* (the number of orbits the body has made since its launch) and *–Longitude/ Right Ascension of the Ascending Node* (or RAAN i.e. the point where an object crosses the equator moving upward from south to north).

We discuss some important simulation parameters such as the *simulation time limit* which stops the simulation when time reaches a given limit. With *fullDuplex* enabled, nodes can simultaneously transmit and receive data, acting as transceivers. Furthermore, there is the *communication range* (also referred as footprint), representing the area in which the reception of transmissions produced by the transmitter is possible. Satellites create their footprint on Earth, i.e. a circular section in which communications are feasible, depending on their altitude from the Earth's surface and their elevation angle, as described in Section B.I. Our satellites are located in a constellation at *600km altitude* with a minimum *elevation angle* of 25° (like the Starlink satellites at this altitude) [116]. According to *Table 1* their (%) coverage is approximately 0,62% of the Earth's surface. Since the radius of the Earth is known ($R_e$=6378x10$^3$m) we conclude that the radius of the satellite's footprint is approximately 1008km. That radius is transformed in the 2D OMNeT++ and due to the scale of the map it represents a communication range of around 100m, a parameter that remains constant for all of our experiments as the satellites' altitude is assumed to remain stable at 600km.

Additionally, there is the *updateInterval* variable, used as granularity to signal mobility state changes. While in the updateInterval time, pings can be sent, received or discarded within the current topology. Subsequently the position of nodes on the map is updated (as MCCs are stable, only satellite positions change), unnecessary routes are deleted and routing tables are reconfigured. A small updateInterval leads to greater accuracy and smoothness of results as more alternative route paths can be identified, with the corresponding negatives including higher complexity (higher routing complexity and larger routing tables) and execution time of the experiment (specifically in denser constellations in which the runtime becomes noticeable, it takes significant time to reconfigure all the routes and build a larger graph of the network topology).

As the topology of the simulation is updated based on the updateInterval, the *Dijkstra's shortest path algorithm* is called repeatedly throughout the simulation. Edges are created for nodes located in a communication range, and edge weights are set to determine whether a connection is feasible. If connections are achievable, the propagation delay is set as the edge weight, elseways an infinite weight is given to the edge. Finally, every route is added to each node's routing tables. When the topology changes (in the next updateInterval), any relevant parameter that requires update is reconfigured (e.g., satellite position and node routing tables), while IP addresses remain untouched as they have already been set. [113]

With the *intersatelliteLink* parameter, communications among satellites can be enabled with some preconditions. Therefore, satellites placed on the same plane can communicate even without an MCC between them. While disabled, connections of the type Satellite<->MCC<->Satellite are feasible, meaning that an MCC must always exist between satellites in order to achieve communication. Finally, intersatelliteLinks only work with the updated NoradA module and not the original Norad from OS3.

The *MCC* (Mission Control Center) nodes represent wireless ground hosts which send pings as PingApps from the INET framework, the number of which is considered with the *numOfMCCs* variable. They generate ping requests and transmit the first ping at *startTime* while the *sendInterval* parameter estimates the time to wait between ping transmissions. We disable MCC to MCC connections as our work aims for a space-terrestrial infrastructure with strict communications among MCCs and satellites, or just among satellites. Regarding the positioning of MCCs, the framework is modified so that the transmission model from INET can encapsulate the real coordinates (*longitude, latitude and altitude*) of the node and calculate the actual distances and propagation delays over the 2D OMNeT++ map. The altitude of MCCs is zero but could be changed in future experiments. At the end of the simulation, we calculate transmitted and received pings, round trip times (RTT), ping losses and frequently displayed values of delay.

The *Satellites* are also wireless hosts, but mobile, orbiting the Earth in an altitude of 600km. We can build constellations using two approaches: 1) with .txt files including the current satellite constellation data and 2) creating a total new constellation based on our standards. The former is provided by the original OS3 framework, using Two-Line Element (TLE) sets of data that can be downloaded from [117] and be imported using the OS3 Norad module (presented in *Table 3*). This approach limits the scalability of the available simulation examples and makes the intersatelliteLinks parameter difficult to implement due to lack of information related to the plane of each satellite. Consequently, we use the NoradA module from the updated OS3 version [4], where we can create our own desirable satellite constellation. Therefore, the *numOfSats, planes, satPerPlane* and *intersatelliteLink* are the primary parameters included in the constellation, while others such as the *meanMotion, eccentricity, inclination* and *aerodynamic drag* are used in the SGP4 algorithm.

Finally we include no buffers in our satellites and as a result every ping that comes in queue is popped out immediately. Consequently, if simultaneously transmissions exist, an increase of the ping/RTT can be observed or even a loss of connection. An example of this issue can be seen in the *Section D.II.a*).

*Table 3: TLE data example and explanation per Line and column* [117]

| Name | NOAA 14 |
|------|---------|
| Line1 | 23455U 94089A   97320.90946019  .00000140  00000-0  10191-3 0  2621 |
| Line2 | 23455  99.0090 272.6745 0008546 223.1686 136.8816 14.11711747148495 |

| Line 1 | |
|--------|-------------|
| Column | Description |
| 01 | Line Number of Element Data |
| 03-07 | Satellite Number |
| 08 | Classification (U=Unclassified) |
| 10-11 | International Designator (Last two digits of launch year) |
| 12-14 | International Designator (Launch number of the year) |
| 15-17 | International Designator (Piece of the launch) |
| 19-20 | Epoch Year (Last two digits of year) |
| 21-32 | Epoch (Day of the year and fractional portion of the day) |
| 34-43 | First Time Derivative of the Mean Motion |
| 45-52 | Second Time Derivative of Mean Motion (Leading decimal point assumed) |
| 54-61 | BSTAR drag term (Leading decimal point assumed) |
| 63 | Ephemeris type |
| 65-68 | Element number |
| 69 | Checksum (Modulo 10)<br>(Letters, blanks, periods, plus signs = 0; minus signs = 1) |

| Line 2 | |
|--------|-------------|
| Column | Description |
| 01 | Line Number of Element Data |
| 03-07 | Satellite Number |
| 09-16 | Inclination [Degrees] |
| 18-25 | Right Ascension of the Ascending Node [Degrees] |
| 27-33 | Eccentricity (Leading decimal point assumed) |
| 35-42 | Argument of Perigee [Degrees] |
| 44-51 | Mean Anomaly [Degrees] |
| 53-63 | Mean Motion [Revs per day] |

| 64-68 | Revolution number at epoch [Revs] |
|-------|-----------------------------------|
| 69    | Checksum (Modulo 10)              |

### a) Simulation pseudocode

*While in sim-time-limit:*

*For every updateInterval window:*

*For every node event check:*

> *If NOT in communication range:*
>
>> *-(ipv4.ip): WARN: unroutable, sending icmp error ICMP_DESTINATION_UNREACHABLE, dropping packet /*
>>
>> *-(configurator): routingTable: delete routes / -Configuring all routing tables / -add routes /*
>>
>> *-(mobility, selfmsg move)* ***[//end of if not in communication range // end]***
>
> *If IN communication range: starting the simulation*
>
>> *If SENDING: the sender broadcasts the ping'x' to every node(Satellite, MCC and Sensor) in the current topology*
>>
>>> *-(app[0]): sending ping request #'x' to lower layer*
>>>
>>> *-(ipv4.ip): received, sending datagram 'ping'x'' with destination 10.0.x.x/receiver*
>>>
>>> *-(wlan[0].mac.queue): pushing packet ping'x' into the queue, popping packet ping'x' from the queue, starting transmission of ping61...*
>>>
>>> *-(wlan[0].radio): -transmission started: ID=x, TRANSMITTERID=y, changing radio transmission state from IDLE to TRANSMITTING, changing radio transmitted signal part from NONE to WHOLE,*
>>>
>>> *-transmission ended: id=x, transmitterID=y, changing radio transmission state from TRANSMITTING to IDLE, changing radio transmitted signal part from WHOLE to NONE* ***[//end]***
>
>> *If RECEIVING:*
>>
>>> *If the node is OUT of communication range (from sender's ping)->*
>>>
>>>> *-Reception started: not attempting, power = UNDETECTABLE*
>>>>
>>>> *-Reception ended: ignoring, power = UNDETECTABLE* ***[/end]***
>>>
>>> *If the node is IN communication range -> power = RECEIVABLE*
>>>
>>>> *-Reception started:, attempting, power=RECEIVABLE, transmissionID=x(same as the previous id=x), receiverID=y+1, changing radio reception state from IDLE to RECEIVING, changing radio received signal part from NONE to WHOLE,*
>>>>
>>>> *-Reception ended: successfully, power=RECEIVABLE, transmissionID=x(same as the previous id=x), receiverID=y+1, sending up, changing radio reception state from RECEIVING to IDLE* *result: if NOT DESTINATION: Not Accepted*

-*(wlan[0].mac): frame 'ping'x'' not destined to us, discarding* **[// end]**

*result: if DESTINATION: Accepted*

-*(wlan[0].mac): passing up contained packet 'ping'x'' to higher layer*

-*(ipv4.ip): received ping'x' from network / received datagram with dest=10.0.x.x*

-*(wlan[0].mac.queue): pushing packet ping'x' into the queue, popping packet ping'x' from the queue,*

*starting transmission of ping'x'...* **[//end]**

*If initial sender receive ping-reply:*

-*(wlan.mac): received ping'x'-reply from network / received datagram with dest=10.0.x.x / passing up to socket 1 / passing up to protocol icmpv4 ...*

-*(app[0]): ping reply #'x' arrived RTT = ... / jump in seq numbers assuming pings since #0 got lost /*

***[//end of If in communication range]***

***[//end of For every updateInterval window]***

***[//end of For every node event check]***

***[//end of While]***

## b) Example of transmitting ping:

- **1st broadcast (on ping61):** MCC[0] sends initial with id=53 → sat[72] receives with transmissionId=53 (accepts) → MCC[1] receives with transmissionId=53 (rejects – direct MCC-to-MCC disabled) – others reject (out of comm.range - unroutable or not destination/next hop - not destined to us)
- **2nd broadcast (on ping61):** sat[72] sends new with id=54 (has received with transmissionId=53 sent from MCC[0] with id=53) → MCC[1] receives with transmissionId=54 (accepts) → MCC[0] receives with transmissionId=54 (rejects – not destined to us)
- **3rd broadcast (on ping61-reply):** MCC[1] sends new with id=55 (has received with transmissionId=54 sent from sat[72] with id=54) → sat[72] receives with transmissionId=55 (acccepts) → MCC[0] receives with transmissionId=55 (rejects – direct MCC-to-MCC disabled)
- **4th broadcast (on ping61-reply):** sat[72] sends new with id=56 (has received with transmissionId=55 sent from MCC[1] with id=55) → MCC[0] receives with transmissionId=56 (accepts) → MCC[1] receives with transmissionId=56 (rejects – not destined to us)
- MCC[0] accepts and sends new... repeat ...

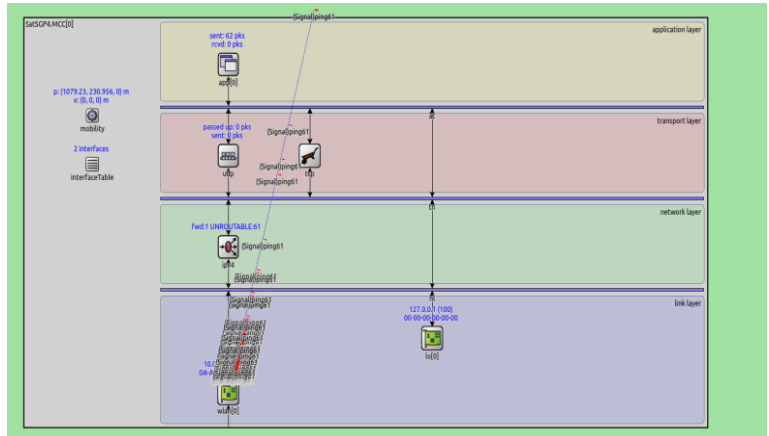

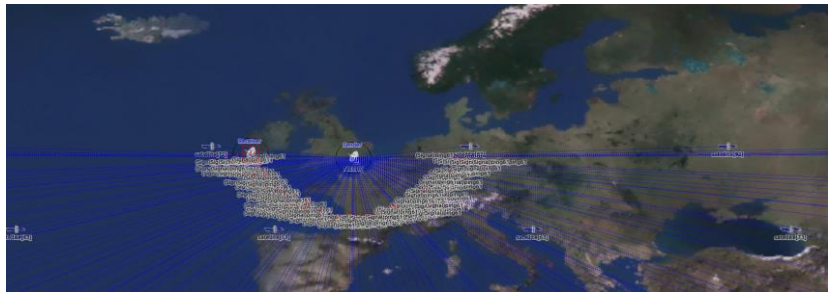

*Figure 5: 1st broadcast (on ping61): MCC[0] sends initial*

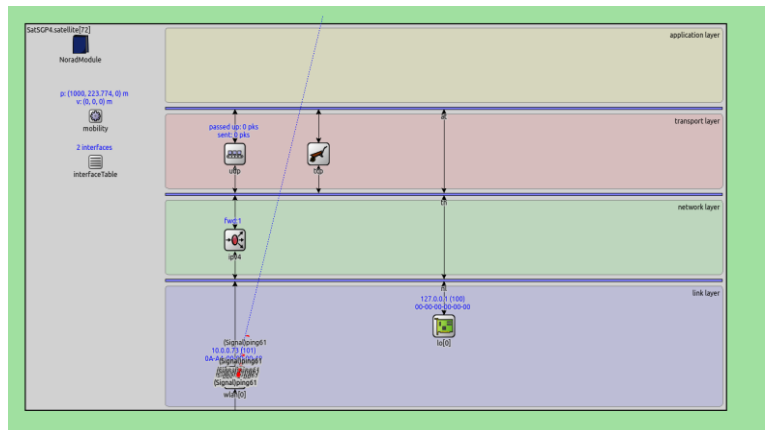

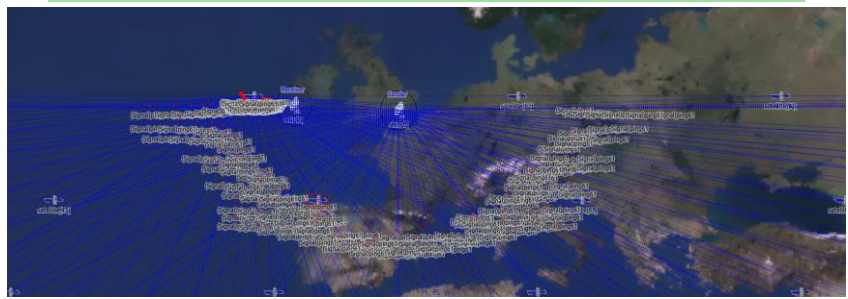

*Figure 6: 2nd broadcast (on ping61): sat[72] sends new with id=54*

## II.    Scenarios

Our experiments are presented in three groups departing from the "*Simple*", proceeding with the "*More Sophisticated*" and ending up with the "*Rewire Case Study*" scenario. In all of these groups the satellites' altitude is constant at 600km above the Earth's surface and consequently their communication range in the 2D OMNeT++ scaled map also remains unchanged with a radius of 100m. We capture the produced RTT and evaluate the reliability of the communications between Sender and Receiver with metrics such as the mean and range RTT, the most frequent presented RTT value and the ping loss.

In the "*Simple*" scenario we describe some experiments with one MCC as Sender and one as Receiver. Initially, we create a topology with these MCCs located at close distance (London-Ireland: 658,64km) and a constellation of 150 satellites (15planes X 10satellites per plane) without intersatelliteLinks. Afterwards, we experimented on a more distant communication (London-New York: 5571,97km) with a constellation of 360 satellites (6planes X 60satellites per plane) as the aforementioned constellation (of the 150 satellites) was considered insufficient, while we also placed 3 additional MCCs, representing relay ships, to strengthen the communication. We tested the functionality of the topology with both interSatelliteLinks disabled and enabled, observing that without interSatelliteLinks no pings are received as the MCC<->Satellite<->MCC pattern of communication is infeasible. On the contrary, interSatelliteLinks enable the Sender-Receiver connection approximately at the half of the simulation time limit. We attempt to improve the latter results by alternating some simulation parameters. Firstly, we add a few more MCCs in our topology (specifically 5 MCCs) and finally we increase the constellation's size. Comparing the results, we concluded that our modifications enhanced the communication's performance regarding all of the presented metrics (mean/range RTT and ping loss).

Subsequently, in the "*More Sophisticated*" scenario, we introduce some MCCs we call Sensors aiming to construct a more realistic scenario in which multiple Senders are positioned close to each other and transmit at their destinations (USA-EU: 5320,82km and EU-EU: 2881,52km). Due to their location, they utilize the same satellites to transmit data to their destination resulting in overlapping signals. We capture this issue in our first experiment in which the simultaneous transmissions of the Sensors cause ping losses, increased RTTs and constantly interrupted communications or even complete loss of connection. We assume that the same startTimes and the lack of buffers in the satellites are the primarily reasons for the mentioned issues and therefore we implement different startTimes to the Sensors so no transmission overlaps. The results confirm our hypothesis as we observe uninterrupted communications. We go one step further and successfully decrease the RTTs and ping loss, by increasing the constellation's size to 600 satellites (10planes X 60satellites per plane) and adding 2 more relay ships.

Finally, we conclude our work with an IoT Case Study. During the ReWire project [1][2][3], we have taken various sensor data from the SmartSantander test-bed. We selected 10 of those sensors and used their real measurements to create a topology and evaluate a space assisted communication infrastructure. At first, we provide a detailed presentation of the sensor's characteristics, i.e. the types of sensors and their measurements (including id, timestamps, coordinates, type and values) and import their timestamps in OMNeT++ as transmission indictors. Afterwards, we describe the topology of the experiment with the selected sensors of SmartSantander transmitting to different test-beds in EU and USA according to the real timestamps derived from a .csv file. We evaluate the mean and range RTT as well as the ping loss regarding various constellation parameters such as the number of satellites and planes, the satellites per plane and the existence of intersatelliteLinks. Eventually, the results of the experiments are compared per sensor with constellations consisting of 360,600 and 900 satellites respectively.

## III.   <u>ReWire</u>

The summary of the ReWire project is highlighted in [4]. As mentioned, *"the challenging, dynamic network conditions of Smart-Cities can be addressed with a relevant multi-protocol solution, as long as the latter deploys and configures on-demand the most appropriate protocol strategy to each node. Here, we target systematic experiments with the REWIRE platform, a relevant SDN-based management solution that improves the adaptability of our previous multi-protocol endeavours, tackling reliability and scalability issues associated with real Smart-City network conditions and IoT application patterns. We utilize existing SDN control and monitoring features, containerized implementations of Non-IP protocol stacks and anomaly detection mechanisms, as well as the novel Fed4FIRE+ test-beds, enabling realistic Smart-City based experimentation."*

Consequently, the publicly available paper regarding the ReWire project is described in [1][2][3], in which authors present an SDN-based solution to *"facilitate NDN adaptability in unstable wireless mesh networking (WMN) conditions"*. They introduce an integrated SDN-NDN deployment over WMNs, which are used to extend the communication range of IoT deployments. As link failure issues regularly arise, (e.g., due to unstable topologies) they utilize the NDN architecture to enhance WMNs and meet the IoT requirements. The proposed system is deployed, experimented and evaluated over the wiLab.1 Fed4FIRE+ test-bed [118].

In our ReWire Case Study (Section D.III), we evaluate the compatibility and reliability of sensors in real conditions, using actual smart city measurements from the SmartSntander test-bed. The purpose of this scenario was to investigate disruption cases in IoT environments (which frequently occur in WSNs) that could potentially be covered by a satellite infrastructure. We also explore the effects of some parameters (such as the number of satellites, orbits and intersatelliteLinks) in the resulted mean & range RTT and the ping loss, using the OMNet++ and the OS3 framework.

# D. Experiments and Results

## I.    "Simple" scenarios

Our experiments describe cases of a single Sender transmitting data to a Receiver. In *Table 4* we present the parameters that remain constant in all of the "Simple" scenarios, such as the simulation time limit (i.e. 1200 seconds or 20 minutes) and the sendInterval valued at 500ms, resulting in a total of 2400 transmitted pings. Additionally the updateInterval, i.e. the update topology parameter, remains unchanged (at 10 seconds), while the communication range for our satellites at 600km above the Earth is converted into a 100m footprint due to the map scale.

*Table 4: General characteristics of the group "simple experiments"*

| 1 sender – 1 receiver | |
|---|---|
| sim-time-limit | 1200s |
| sendInterval | 500ms |
| pingTx count (transmitted pings) | 1200s x 2pings/s = 2400pings |
| updateInterval | 10s |
| communicationRange | 100m |

### a)   London- SW Ireland / Without interSatelliteLinks / 150satellites

We describe an "entry level" experiment with the Sender and the Receiver located relatively close (658,64km distance). The satellite constellation consists of 150 satellites without interSatelliteLinks, as we consider the communication (between Sender & Receiver) though one satellite satisfactory for the purposes of this scenario.

*Table 5: Simulation's characteristics of the "Simple" (a) scenario*

| numOfMCCs | 2 |
|---|---|
| Sender | London |
| Receiver | SW Ireland |
| Ships | 0 |
| numOfSats | 150 |
| planes | 15 |
| satPerPlane | 10 |
| enableInterSatelliteLinks | false |

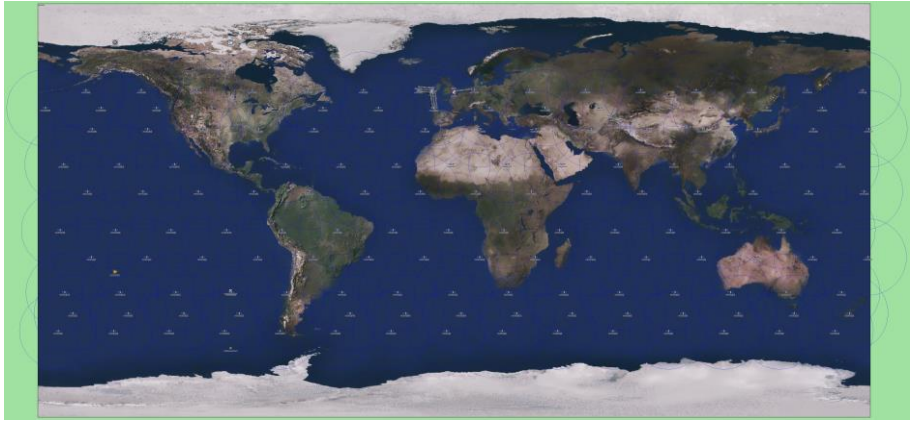

*Figure 7: Full constellation snapshot of the "Simple" (a) scenario*

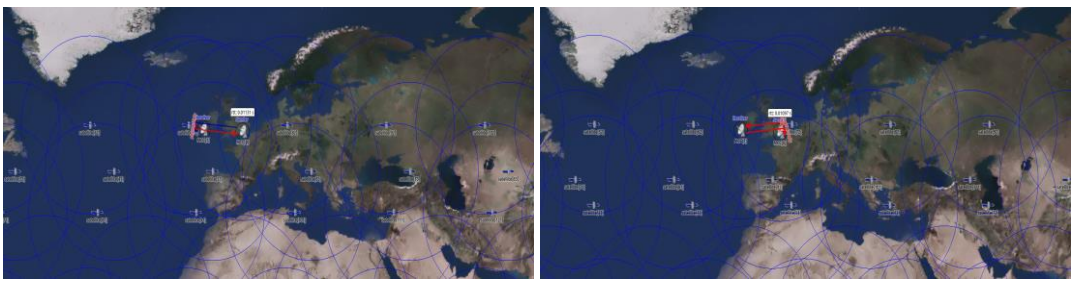

*Figure 8: Simulation snapshots of the "Simple" (a) scenario*

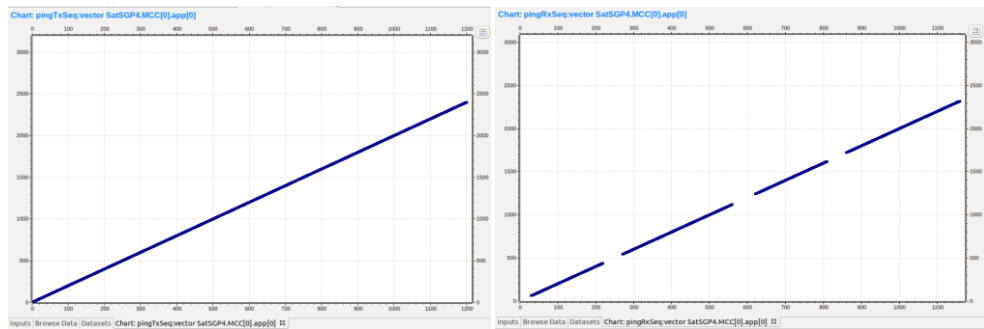

*Figure 9: a) pingRxSeq:vector (Transmitted pings/time), b) pingRxSeq:vector (Received pings/time) of the "Simple" (a) scenario*

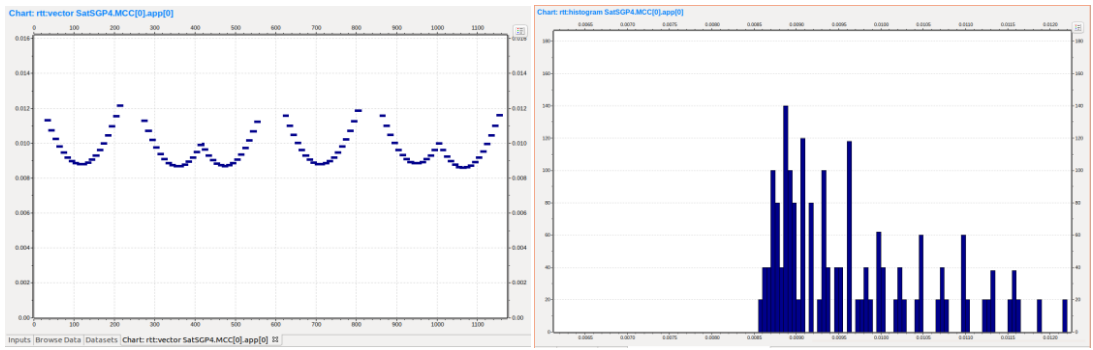

*Figure 10: a) rtt:vector (rtt/time), b) rtt:histogram (frequency/rtt) of the "Simple" (a) scenario*

*Table 6: Statistics / MCC of the "Simple" (a) scenario*

|  | Pings Transmitted | Pings Received | Range rtt (ms) | Mean rtt (ms) | Ping loss (%) | Freq / Values (times) / (ms) |
|---|---|---|---|---|---|---|
| **MCC[0]** | 2400 | 1936 | 8,15-12,2 | 9,65 | 19,33 | 140 / 8,9 |

**<u>Results:</u>** While MCCs are in the satellites' footprint range, communications of the pattern MCC(Sender)<->Satellite<->MCC(Receiver) are created. During that connection, a new satellite in the communication range can interfere and replace the previous if it provides a lower RTT (due to its position on the map). Moreover, we observe interruptions of the communication in *Figures 9.b & 10.a* (Received pings/time and RTT/time), primarily explained by the lost of coverage. The resulted ping loss is calculated by dividing the Transmitted with the Received pings and valued at 19,33%. Examining the rtt:vector (*Figure 10.a*) we notice a periodicity with two convex curves repeating. The first represents the path of the satellite as it travels between the Sender and the Receiver. Therefore, when the satellite is located at the middle of the two MCCs, the minimum RTT value is presented. The second curve follows the first, until another satellite appears, with which connection will be achieved after creating conditions of smaller delay. The two curves are repeated until the end of the simulation. In *Table 6* we observe an RTT range between 8,15-12,2ms and a mean RTT of 9,65ms, while the 8,9ms of RTT is the most frequently presented value ( existing 140 times).

b) <u>**London-New York / With&Without interSatelliteLinks / With Ships / 360satellites**</u>

The current simulation example aims to connect Europe with USA using a satellite constellation and some "Ships" that operate as relay MCCs. The constellation consists of 360 satellites as we considered the aforementioned 150 satellites constellation insufficient for feasible communications. In addition, a total of 5 MCCs (with the 3 MCCs working as Ship relays) is used, including the Sender and the Receiver (5571,97km distance). The scenario is evaluated with interSatelliteLinks both disabled and enabled.

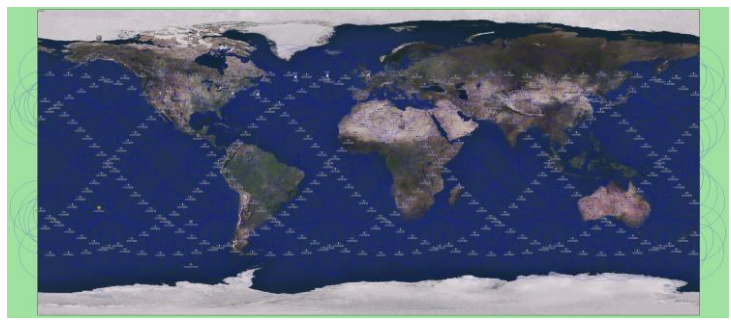

*Figure 11: Full constellation snapshot of the "Simple" (b) scenario*

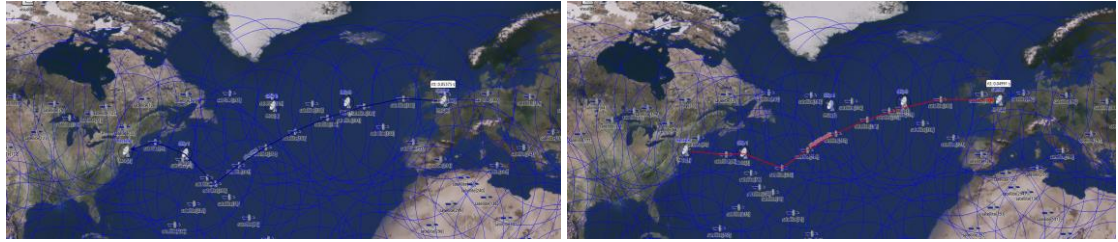

*Figure 12: Simulation snapshots of the "Simple" (b) scenario*

*Table 7: Simulation's characteristics of the "Simple" (b) scenario*

| numOfMCCs | 5 |
|---|---|
| Sender | London |
| Receiver | New York |
| Ships | 3 |
| numOfSats | 360 |
| planes | 6 |
| satPerPlane | 60 |
| enableInterSatelliteLinks | true |

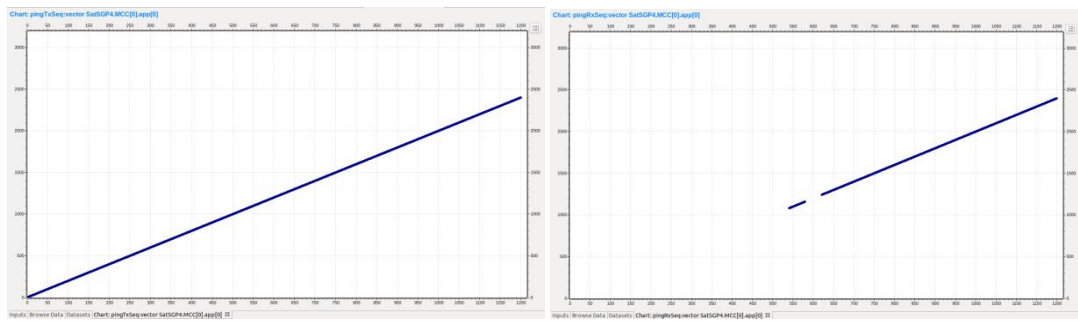

*Figure 13: a) pingRxSeq:vector (Transmitted pings/time), b) pingRxSeq:vector (Received pings/time) of the "Simple" (b) scenario*

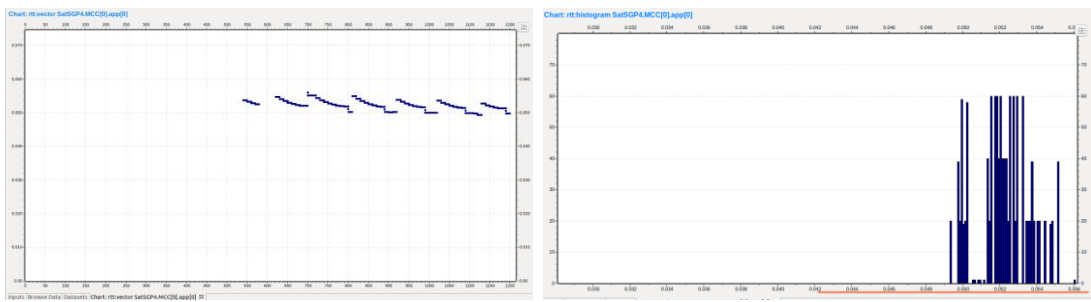

*Figure 14: a) rtt:vector (rtt/time), b) rtt:histogram (frequency/rtt) of the "Simple" (b) scenario*

*Table 8: Statistics / MCC of the "Simple" (b) scenario*

| | Pings Transmitted | Pings Received | Range rtt (ms) | Mean rtt (ms) | Ping loss (%) | Freq / Values (times) / (ms) |
|---|---|---|---|---|---|---|
| **MCC[0]** | 2400 | 1238 | 49,3-56,1 | 52,22 | 48,41 | 60 / 52-53 |

**Results:** When we disable interSatelliteLinks, we observe the simulation (1200s) end without any communication interactions – no pings transmitted – and therefore no produced RTT. We justify the results by the lack of available MCC<->Satellite<->MCC routes, as intersatelliteLinks demand an MCC between each satellite connection (see Section C.I).

With the interSatelliteLink parameter enabled, we also notice no results at the beginning, until approximately the 550[th] second, after which a valid route between Sender and Receiver is created. We should highlight that Ship1/MCC[2] is used as a relay for the functionality of intersatelliteLinks, connecting satellites from different planes and achieving communication (see *Figure 12* in which although satellite[180] and satellite[73] are located in different planes they are connected through MCC[2]). In the rtt:vector (*Figure 14.b*) we notice similar curves with their differences explained by the position of satellites. In *Table 8* we observe the ping loss (at 48,41%), the RTT range (49,3-56,1ms), the mean RTT (52,22ms) and the maximum RTT frequency (60 times) for multiple RTT values between 52-53ms.

### c) London-New York / With&Without interSatelliteLinks / With more Ships / 360satellites

In "c" we run two simulations using the same satellite constellation and Sender-Receiver as in "b", adding more MCCs between them. We aim to increase the received pings and enhance the duration of feasible connections while we attempt to achieve communication with intersatelliteLinks disabled and highlight some observations.

#### 1. *Without intersatelliteLinks*

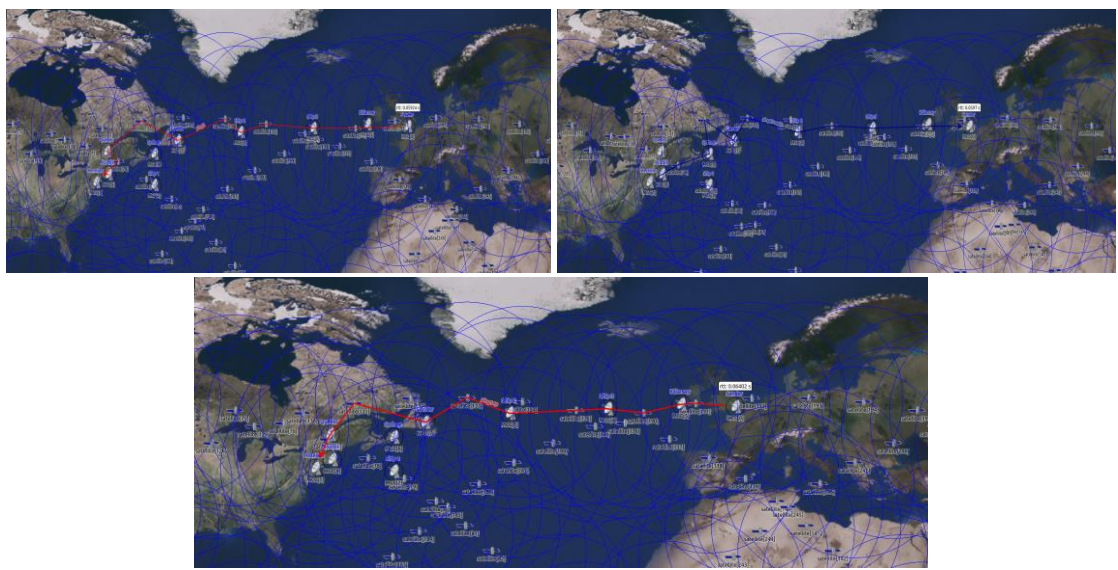

*Figure 15: Simulation snapshots of the "Simple" (c.1) scenario*

*Table 9: Simulation's characteristics of the "Simple" (c.1) scenario*

| numOfMCCs | 10 |
|---|---|
| Sender | London |
| Receiver | New York |
| Ships | 3 |
| numOfSats | 360 |
| planes | 6 |
| satPerPlane | 60 |
| enableInterSatelliteLinks | false |

*Table 10: Statistics / MCC of the "Simple" (c.1) scenario*

| | Pings Transmitted | Pings Received | Range rtt (ms) | Mean rtt (ms) | Ping loss (%) | Freq / Values (times) / (ms) |
|---|---|---|---|---|---|---|
| **MCC[0]** | 2400 | 2400 | 50,4-65,4 | 61,24 | 0 | 500 / 64 |

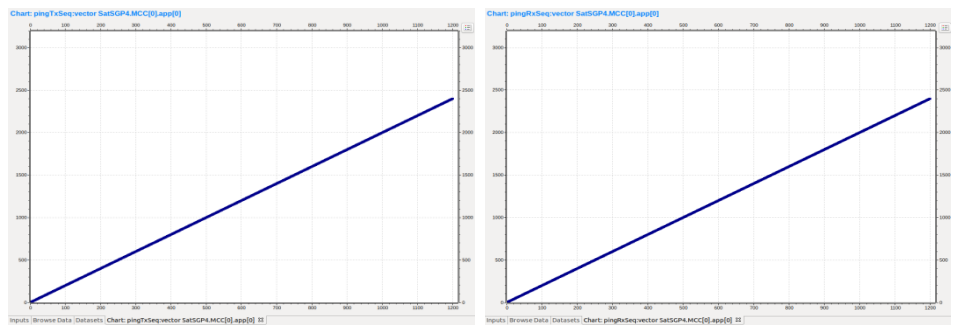

*Figure 16: a) pingRxSeq:vector (Transmitted pings/time), b) pingRxSeq:vector (Received pings/time) of the "Simple" (c.1) scenario*

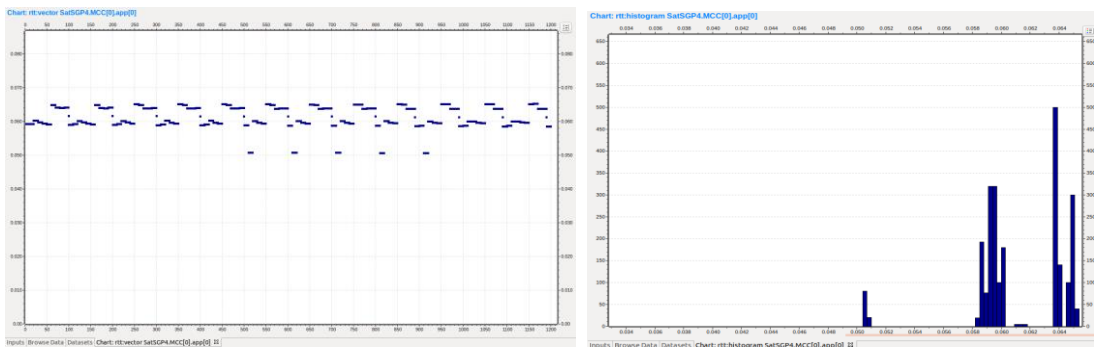

*Figure 17: a) rtt:vector (rtt/time), b) rtt:histogram (frequency/rtt) of the "Simple" (c.1) scenario*

**Results:** The continuous communication between Sender & Receiver is achieved without any missing pings. We notice an improvement of the communication with minimized ping loss, which was the goal of the experiment. We also observe a periodicity of the RTT curves (*Figure 17.a*) with shorter periods and more repetitions, while most values are between 60 and 65ms. In *Table 11* we notice an RTT range between 50,5-65,4ms and a mean RTT of 61,24ms. The maximum RTT frequency is 500 at 64ms meaning that this RTT value is presented the most times.

Comparing the results of "b" with "c" (where more MCCs exist but intesatelliteLinks) we observe uninterrupted communication even before the 550th second in contrast with "b". As a tradeoff of the continuous communication, we detect increased values both in the range and mean RTT but also in the frequency of the values as presented in the *Table 11*.

*Table 11: Comparison Statistics / MCC of the "Simple" (b & c.1) scenarios*

|  | Pings Transmitted | Pings Received | Range rtt (ms) | Mean rtt (ms) | Ping loss (%) | Freq / Values (times) / (ms) |
|---|---|---|---|---|---|---|
| "b" | 2400 | 1238 | 49,3-56,1 | 52,22 | 48,41 | 60 / 52-53 |
| "c.1" | 2400 | 2400 | 50,4-65,4 | 61,24 | 0 | 500 / 64 |

### 2. *With intersatelliteLinks*

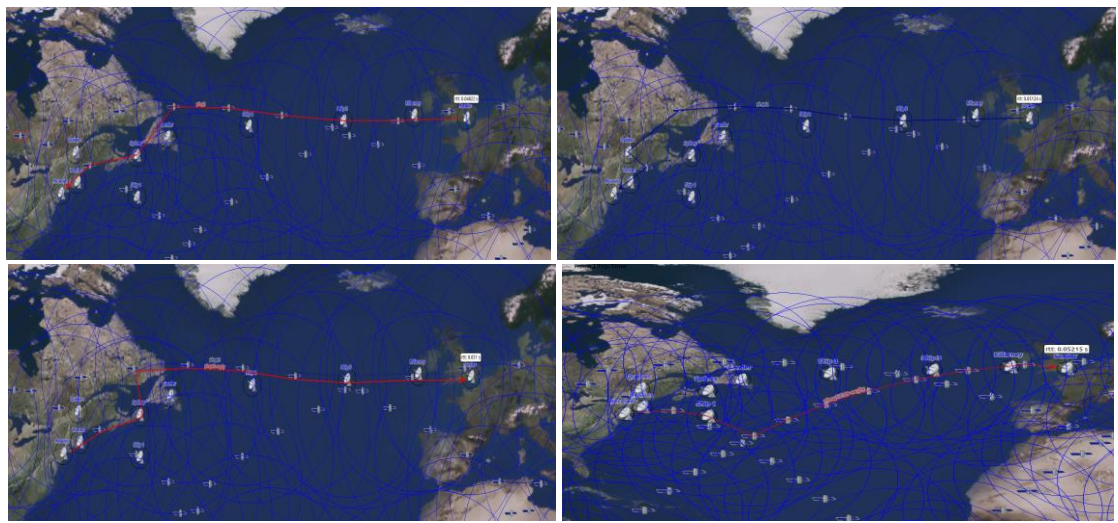

*Figure 18: Simulation snapshots of the "Simple" (c.2) scenario*

*Table 12: Simulation's characteristics of the "Simple" (c.2) scenario*

| numOfMCCs | 10 |
|---|---|
| Sender | London |
| Receiver | New York |
| Ships | 3 |
| numOfSats | 360 |
| Planes | 6 |
| satPerPlane | 60 |
| enableInterSatelliteLinks | true |

*Table 13: Statistics / MCC of the "Simple" (c.2) scenario*

| | Pings Transmitted | Pings Received | Range rtt (ms) | Mean rtt (ms) | Ping loss (%) | Freq / Values (times) / (ms) |
|---|---|---|---|---|---|---|
| **MCC[0]** | 2400 | 2400 | 47,4-52,4 | 50,6 | 0 | 230 / 52,2 |

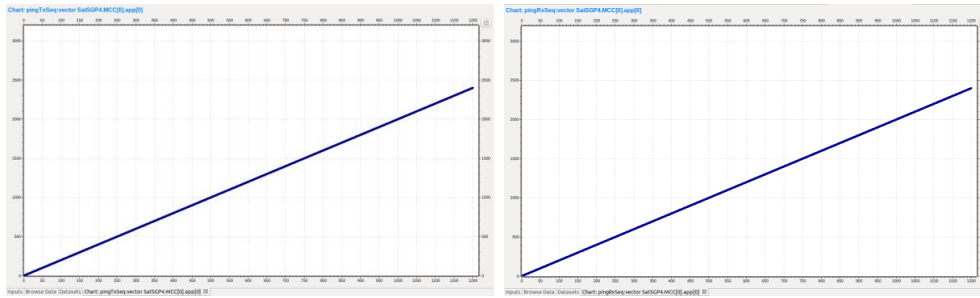

*Figure 19: a) pingRxSeq:vector (Transmitted pings/time), b) pingRxSeq:vector (Received pings/time) of the "Simple" (c.2) scenario*

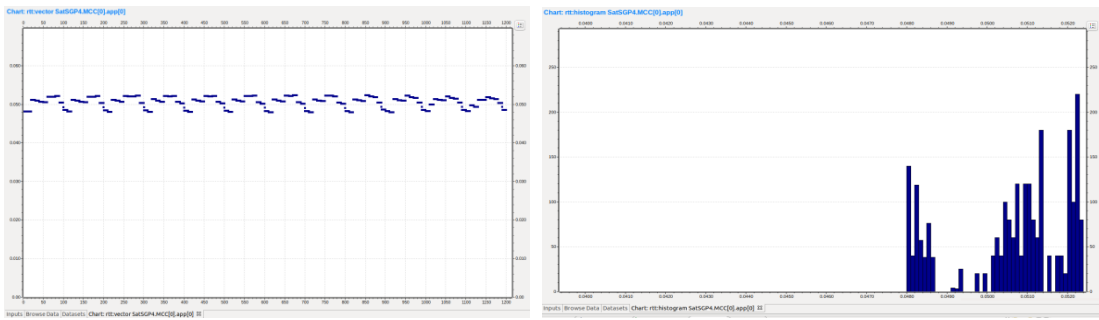

*Figure 20: a) rtt:vector (rtt/time), b) rtt:histogram (frequency/rtt) of the "Simple" (c.2) scenario*

**Results:** Enabling the intersatelliteLink parameter also results in a continuous communication. There are even shorter periods of RTT and more repetitions while most values are close to 50ms (see *Figure 20.a*) and the mean RTT is 50,6ms (see *Table 13*). Additionally, in the histogram tab (*Figure 20.b*) we observe the RTT range between 47,4-52,4ms and the maximum RTT frequency (230) with the RTT value at 52,2ms.

We capture the differences in the range and mean RTT and in the frequency of the values (*Table 14*). Comparing the outcome of the experiments of "c" we notice that in every metric we have improved results when intersatelliteLinks are enabled. We observe more compressed RTT ranges with differences of the order of 5ms (while in MYExperiment1_1 it reaches 15ms), significantly lower mean RTT (50,6ms from 61,24ms) and the most frequent displayed value to be 11,8ms smaller. The aforementioned results are explained considering the way intersatelliteLinks function, as they connect satellites of the same plane and minimize the satellite<->MCC exchanges.

*Table 14: Comparison Statistics / MCC of the "Simple" (b,c.1 & c.2) scenarios*

|        | Pings Transmitted | Pings Received | Range rtt (ms) | Mean rtt (ms) | Ping loss (%) | Freq / Values (times) / (ms) |
|--------|-------------------|----------------|----------------|---------------|---------------|------------------------------|
| "b"    | 2400              | 1238           | 49,3-56,1      | 52,22         | 48,41         | 60 / 52-53                   |
| "c.1"  | 2400              | 2400           | 50,4-65,4      | 61,24         | 0             | 500 / 64                     |
| "c.2"  | 2400              | 2400           | 47,4-52,4      | 50,6          | 0             | 230 / 52,2                   |

### d) **London-New York / With interSatelliteLinks / Without Ships / 600satellites**

This simulation investigates an alternative way to replace the extra MCCs we used in our previous experiments (considering availability and cost issues) which involves an increase of the satellites' number. We build a constellation of 600 satellites, connecting the Sender and the Receiver only through satellites (no intermittent MCCs) with intersatelliteLinks enabled.

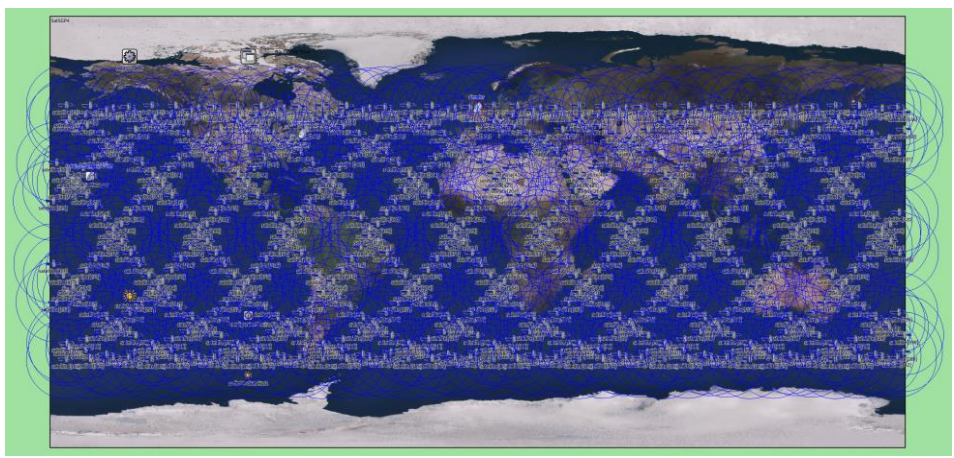

*Figure 21: Full constellation snapshot of the "Simple" (d) scenario*

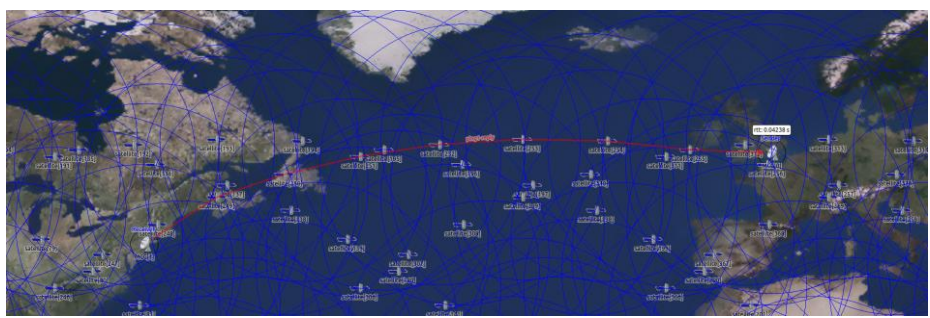

*Figure 22: Simulation snapshots of the "Simple" (d) scenario*

*Table 15: Simulation's characteristics of the "Simple" (d) scenario*

| numOfMCCs | 2 |
|---|---|
| Sender | London |
| Receiver | New York |
| Ships | 0 |
| numOfSats | 600 |
| Planes | 10 |
| satPerPlane | 60 |
| enableInterSatelliteLinks | true |

*Table 16: Statistics / MCC of the "Simple" (d) scenario*

| | Pings Transmitted | Pings Received | Range rtt (ms) | Mean rtt (ms) | Ping loss (%) | Freq / Values (times) / (ms) |
|---|---|---|---|---|---|---|
| **MCC[0]** | 2400 | 2400 | 42-44,65 | 43,4 | 0 | 195 / 42,3 |

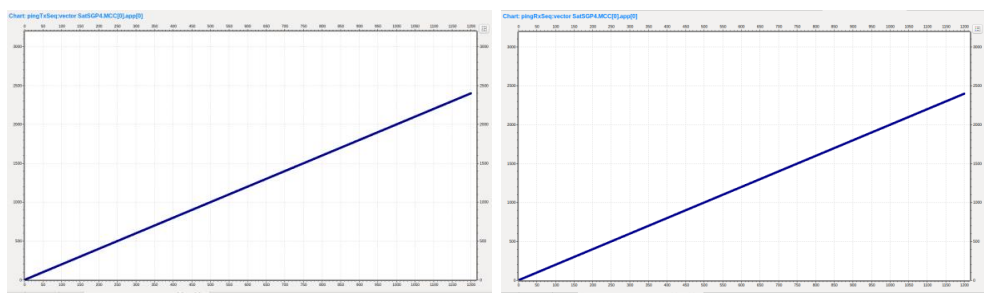

*Figure 23: a) pingRxSeq:vector (Transmitted pings/time), b) pingRxSeq:vector (Received pings/time) of the "Simple" (d) scenario*

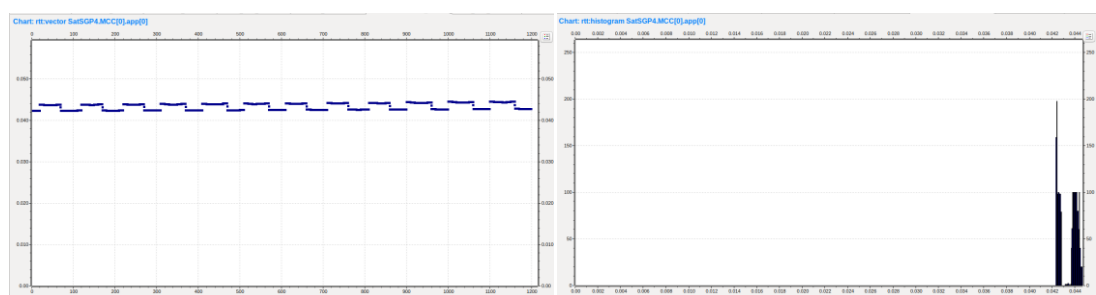

*Figure 24: a) rtt:vector (rtt/time), b) rtt:histogram (frequency/rtt) of the "Simple" (d) scenario*

*Table 17: Comparison Statistics / MCC of the "Simple" (b, c.1, c.2 & d) scenarios*

| | Pings Transmitted | Pings Received | Range rtt (ms) | Mean rtt (ms) | Ping loss (%) | Freq / Values (times) / (ms) |
|---|---|---|---|---|---|---|
| **"b"** | 2400 | 1238 | 49,3-56,1 | 52,22 | 48,41 | 60 / 52-53 |
| **"c.1"** | 2400 | 2400 | 50,4-65,4 | 61,24 | 0 | 500 / 64 |
| **"c.2"** | 2400 | 2400 | 47,4-52,4 | 50,6 | 0 | 230 / 52,2 |
| **"d"** | 2400 | 2400 | 42-44,65 | 43,4 | 0 | 195 / 42,3 |

**Results:** The continuous communication is preserved as in experiment "c". Observing the rtt:vector (*Figure 24.a*) we notice the smoothest graphic representation so far, with similar RTTs close to 43ms. In *Table 17* the range RTT is between 42ms and 44,65ms, the most compressed range in our scenarios and the mean RTT counted at 43,4ms while the maximum RTT frequency is 195 with the RTT value at 42,3ms.

Comparing the results of "d" and "c" (*Table 17*) we notice that the increase of the satellites in the constellation even without the intermediate MCCs offers the best measurements. We observe a 2,65ms difference in the RTT range compared to the previous 5ms, as well as 7,2ms lower mean RTT and the most frequently displayed RTT value being approximately 10ms smaller.

## II.   "More sophisticated" scenarios

In this group of experiments we introduce some MCCs we call "Sensors" attempting to simulate and evaluate the communication of real sensors that sense, record and communicate data to ground stations around the globe through a space-terrestrial infrastructure. The difference with the aforementioned "Simple" scenarios includes the placement of multiple Sensors/Senders in a specific area. Considering that sensors may be located in a city or even closer, they primarily use the same satellite(s) to transmit data to their destination(s). As we didn't incorporate any queue buffers in our satellites, received pings are popped immediately and forwarded to the next satellite or MCC of the route. Although until now there were no problems in this procedure, now multiple Senders exist, and issues regarding the increased delay and the number of lost pings arise.

### a) EU-USA / 10MCCs (Sensors) / With intersatelliteLinks / With Ships / 360satellites

The topology of the scenario contains 10 Sensors, 5 in EU and 5 in USA which transmit pings to corresponding MCCs, with slight different startTimes in order to visually distinguish the source of the different pings (5320,82km average distance). We observe the issue of simultaneous transmissions in *Figure 25* in which multiple overlapping pings occur, primarily due to the utilization of the same satellites.

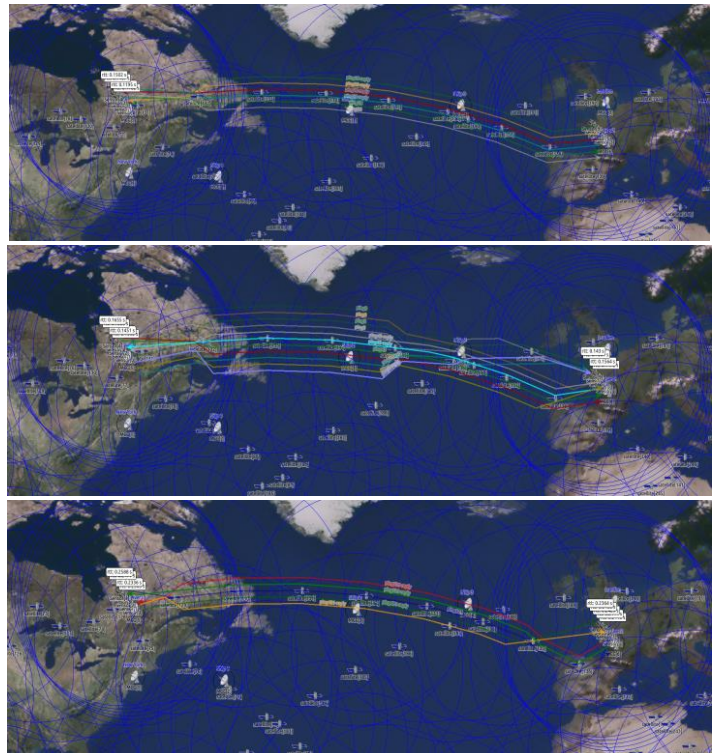

*Figure 25: Simulation snapshots of the "More Sophisticated" (a) scenario*

*Table 18: Simulation's characteristics of the "More Sophisticated" (a) scenario*

| numOfMCCs | 7 |
|---|---|
| Senders | Sensor[0..4], Sensor[5..9] |
| Receivers | MCC[6], MCC[5] |
| Ships | 3 |
| numOfSens | 10 |
| numOfSats | 360 |
| Planes | 6 |
| satPerPlane | 60 |
| enableInterSatelliteLinks | true |
| updateInterval | 10s |
| startTime | 0s, 20s |
| sendInterval | 500ms |
| sim-time-limit | 307s |

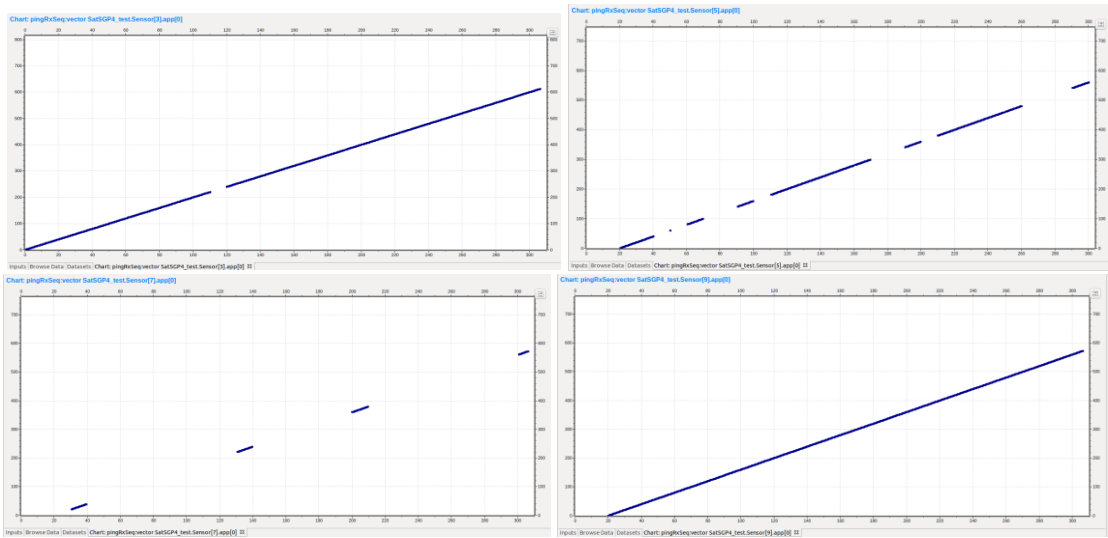

*Figure 26: pingRxSeq:vector (Received pings/time) for sensors 3,5,7,9 of the "More Sophisticated" (a) scenario*

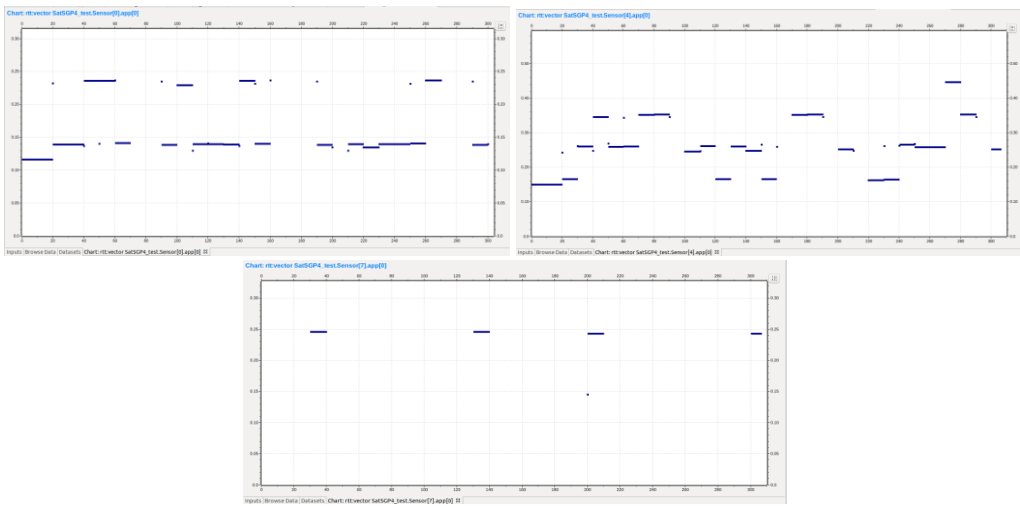

*Figure 27: rtt:vector (rtt/time) for sensors 0,4,7 of the "More Sophisticated" (a) scenario*

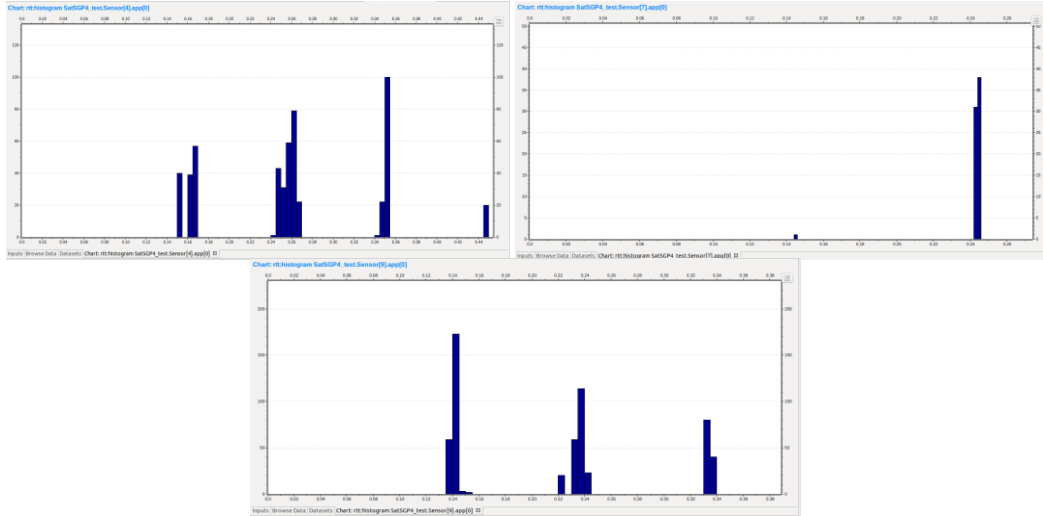

*Figure 29: rtt:histogram (frequency/rtt) for sensors 4,7,9 of the "More Sophisticated" (a) scenario*

**Results:** As highlighted in *Figures 26 & 27* and *Table 19*, the simultaneous transmissions of multiple Sensors result in interrupted communications with ping losses, lost of connectivity and high RTTs. Although Sensors are not significantly distant (in reference to the complete map), they present different results (ping loss and range & mean RTT) which cannot be justified from the distance factor (which influences the RTT outcomes) but only from the multiple and simultaneous transmissions. We observe ping losses varying from 3,75% to 87,78%, with the most impressive result the connection loss of Sensor[2] throughout the entire simulation. In addition, the range and mean RTTs are unusually increased, even quadrupled in certain cases (e.g., Sensor[4]). The "chaotic" results can be spotted in the rtt:vector (Figure 27) and the summarized *Table 19*.

To verify the effect of the multiple transmissions we run a simulation with two Sensors at the exact locations of Sensor[2] and Sensor[5], resulting in continuous connection and an average RTT of 42,5ms&41,6ms respectively (see *Figures 30&31*).

*Table 19: Statistics / MCC of the "More Sophisticated" (a) scenario*

|           | Pings sent | Pings received | Range rtt (ms) | Mean rtt (ms) | Ping loss (%) |
|-----------|------------|----------------|----------------|---------------|---------------|
| **Sensor[0]** | 613 | 442 | 86-268 | 159,5 | 27,89 |
| **Sensor[1]** | 613 | 317 | 88-272 | 169,4 | 48,28 |
| **Sensor[2]** | 613 | 0 | 0 | 0 | 100,00 |
| **Sensor[3]** | 613 | 590 | 95-395 | 243,6 | 3,75 |
| **Sensor[4]** | 613 | 514 | 100-450 | 261,4 | 16,15 |
| **Sensor[5]** | 573 | 335 | 118-342 | 221,1 | 41,53 |
| **Sensor[6]** | 573 | 325 | 95-390 | 247,4 | 43,28 |
| **Sensor[7]** | 573 | 70 | 118-272 | 243,0 | 87,78 |
| **Sensor[8]** | 573 | 350 | 124-276 | 196,2 | 38,91 |
| **Sensor[9]** | 573 | 573 | 90-385 | 217,8 | 0 |

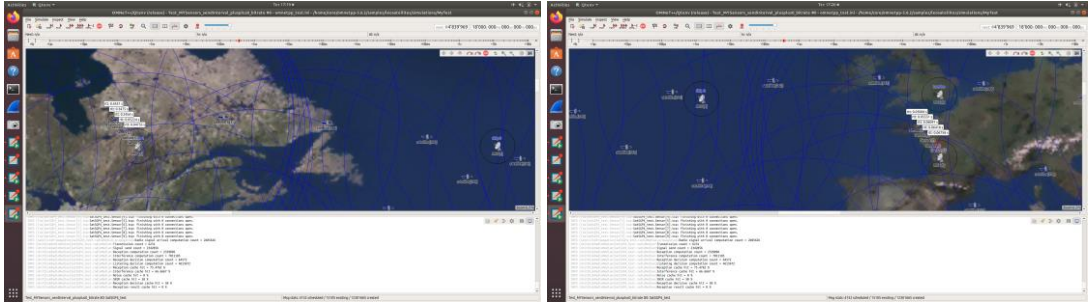

*Figure 30: RTT simultaneous transmission problem snapshot*

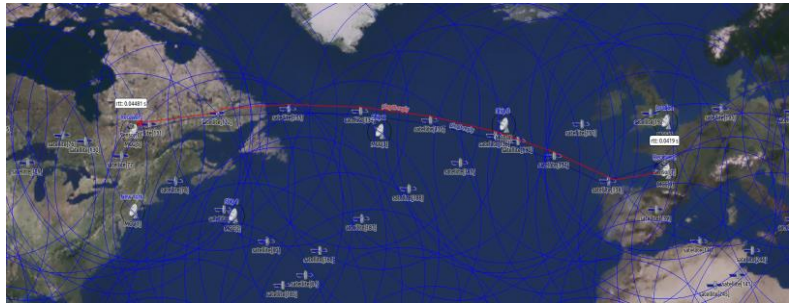

*Figure 31: Simulation snapshots of "normal" RTT tested*

**b)** **EU-USA & EU-EU / 16MCCs (Sensors) / With intersatelliteLinks / With Ships / Different startTimes / 360 & 600 satellites**

We attempt to solve the aforementioned issues of the simultaneous transmissions (described in "a") by implementing different startTimes to the Sensors so that no transmission overlap. Thus, every Sensor transmits with 10 seconds difference without interrupting the other Sensors, even if they use the same satellites.

We run two examples differentiating only the number of satellites in the constellation, the first with 360 satellites (6 satellites X 60satsperplane) and the second with 600 (10 satellites X 60satsperplane). We place 16 Sensors in the topology – 5 in USA transmitting pings to an MCC in EU, 5 in EU following the reverse path and 6 in an EU<–>EU communication. Additionally, we include 2 more MCCs/Ships – 5 in total, to enable some alternative routes among satellites of different planes with the intersatelliteLink parameter. We also set the sendInterval to 300 seconds (or 5 minutes) and the updateInterval to 100 seconds as a more fitting and less time consuming setting. Finally, the simulation time limit is expanded to 24hours to demonstrate a complete day of transmissions (EU–USA: 5320,82km average distance / EU–EU: 2881,52km distance).

**1. Satellite constellation with 360 satellites (6 satellites X 60satsperplane)**

*Figure 32: Full constellation snapshot of the "More Sophisticated" (b.1) scenario*

*Figure 33: Simulation snapshots of the "More Sophisticated" (b.1) scenario*

*Table 20: Simulation's characteristics of the "More Sophisticated" (b.1) scenario*

| | |
|---|---|
| numOfMCCs | 11 |
| Senders | Sensor[0..4], Sensor[5..9], Sensor[10..12], Sensor[13..15] |
| Receivers | MCC[6], MCC[5] MCC[7], MCC[8] |
| Ships | 5 |
| numOfSens | 16 |
| numOfSats | 360 |
| Planes | 6 |
| satPerPlane | 60 |
| enableInterSatelliteLinks | true |
| updateInterval | 100s |
| startTime | [0s, 10s, 20s, 30s, 40s], [200s, 210s, 220s, 230s, 240s], [610s, 620s, 630s], [810s, 820s, 830s] |
| sendInterval | 300s |
| sim-time-limit | 24h |

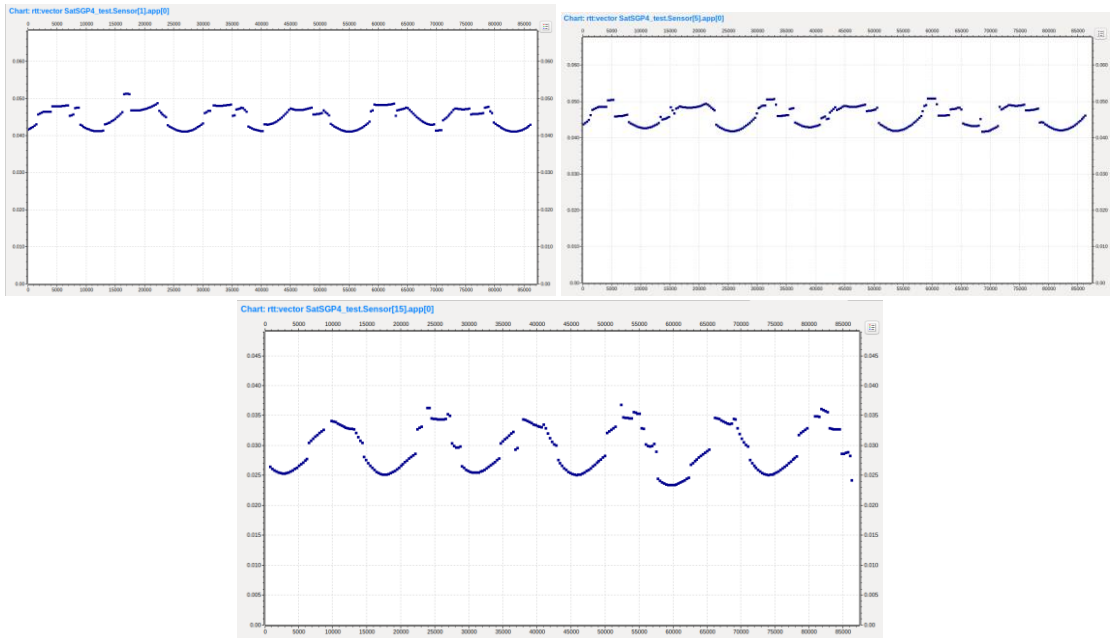

*Figure 34: rtt:vector (rtt/time) for sensors 1,5,15 of the "More Sophisticated" (b.1) scenario*

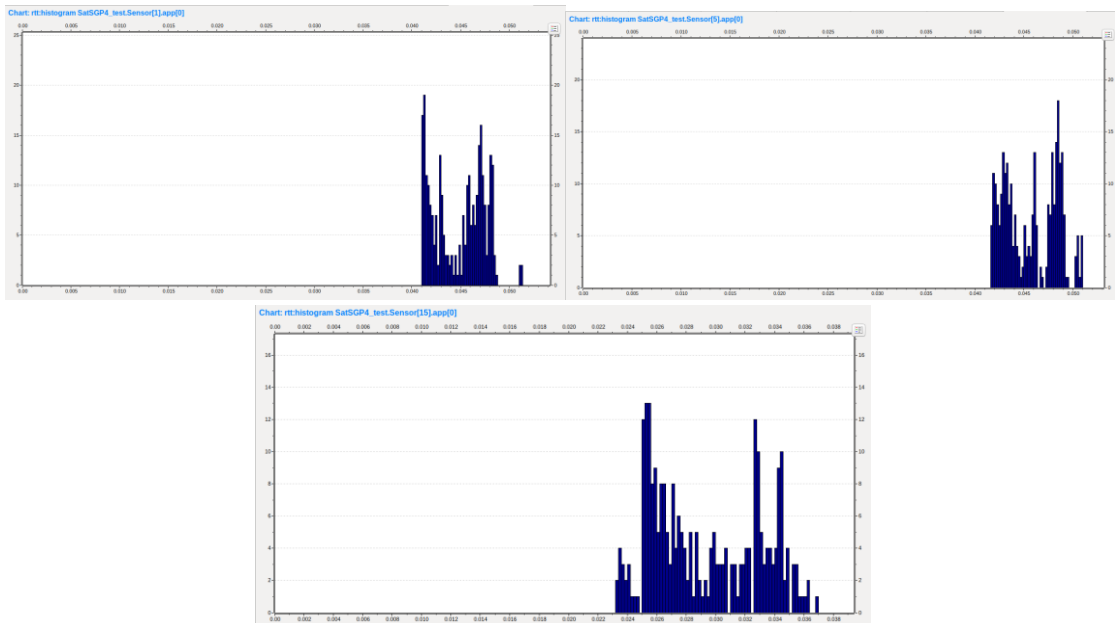

*Figure 35: rtt:histogram (frequency/rtt) for sensors 1,5,15 of the "More Sophisticated" (b.1) scenario*

*Table 21: Statistics / MCC of the "More Sophisticated" (b.1) scenario*

|  | Pings sent | Pings received | Range rtt (ms) | Mean rtt (ms) | Ping loss (%) |
|---|---|---|---|---|---|
| Sensor[0] | 288 | 85 | 38,6-49,8 | 44,7 | 70,48 |
| Sensor[1] | 288 | 287 | 38,4-54 | 44,8 | 0,34 |
| Sensor[2] | 288 | 288 | 38,6-54 | 45,2 | 0 |
| Sensor[3] | 288 | 288 | 38,8-54,2 | 45,6 | 0 |
| **Sensor[4]** | 288 | 288 | 39-54,8 | 46 | 0 |
| **Sensor[5]** | 288 | 288 | 39,4-53 | 45,7 | 0 |
| **Sensor[6]** | 288 | 288 | 37-49,4 | 43 | 0 |
| **Sensor[7]** | 288 | 288 | 36,8-48,2 | 42,4 | 0 |
| **Sensor[8]** | 288 | 288 | 36,6-47,2 | 41,8 | 0 |
| **Sensor[9]** | 288 | 288 | 36,4-46,8 | 41,4 | 0 |
| **Sensor[10]** | 286 | 277 | 19,5-42 | 29,6 | 3,14 |
| **Sensor[11]** | 286 | 284 | 19,5-42 | 29,6 | 0,69 |
| **Sensor[12]** | 286 | 286 | 19,5-42 | 29,5 | 0 |
| **Sensor[13]** | 286 | 272 | 19,5-41,5 | 29,3 | 4,89 |
| **Sensor[14]** | 286 | 272 | 20-39,8 | 29,3 | 4,89 |
| **Sensor[15]** | 286 | 272 | 22,2-39,2 | 29,3 | 4,89 |

**Results:** Observing the results of this scenario we confirm that we resolved the corresponding issues of the simultaneous transmissions by alternating the parameters described at the beginning of the experiment. Not only is the communication uninterrupted and the ping loss minimized but also the rtt:vector diagrams (*Figure 34*) look much smoother, albeit discontinuous, with fluctuations and curves that tend to show periodicity, in contrast to the "chaotic" results of "a". Finally, the range & mean RTT (see *Table 21*) present normal values especially in comparison with "a", where we detected tripled or even quadrupled RTT values.

2. *Satellite constellation with 600 satellites (10planes X 60satperplane)*

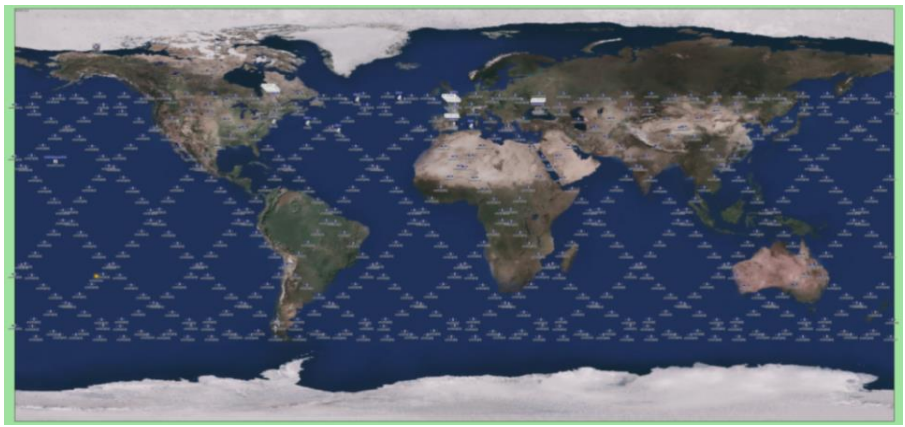

*Figure 36: Full constellation snapshot of the "More Sophisticated" (b.2) scenario*

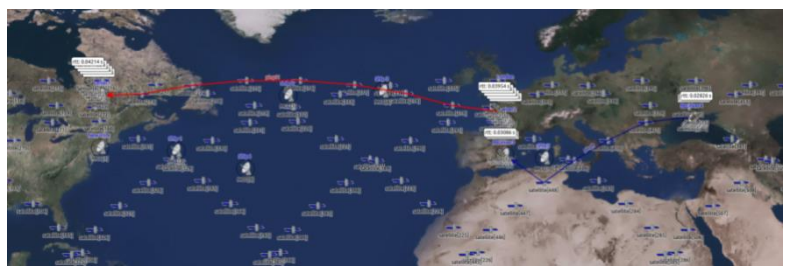

*Figure 37: Simulation snapshots of the "More Sophisticated" (b.2) scenario*

*Table 22: Simulation's characteristics of the "More Sophisticated" (b.2) scenario*

| numOfMCCs | 11 |
|---|---|
| Senders | Sensor[0..4], Sensor[5..9], Sensor[10..12], Sensor[13..15] |
| Receivers | MCC[6], MCC[5] MCC[7], MCC[8] |
| Ships | 5 |
| numOfSens | 16 |
| numOfSats | 600 |
| Planes | 10 |
| satPerPlane | 60 |
| enableInterSatelliteLinks | true |
| updateInterval | 100s |
| startTime | [0s, 10s, 20s, 30s, 40s], [200s, 210s, 220s, 230s, 240s], [610s, 620s, 630s], [810s, 820s, 830s] |
| sendInterval | 300s |
| sim-time-limit | 24h |

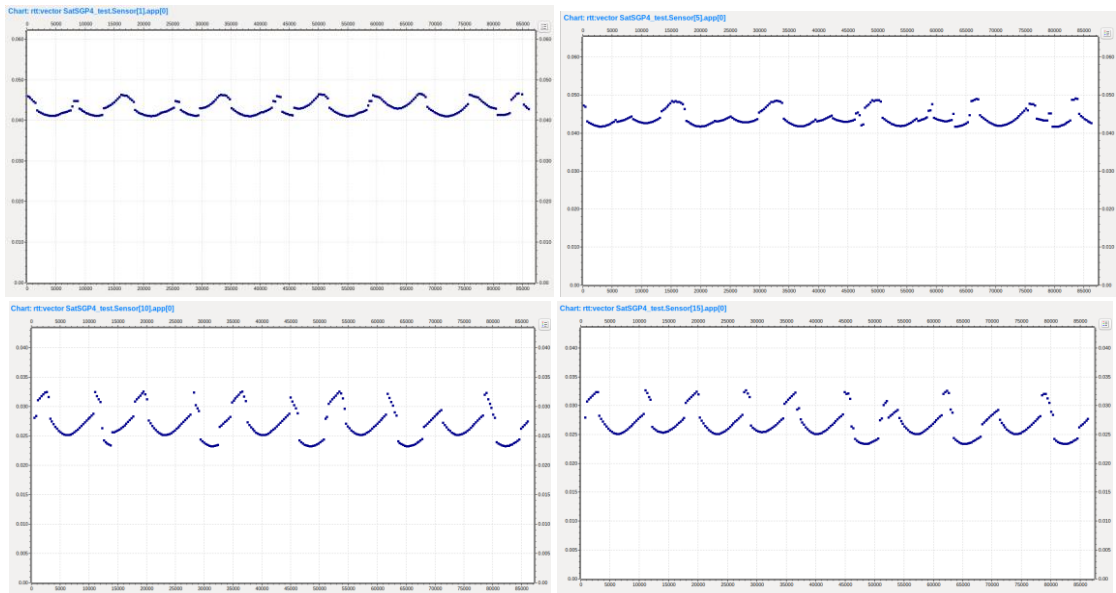

*Figure 38: rtt:vector (rtt/time) for sensors 1,5,10,15 of the "More Sophisticated" (b.2) scenario*

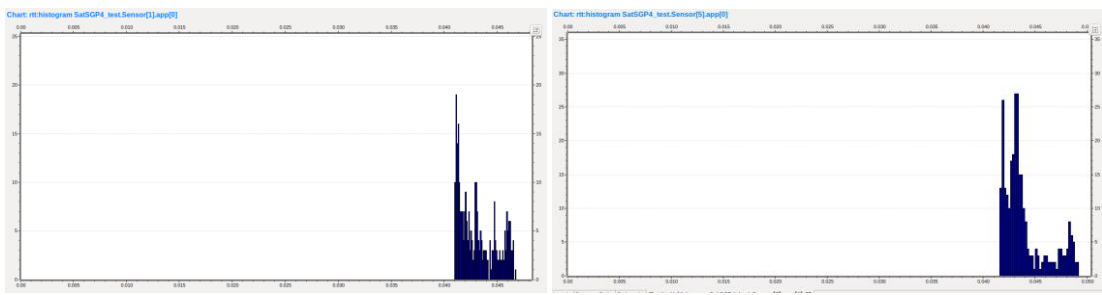

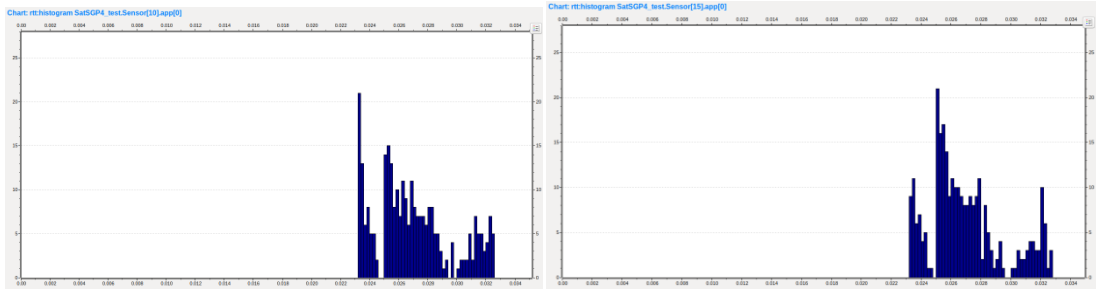

*Figure 39: rtt:histogram (frequency/rtt) for sensors 1,5,10,15 of the "More Sophisticated" (b.2) scenario*

*Table 23: Statistics / MCC of the "More Sophisticated" (b.2) scenario*

| | Pings sent | Pings received | Range rtt (ms) | Mean rtt (ms) | Ping loss (%) |
|---|---|---|---|---|---|
| **Sensor[0]** | 288 | 89 | 39,2-47,3 | 42,9 | 69,09 |
| **Sensor[1]** | 288 | 287 | 39,6-48,2 | 43,1 | 0,34 |
| **Sensor[2]** | 288 | 286 | 39,8-48,3 | 43,5 | 0,69 |
| **Sensor[3]** | 288 | 288 | 40-48,4 | 43,9 | 0 |
| **Sensor[4]** | 288 | 288 | 40,3-48,5 | 44,2 | 0 |
| **Sensor[5]** | 288 | 288 | 40-50,2 | 43,9 | 0 |
| **Sensor[6]** | 288 | 288 | 37,4-47,2 | 41,5 | 0 |
| **Sensor[7]** | 288 | 288 | 37,2-46,6 | 41 | 0 |
| **Sensor[8]** | 288 | 288 | 36,9-45,9 | 40,6 | 0 |
| **Sensor[9]** | 288 | 288 | 37,1-44,5 | 40,3 | 0 |
| **Sensor[10]** | 286 | 285 | 20,8-35 | 26,9 | 0,34 |
| **Sensor[11]** | 286 | 286 | 21,2-34,8 | 26,9 | 0 |
| **Sensor[12]** | 286 | 284 | 21,4-34,6 | 26,9 | 0,69 |
| **Sensor[13]** | 286 | 286 | 20,8-35,2 | 26,9 | 0 |
| **Sensor[14]** | 286 | 286 | 20,8-35,4 | 26,9 | 0 |
| **Sensor[15]** | 286 | 286 | 23-34,8 | 26,9 | 0 |

**Results:** In the 600 satellite constellation we notice close to continuous curves with distinct periodicities (see *Figure 38*) and compressed RTT ranges (also see *Figure 39 & Table 23*). Comparing the two experiments in "b" (see *Figure 40*), we observe a lower range and mean RTT and decreased (or even eliminated) ping losses for every Sensor within the 600 satellite experiment. Specifically, the range of ping losses (apart from the Sensor[0]) in the 360 satellite constellation is between 0-4,89%, while with 600 satellites it is almost eliminated, valued between 0-0,69%. Finally for Sensor[0], in which the received pings are limited probably due to collision issues, the ping loss is decreased from 70,48% to 69,09%.

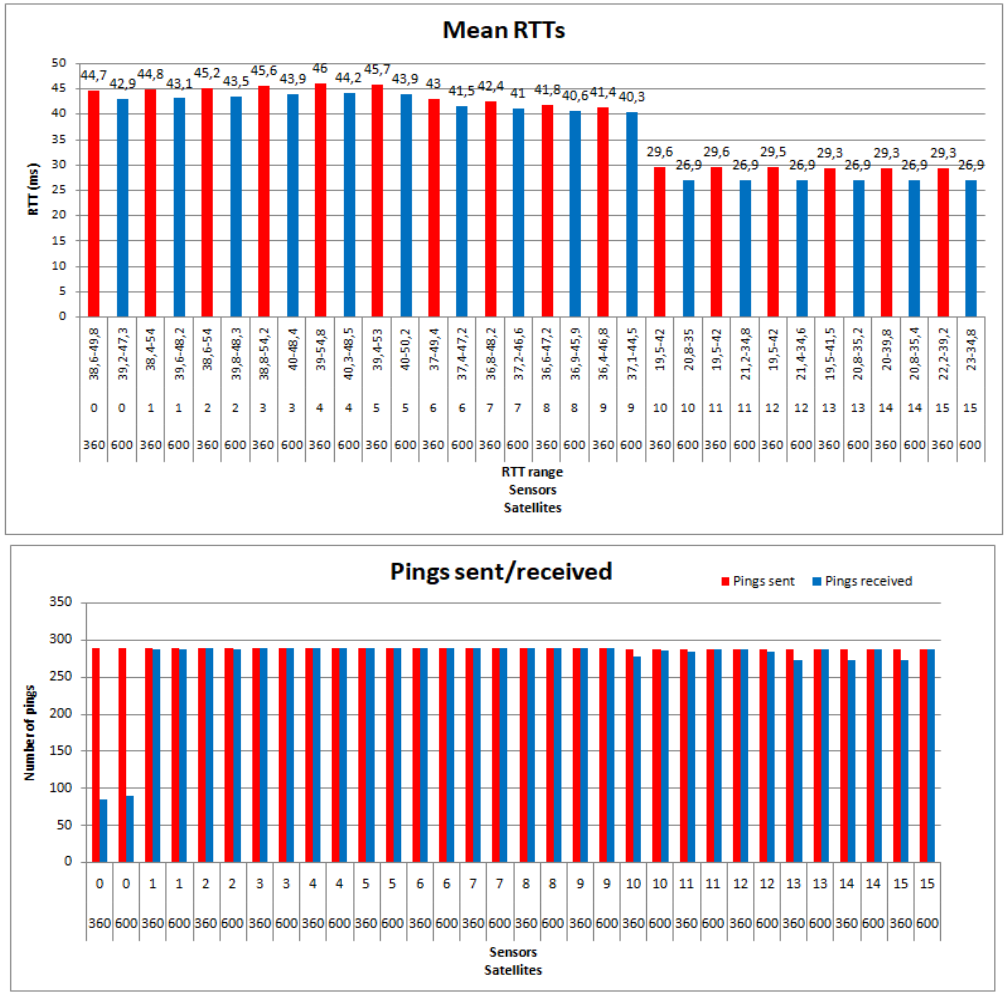

*Figure 40: Comparison Statistics / MCC / constellation of the "More Sophisticated" (b.1 & b.2) scenarios*

## III.   "ReWire Case Study" scenarios

Regarding the EC H2020-Rewire project [1][2][3][118] we have taken some measurements of a set of sensors from the Smart Santander test-bed. As mentioned in [5], SmartSantander proposes a *"unique city-scale experimental research facility in support of typical applications and services for a smart city, stimulating the development of new applications including experimental advanced research on IoT technologies and realistic assessment of users' acceptability tests"*. The project *"envisions the deployment of 20,000 sensors in Belgrade, Guildford, Lübeck and Santander (12,000)", with the Santander testbed composed currently of around 2000 IEEE 802.15.4 devices deployed*.

Among the 2.000 available sensors, both static and mobile, we have drawn data for 310 of them (including id, timestamps, coordinates: latitude/longitude and measurements), some of which we utilize by introducing them in the present case study. From the 310 sensors the 263 are stable, either positioned in parking slots (with the ":np" symbol in their data identification) or scattered in the city recording different measurements (distinguished by the ":t" symbol). The remaining 47 sensors are mobile (with the ":ar" symbol), verified from their variable coordinates during their transmissions. The stable ":np" sensors transmit signals (0 or 1) when they detect (with a magnetic plate) changes in the presence of a parked car. Therefore when a vehicle arrives or departs, a signal is communicated with its respective timestamp and id. The stable ":t" sensors consider a variety of measurements such as the batteryLevel, temperature, illuminance, soundPressureLevel or the electricField. Finally, the remaining ":ar" mobile sensors are positioned on cars, buses and smartphones transmitting information about the temperature, relativeHumidity, atmospheric concentration (CO, airParticles, $O_3$, $NO_2$), position, speed, direction and mileage or the device model and OS (in case of smartphone usage).

Ten (10) stable sensors have been selected and used in our simulations with their real measurements taken for the date 3/12/2022, specifically the sensors with id: 1, 9, 19, 24, 53, 90, 110, 135, 146 and 213. Even with the selection of the most distant remote sensors, separating their location on the map is almost impossible (maybe with the maximum zoom in *Figure 41*) due to the small (proportional to the map) differentiation of their coordinates. Consequently, sensors look as if they were positioned exactly at the same point.

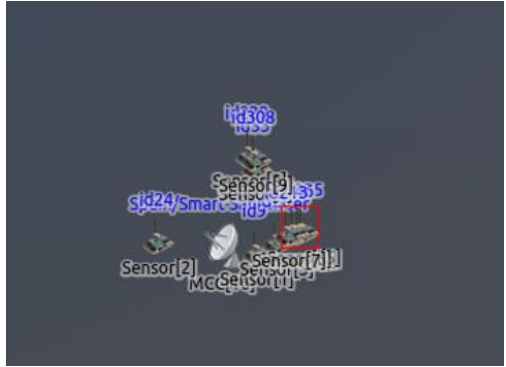

*Figure 41: Maximum zoom of the Smart Santander sensors in OMNeT++*

*Table 24: Selected sensors and their characteristics of the "ReWire Case Study" scenarios*

| id | urn_data | location_type | coordinates | type | |
|----|----------|---------------|-------------|------|---|
| 1 | urn:x-iot:smartsantander:u7jcfa:**np**3870 | Point | -3.7974231243134,43.464595794678 | stable | parking |
| 9 | urn:x-iot:smartsantander:u7jcfa:**t**258 | Point | -3.80161,43.46262 | stable | |
| 19 | urn:x-iot:smartsantander:u7jcfa:**np**3790 | Point | -3.799674987793,43.463623046875 | stable | parking |
| 24 | urn:x-iot:smartsantander:u7jcfa:**t**370 | Point | -3.81114,43.46367 | stable | |
| 53 | urn:x-iot:smartsantander:u7jcfa:**t**51 | Point | -3.80174,43.47092 | stable | |
| 90 | urn:x-iot:smartsantander:u7jcfa:**t**4074 | Point | -3.796732,43.463869 | stable | |
| 110 | urn:x-iot:smartsantander:u7jcfa:**t**506 | Point | -3.80545,43.46385 | stable | |
| 135 | urn:x-iot:smartsantander:u7jcfa:**np**3873 | Point | -3.7970464229584,43.464653015137 | stable | parking |
| 146 | urn:x-iot:smartsantander:u7jcfa:**np**3864 | Point | -3.7979302406311,43.464511871338 | stable | parking |
| 213 | urn:x-iot:smartsantander:u7jcfa:**np**3856 | Point | -3.7985026836395,43.464431762695 | stable | parking |

Since each sensor transmits at different times, with some of them not following a periodicity, we should choose the appropriate sendInterval. Unlike the scenarios where MCCs send periodically, in this case the sendInterval must follow the .csv file measurements of each Sensor. We also need to consider about the updateInterval parameter. We should have a tradeoff of efficiency and complexity, i.e. not a very small updateInterval (which would require more time than needed), but also not a too large value (in which case the accuracy of results is decreased). We prefer to find and apply the greatest common divisor (GCD) among the sendIntervals of the sensors, which is in our case 60 seconds.

In *Figure 42* we present the sensors' time difference (in minutes) per transmitted time (time:minutes). We create this diagram to evaluate the periodicity of the transmitted data as a distribution of the different time intervals in which each sensor takes a measurement. The more periodical the data, the fewer the changes in the charts and therefore the sendInterval parameter will change less times. Thus, the stable sensors placed in parking slots have limited to no periodicity (see sensors 1, 19, 135, 213) while the other stable sensors present close to constant sendIntervals (see sensors 9, 24, 53, 110) with some appearing to transmit in proportional changes of the initial sendInterval (see Sensor[90] in which the sendIntervals change among 10min, 20min and 30min).

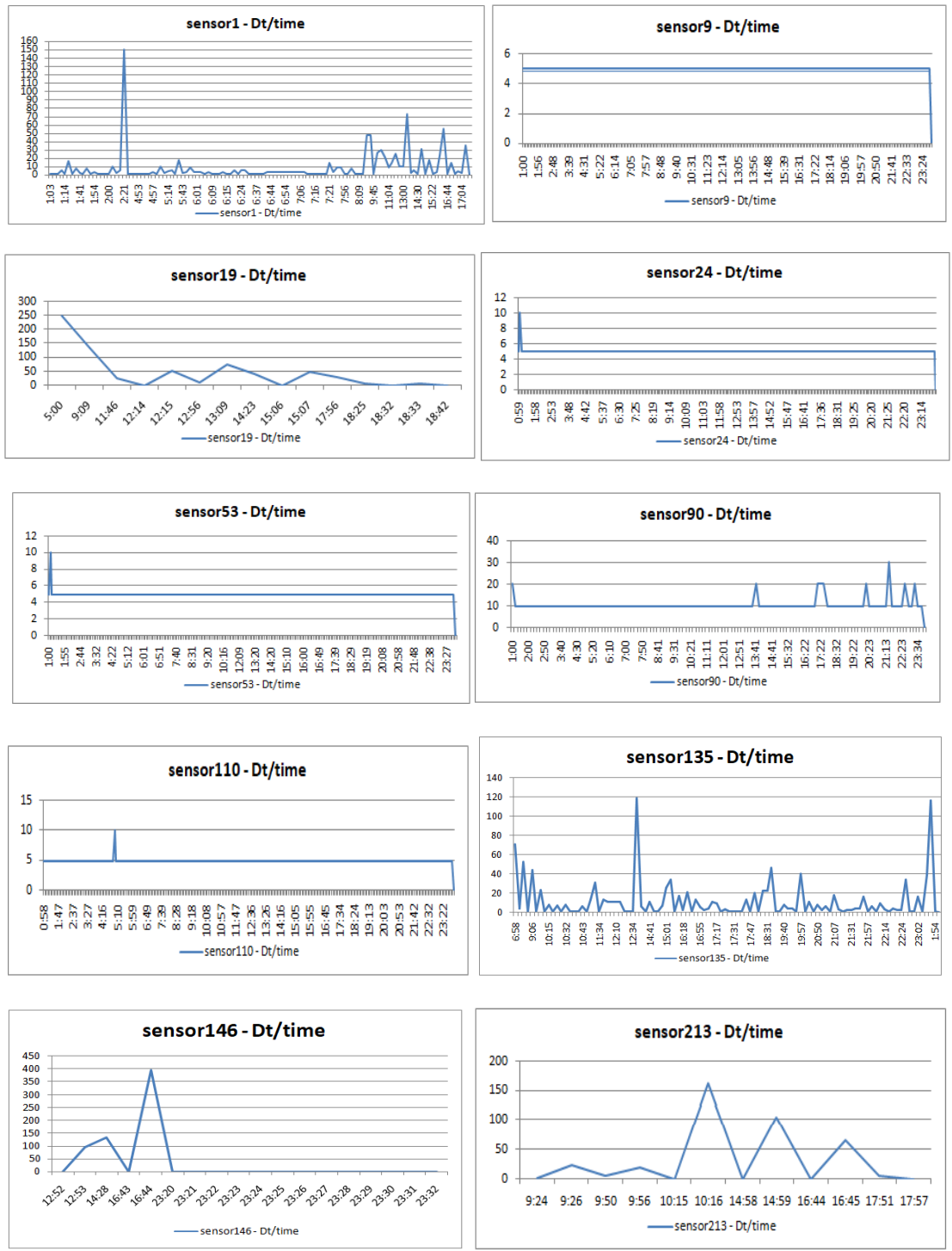

*Figure 42: Sensors Dt/time of the "ReWire Case Study" scenarios*

*Table 25* describes the characteristics of the 10 selected sensors in relation to the simulation parameters. Sensors 0-9 correspond to the actual sensors as shown by their id. The smallest start time of the data transmission was taken as reference point (time 0), i.e. time 0:58 which represents the first measurement from Sensor[4] with id110. Thus, all the other Sensors start by following the first one, for instance Sensor[1] has a startTime = 1min since it actually starts at 0:59.

*Table 25: Sensors' characteristics and simulation variables of the "ReWire Case Study" scenarios*

| Sensors_sim | Sensors_id | Start_Time_sim (min) | Start_Time (h:min) | Last_Time (h:min) | End_Time_sim (h) |
|---|---|---|---|---|---|
| Sensor[0] | sensor_id9 | 2 | 1:00 | 23:55 | |
| Sensor[1] | sensor_id24 | 1 | 0:59 | 23:59 | |
| Sensor[2] | sensor_id53 | 2 | 1:00 | 23:57 | |
| Sensor[3] | sensor_id90 | 2 | 1:00 | 23:54 | |
| Sensor[4] | sensor_id110 | 0 | 0:58 | 23:56 | 24 |
| Sensor[5] | sensor_id213 | 506 | 9:24 | 17:57 | |
| Sensor[6] | sensor_id146 | 714 | 12:52 | 23:32 | |
| Sensor[7] | sensor_id19 | 242 | 5:00 | 18:42 | |
| Sensor[8] | sensor_id1 | 5 | 1:03 | 17:41 | |
| Sensor[9] | sensor_id135 | 358 | 6:58 | 23:58 | |

In *Table 26* we present the variables and the characteristics of the simulation we will keep unchanged in all experiments of this Case Study. Therefore, each Sensor transmits to a different MCC with the updateInterval being 60 seconds and the sendInterval and startTimes following the .csv file.

*Table 26: General characteristics of the group "ReWire Case Study" of the "ReWire Case Study" scenarios*

| | |
|---|---|
| numOfMCCs | 26 |
| Senders | Sensor[0], Sensor[1], Sensor[2], Sensor[3], Sensor[4], Sensor[5], Sensor[6], Sensor[7], Sensor[8], Sensor[9] |
| Receivers | MCC[76], MCC[9], MCC[14], MCC[17], MCC[13], MCC[5], MCC[18], MCC[21], MCC[22], MCC[24] |
| Ships | 7 |
| numOfSens | 10 |
| updateInterval | 60s |
| startTime | 2min, 1min, 2min, 2min, 0min, 506min, 714min, 242min, 5min, 358min |
| sendInterval | .csv timestamps |
| sim-time-limit | 24h |

The different MCCs/destination nodes are actual Europe's test-beds found in the literature. Their names and location (coordinates) are presented in *Table 27*, while the real distances between the sender Sensor and its respective MCC destination is calculated in *Table 28*. We should mention that MCC[5] and MCC[9] do not represent test-beds, but are included for the purpose of the experiment.

*Table 27: MCC test-beds and features of the "ReWire Case Study" scenarios*

| MCC | Location | Coordinates | Name |
|---|---|---|---|
| MCC[5] | USA | 50.334241, -74.0060 | - |
| MCC[8] | Athens/Greece | 37.970833, 23.725110 | NETMODE |
| MCC[9] | Xanthi/Greece | 41.130036, 24.886490 | - |
| MCC[13] | Barcelona/Spain | 41.346176, 2.168365 | OFELIA |
| MCC[14] | Malaga/Spain | 36.719444, -4.420000 | Triangle |
| MCC[15] | Antwerp/Belgium | 51.260197, 4.402771 | CityLab |
| MCC[16] | Smart Santander/ Spain | 43.462776, -3.805000 | smartSantander |
| MCC[17] | Dublin/Ireland | 53.342686, -6.267118 | IRIS |
| MCC[18] | Amsterdam/Netherlands | 52.377956, 4.897070 | ExoGENI |
| MCC[19] | Volos/Greece | 39.366669, 22.933332 | NITOS |
| MCC[20] | Geneva/Switzerland | 46.204391, 6.143158 | IoT Lab |
| MCC[21] | Ljubljiana/Slovenia | 46.056947, 14.505751 | LOG-a-TEC |
| MCC[22] | Warsaw/Poland | 52.237049, 21.017532 | PL-LAB |
| MCC[23] | Paris/France | 48.864716, 2.349014 | EdgeNET&Grid'5000 |
| MCC[24] | Monaco/France | 43.6155, 7.0550 | R2lab |

*Table 28: Distances between Sender-Receiver of the "ReWire Case Study" scenarios*

| Sender-Receiver | Distance (km) |
|---|---|
| Sensor[0] → MCC[20] | 840,84 |
| Sensor[1] → MCC[9] | 2232,52 |
| Sensor[2] → MCC[14] | 947,06 |
| Sensor[3] → MCC[17] | 1029,30 |
| Sensor[4] → MCC[13] | 543,94 |
| Sensor[5] → MCC[5] | 5320,82 |
| Sensor[6] → MCC[18] | 935,35 |
| Sensor[7] → MCC[21] | 1196,81 |
| Sensor[8] → MCC[22] | 1794,49 |
| Sensor[9] → MCC[24] | 874,42 |

a) **Smart Santander-Other Test-beds / With & Without intersatelliteLinks / With Ships / .CSV startTimes / 360satellites**

We run our experiment of 10 Sensors with the simulation variables presented in *Tables 26 & 29*, i.e. 360 satellites in the constellation and interSatelliteLinks both enabled and disabled.

*Table 29: Simulation's characteristics of the "ReWire Case Study" (a) scenarios*

| numOfSats | 360 |
|---|---|
| Planes | 6 |
| satPerPlane | 60 |
| enableInterSatelliteLinks | True / False |

## 1. With interSatelliteLinks

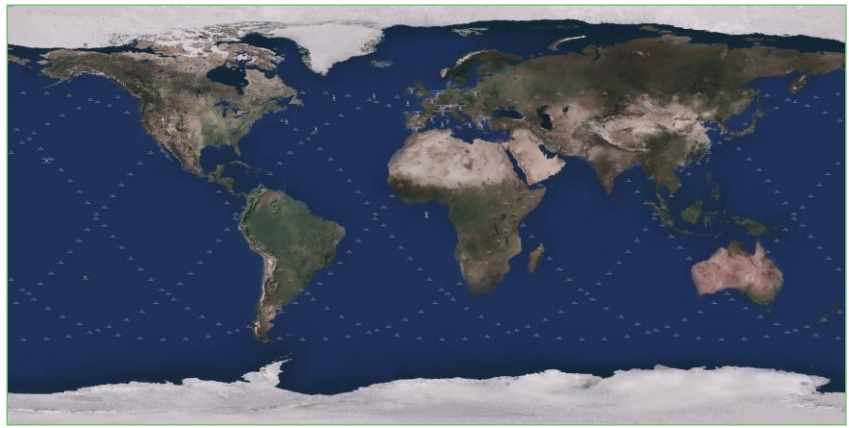

*Figure 43: Full constellation snapshot of the "ReWire Case Study" (a.1) scenario*

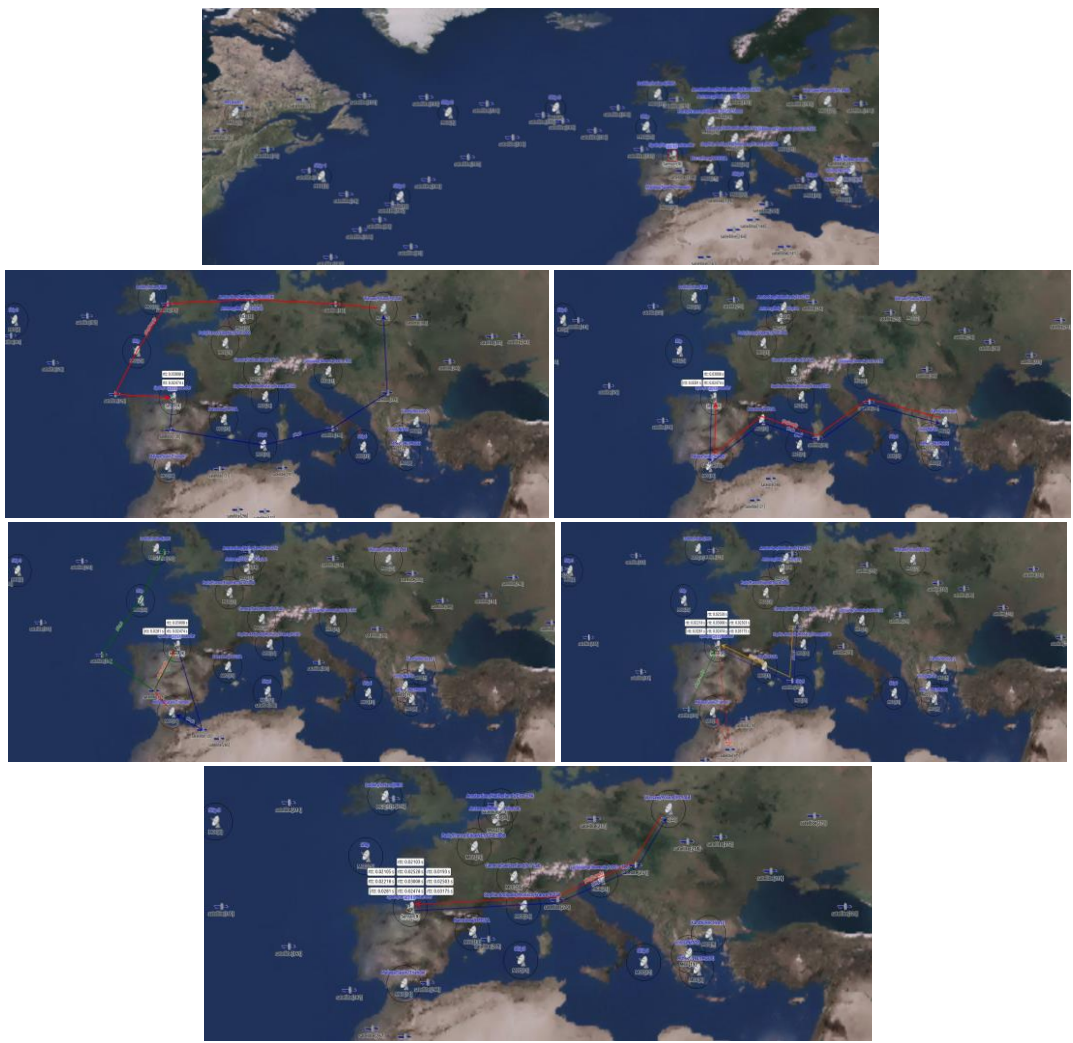

*Figure 43: Simulation snapshots of the "ReWire Case Study" (a.1) scenario*

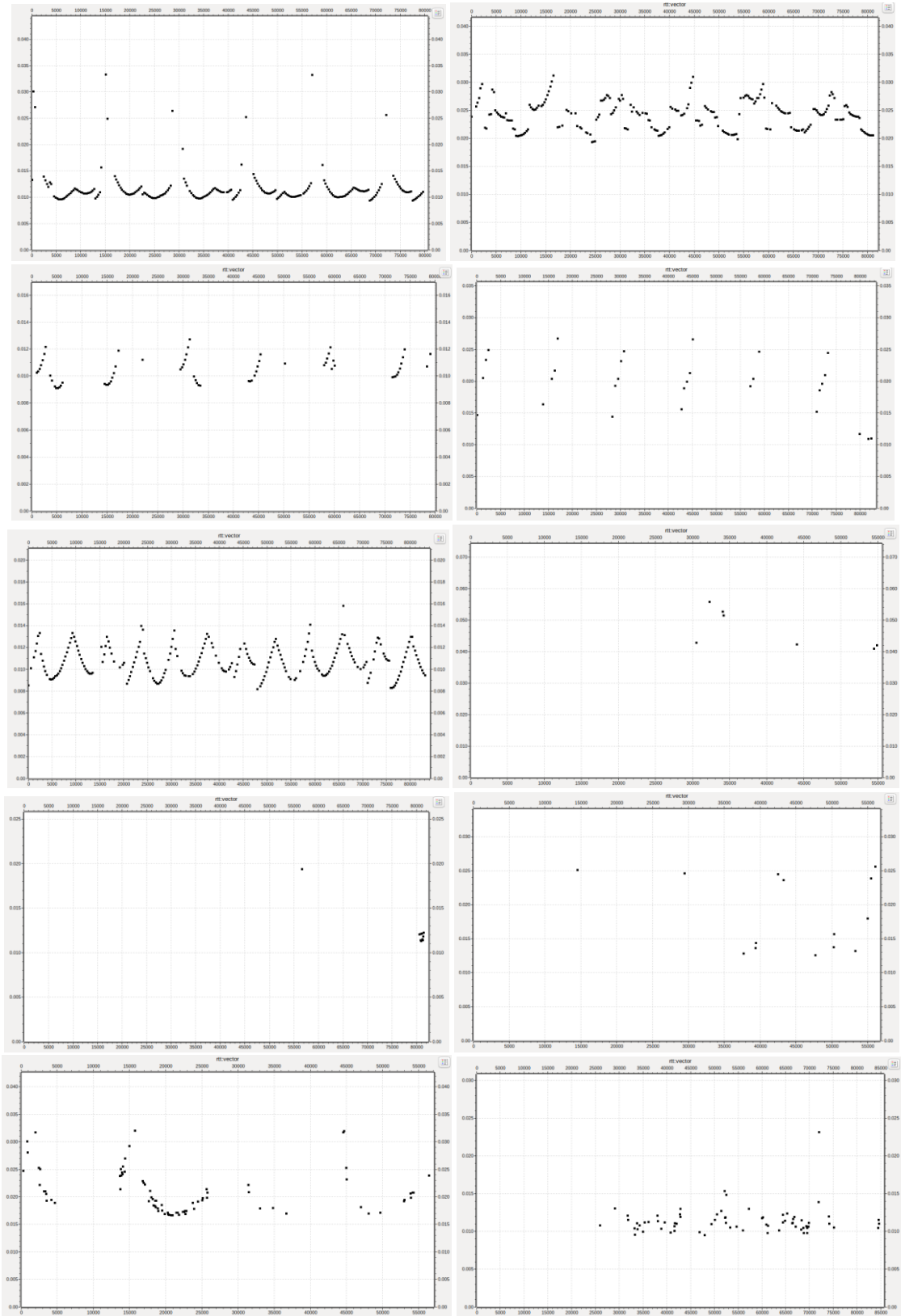

*Figure 44: rtt:vector (rtt/time) for the sensors 0-9 of the "ReWire Case Study" (a.1) scenario*

*Table 30: Statistics / MCC of the "ReWire Case Study" (a.1) scenario*

|  | Pings sent | Pings received | Range rtt (ms) | Mean rtt (ms) | Ping loss (%) |
|---|---|---|---|---|---|
| **Sensor[0]** | 266 | 227 | 9,37-33,31 | 12,04 | 14,66 |
| **Sensor[1]** | 270 | 230 | 19,31-31,22 | 24,36 | 14,81 |
| **Sensor[2]** | 269 | 65 | 9,11-12,72 | 10,39 | 75,84 |
| **Sensor[3]** | 128 | 29 | 10,93-26,72 | 19,93 | 77,34 |
| **Sensor[4]** | 277 | 235 | 8,19-15,81 | 10,82 | 15,16 |
| **Sensor[5]** | 13 | 7 | 40,9-55,79 | 47,88 | 46,15 |
| **Sensor[6]** | 19 | 10 | 11,3-19,38 | 11,82 | 47,37 |
| **Sensor[7]** | 15 | 14 | 12,6-25,61 | 17,42 | 6,67 |
| **Sensor[8]** | 115 | 82 | 16,68-32,04 | 20,2 | 28,69 |
| **Sensor[9]** | 102 | 71 | 9,5-23,15 | 10,6 | 30,39 |

**Results:** With the 360 satellite constellation and intersatelliteLinks enable we capture the result of the simulation in *Figure 44* & *Table 30*. As the destination nodes differ from Sensor to Sensor, the presented measurement per Sensor don't show any similarity. The minimum mean RTT is observed for Sensor[2] at 10,39ms, while the Sensor with the more distant destination i.e. Sensor[5] shows a mean RTT of 47,88ms. Also, the largest ping loss is present in Sensor[3] with only 29 pings received out of the total 128 transmitted pings. Finally, we notice the second lowest ping loss (after the 6,67%, i.e. 14/15 arrived pings, of Sensor[7]) in Sensor[0], with 227 out of 266 pings received resulting in 14,66% of loss.

2. *Without interSatelliteLinks*

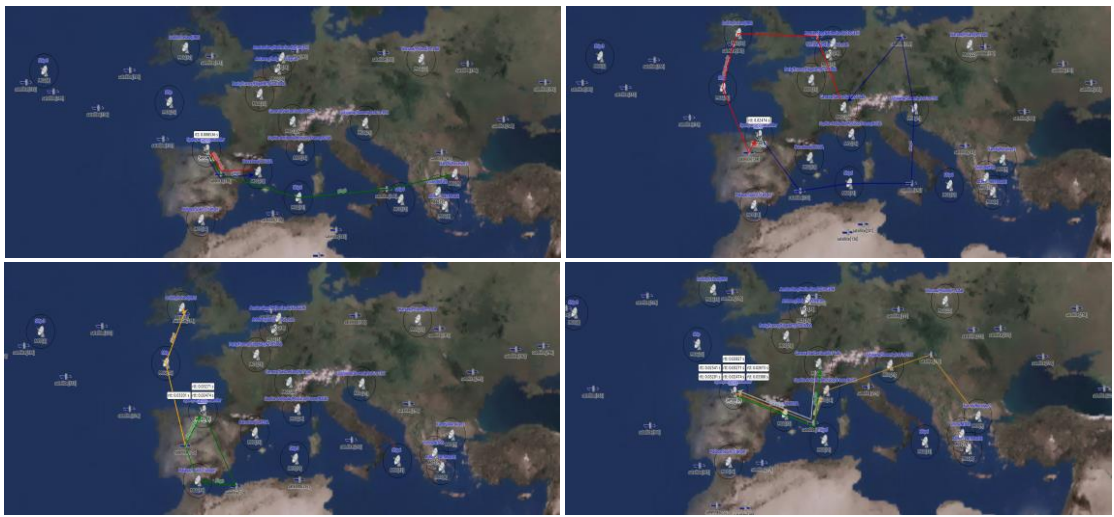

*Figure 45: Simulation snapshots of the "ReWire Case Study" (a.2) scenario*

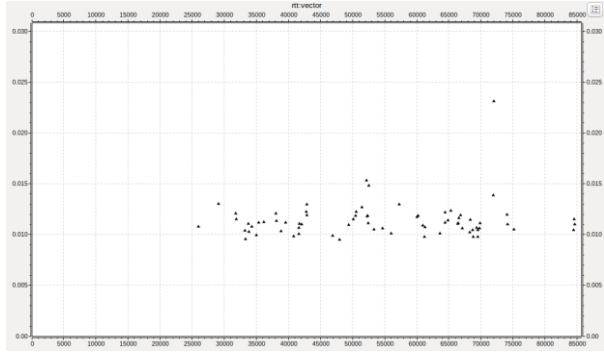

*Figure 46: rtt:vector (rtt/time) for the sensors 0-9 of the "ReWire Case Study" (a.2) scenario*

*Table 31: Statistics / MCC of the "ReWire Case Study" (a.2) scenario*

|  | Pings sent | Pings received | Range rtt (ms) | Mean rtt (ms) | Ping loss (%) |
|---|---|---|---|---|---|
| **Sensor[0]** | 266 | 229 | 9,37-37,92 | 12,05 | 13,91 |
| **Sensor[1]** | 270 | 229 | 23,74-35,37 | 27,01 | 15,19 |
| **Sensor[2]** | 269 | 67 | 9,11-14,03 | 10,52 | 75,09 |
| **Sensor[3]** | 128 | 29 | 10,93-31,57 | 20,73 | 77,34 |
| **Sensor[4]** | 277 | 237 | 8,19-17,02 | 10,71 | 14,44 |
| **Sensor[5]** | 13 | 0 | 0 | 0 | 100 |
| **Sensor[6]** | 19 | 10 | 11,3-19,38 | 12,1 | 47,37 |
| **Sensor[7]** | 15 | 15 | 12,59-25,61 | 17,96 | 0 |
| **Sensor[8]** | 115 | 83 | 18,55-35,10 | 21,51 | 27,83 |
| **Sensor[9]** | 102 | 71 | 9,50-23,15 | 11,85 | 30,39 |

**Results:** While intersatelliteLinks are disable in the 360 satellite constellation, the minimum mean RTT is also observed for Sensor[2] at 10,52ms (as with intersatelliteLinks on), while the Sensor with the more distant destination (Sensor[5]) achieves no communication throughout the simulation due to the lack of MCC<->Satellite<->MCC connection patterns. Thus, the largest ping loss (100%) is presented there (in Sensor[5]). Finally, the lowest ping loss after the 0% (i.e. 15/15 arrived pings) of Sensor[7] is observed in Sensor[0] (as with intersatelliteLinks enable) with 229 out of 266 pings received and 13,91% of loss.

Comparing the experiments of the 360 satellite constellation, in the former with intersatelliteLinks enabled the mean RTTs are lower and we observe more compressed RTT ranges with at least the same or even lower minimum and maximum RTTs. Additionally, there are differences per Sensor regarding the number of received pings, with the most significant occurring in Sensor[5] where communication is not achieved without intersatelliteLinks. Finally we observe some Sensors receive more pings with the intersatelliteLinks parameter disable (e.g., Sensors[0],[7],[8]) due to their location on the map and compared to the satellites' positions.

**b) Smart Santander-Other Test-beds / With & Without intersatelliteLinks / With Ships / .CSV startTimes / 600satellites**

The only difference in the present scenario (compared to "a") is the number of satellites in the constellation (600 satellites). We run two experiments with and without intersatelliteLinks.

*Table 32: Simulation's characteristics of the "ReWire Case Study" (b) scenarios*

| numOfSats | 600 |
|---|---|
| Planes | 10 |
| satPerPlane | 60 |
| enableInterSatelliteLinks | True / False |

*1.  With interSatelliteLinks*

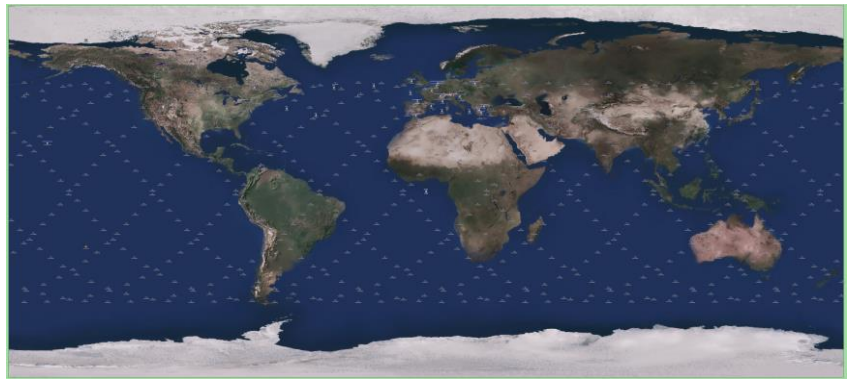

*Figure 47: Full constellation snapshot of the "ReWire Case Study" (b.1) scenario*

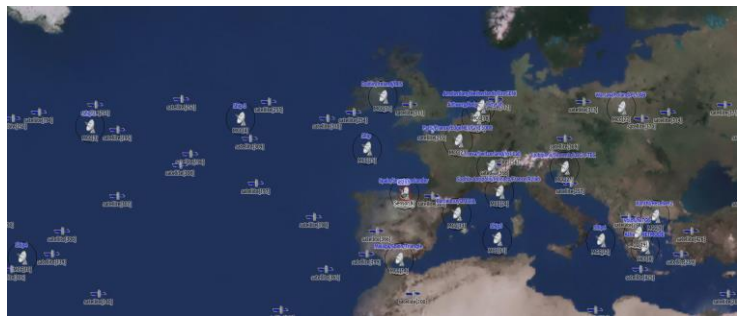

*Figure 48: Simulation snapshots of the "ReWire Case Study" (b.1) scenario*

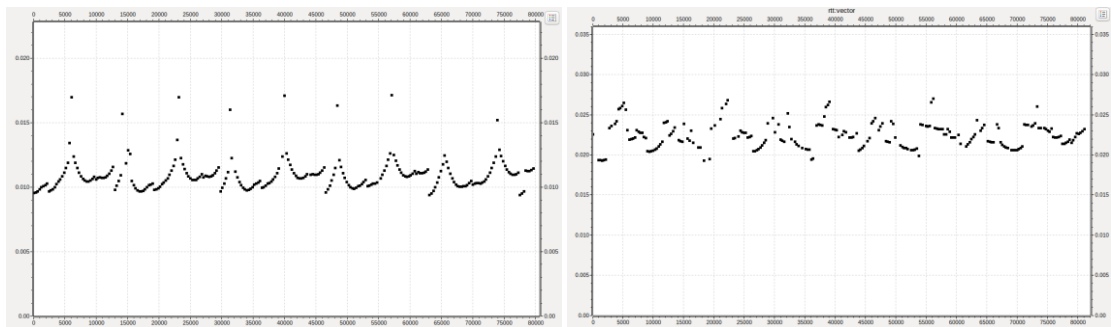

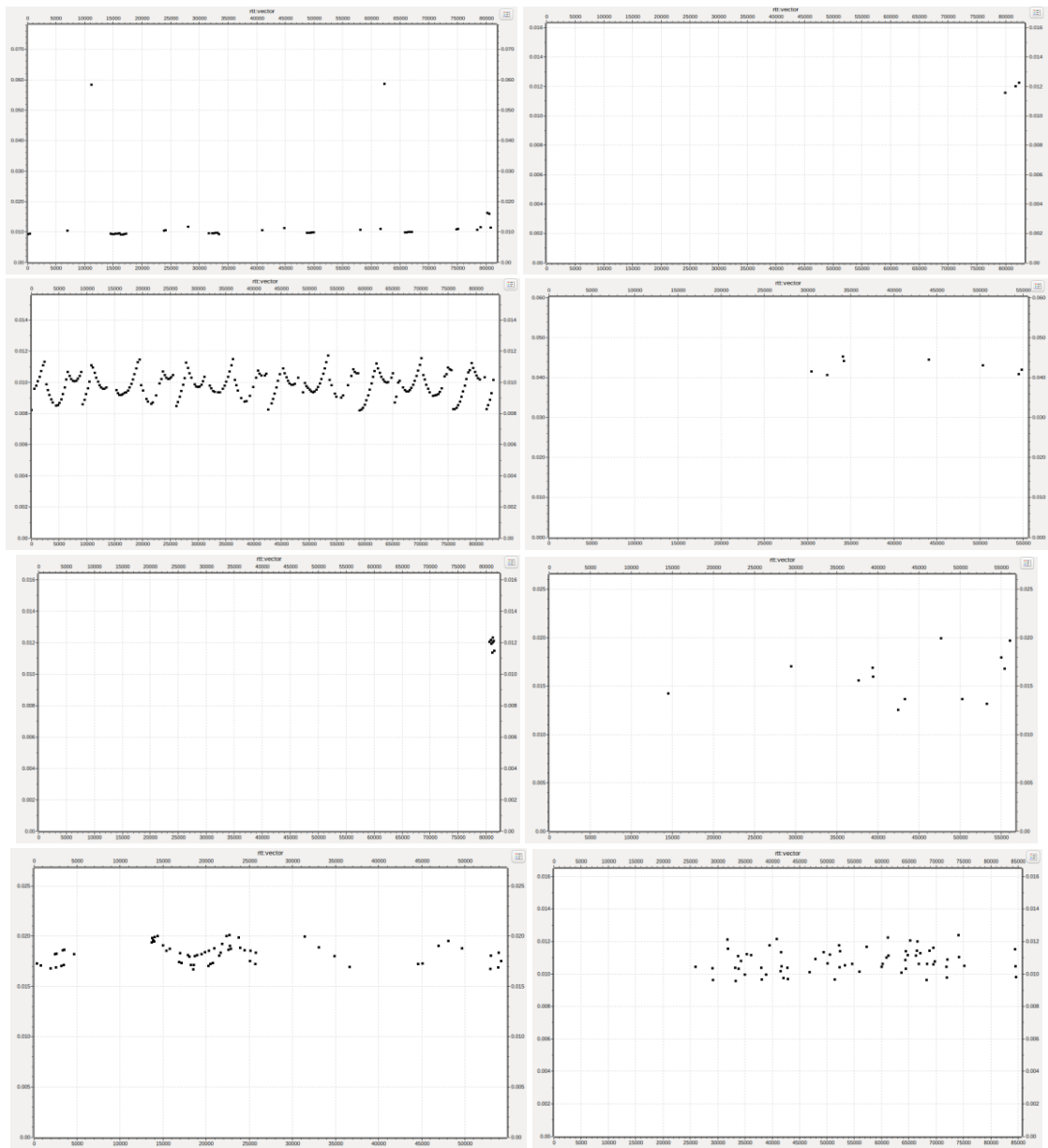

*Figure 49: rtt:vector (rtt/time) for the sensors 0-9 of the "ReWire Case Study" (b.1) scenario*

*Table 33: Statistics / MCC of the "ReWire Case Study" (b.1) scenario*

|  | Pings sent | Pings received | Range rtt (ms) | Mean rtt (ms) | Ping loss (%) |
|---|---|---|---|---|---|
| Sensor[0] | 266 | 261 | 9,37-17,14 | 10,90 | 1,88 |
| Sensor[1] | 270 | 225 | 19,29-27,01 | 22,48 | 16,67 |
| Sensor[2] | 269 | 45 | 9,18-58,67 | 12,84 | 83,27 |
| Sensor[3] | 128 | 3 | 11,56-12,26 | 11,88 | 97,66 |
| Sensor[4] | 277 | 238 | 8,21-11,75 | 9,85 | 14,08 |
| Sensor[5] | 13 | 8 | 40,07-45,3 | 42,83 | 38,46 |
| Sensor[6] | 19 | 8 | 11,37-12,32 | 12,02 | 57,89 |
| Sensor[7] | 15 | 13 | 12,55-19,94 | 16,68 | 13,33 |
| Sensor[8] | 115 | 67 | 16,71-20,11 | 17,89 | 41,74 |
| Sensor[9] | 102 | 68 | 9,55-12,39 | 10,84 | 33,33 |

**Results:** In the 600 satellites constellation with intersatelliteLinks enabled, we observe the smallest mean RTT in Sensor[4] at 9,85ms, while the Sensor with the more distant destination i.e. Sensor[5] presents a mean RTT of 42,83ms. Also, the largest ping loss (97,66%) is detected again in Sensor[3] (as in "a.1") with only 3 pings received out of the total 128 transmitted. Finally, we notice the lowest ping loss in Sensor[0] with 261 out of 266 pings received and 1,88% of loss.

### 2. Without interSatelliteLinks

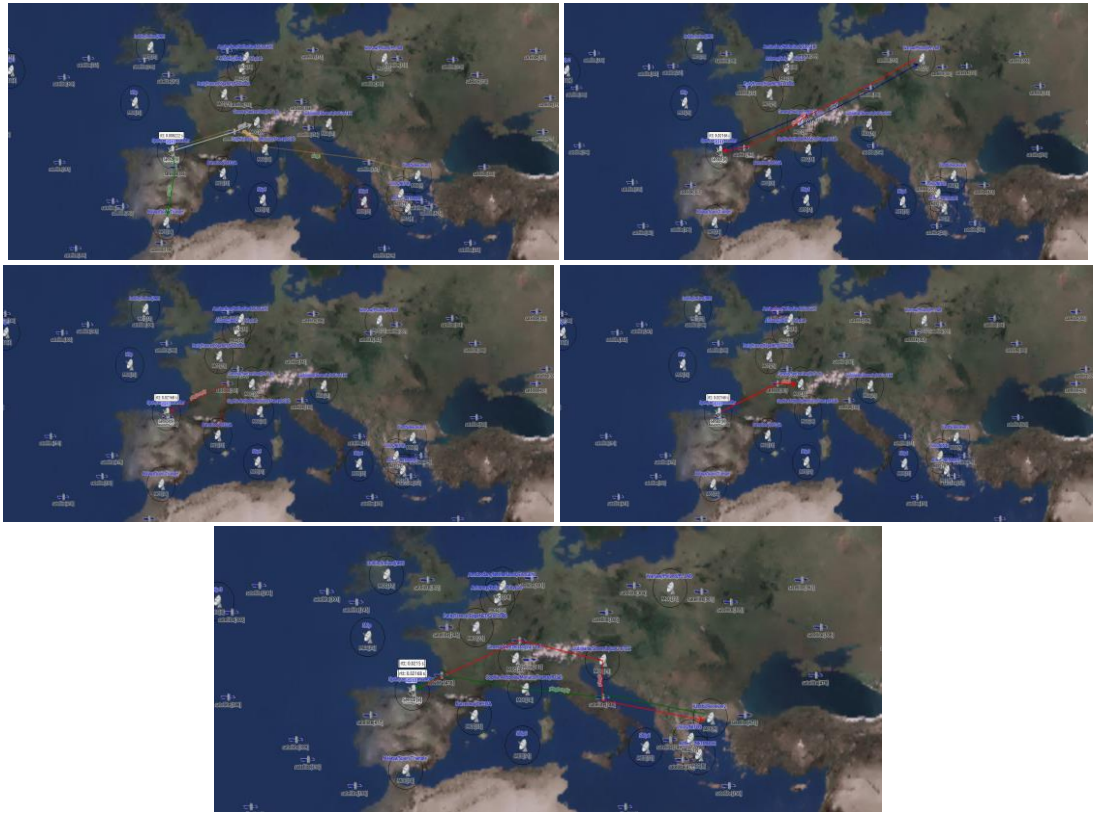

*Figure 50: Simulation snapshots of the "ReWire Case Study" (b.2) scenario*

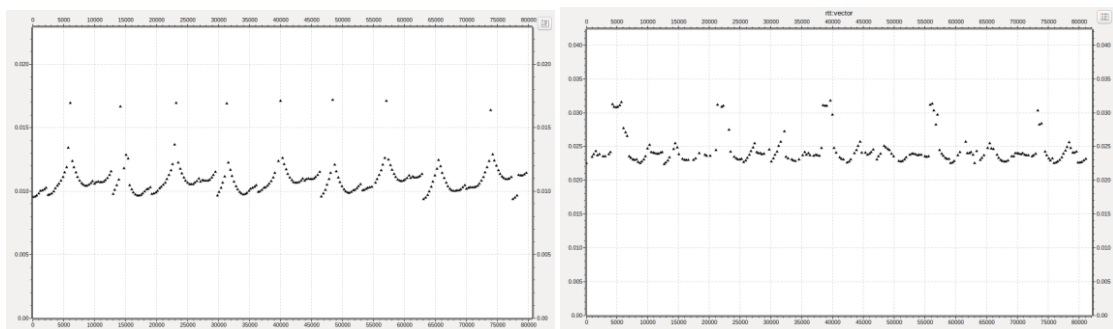

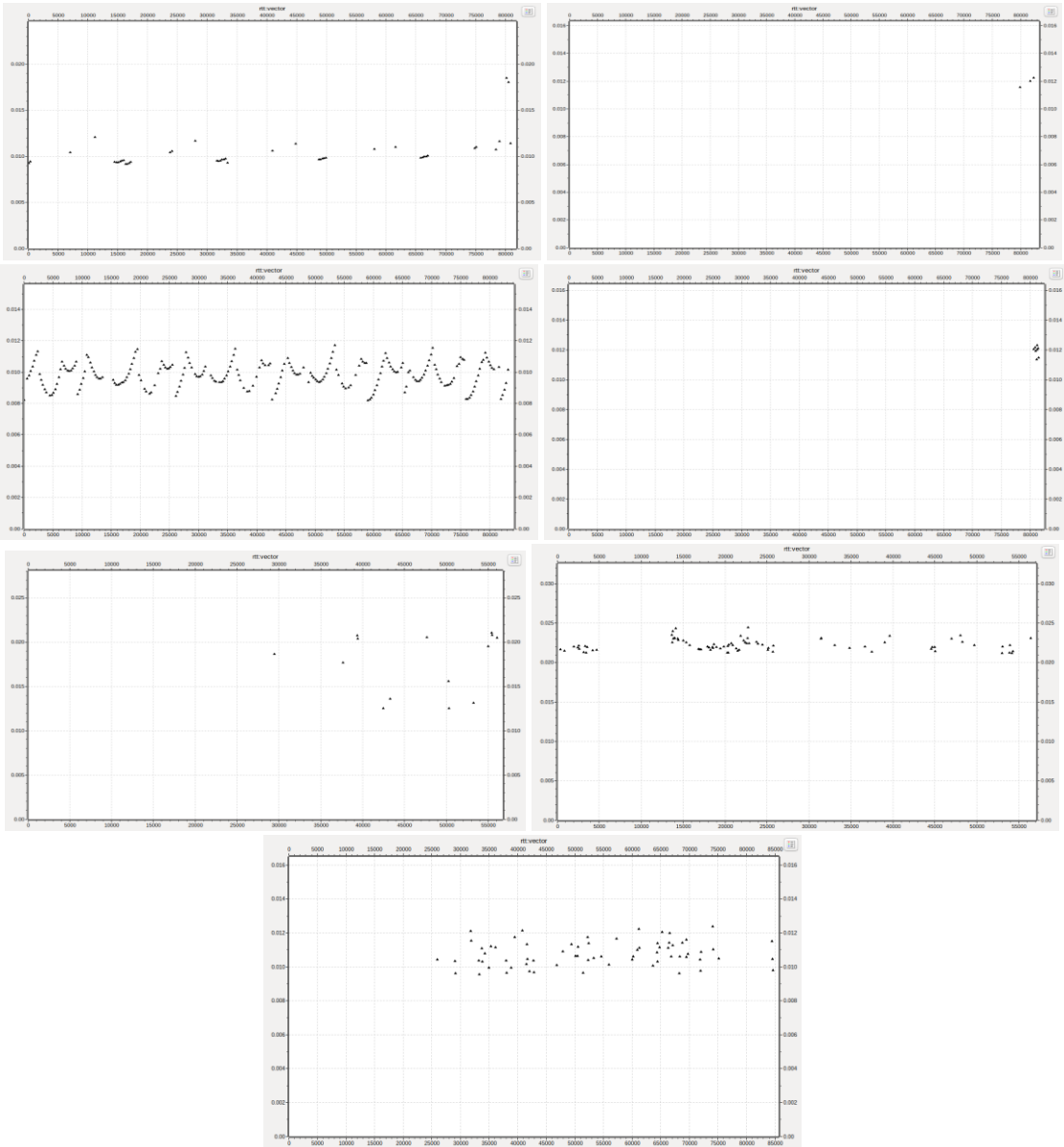

*Figure 51: rtt:vector (rtt/time) for the sensors 0-9 of the "ReWire Case Study" (b.2) scenario*

*Table 34: Statistics / MCC of the "ReWire Case Study" (b.2) scenario*

|  | Pings sent | Pings received | Range rtt (ms) | Mean rtt (ms) | Ping loss (%) |
|---|---|---|---|---|---|
| Sensor[0] | 266 | 263 | 9,37-17,20 | 10,91 | 1,12 |
| Sensor[1] | 270 | 226 | 22,39-31,82 | 24,58 | 16,29 |
| Sensor[2] | 269 | 45 | 9,18-12,07 | 10,31 | 83,27 |
| Sensor[3] | 128 | 3 | 11,56-12,26 | 12,02 | 97,66 |
| Sensor[4] | 277 | 240 | 8,21-11,73 | 9,85 | 13,36 |
| Sensor[5] | 13 | 0 | 0 | 0 | 100 |
| Sensor[6] | 19 | 8 | 11,37-12,32 | 11,92 | 57,89 |
| Sensor[7] | 15 | 14 | 12,55-21,08 | 17,17 | 6,67 |
| Sensor[8] | 115 | 82 | 21,23-24,47 | 21,85 | 28,69 |
| Sensor[9] | 102 | 70 | 9,55-12,39 | 10,9 | 13,37 |

**Results:** While intersatelliteLinks are disable in the 600 satellite constellation, the minimum mean RTT is also observed for Sensor[4] (as with intersatelliteLinks) at 9,85ms, while the Sensor with the most distant destination (Sensor[5]) does not achieve communication due to the lack of MCC<->Satellite<->MCC connection patterns. Thus, the largest ping loss (100%) is presented in Sensor[5]. Finally, we notice the lowest ping loss in Sensor[0] (as with intersatelliteLinks enable) with 263 out of 266 pings received and 1,12% of loss.

Comparing the experiments of the 600 satellite constellation (scenario "b"), in the former with intersatelliteLinks the average RTTs are lower and the range RTT more compressed, with at least the same or smaller minimum and maximum values. Sensor[2] is an exception in which an extreme value (58,67ms) is displaced, increasing the mean RTT and the highest presented value. While both of the range and mean RTTs are higher in the experiment without intersatelliteLinks, there are some Sensors in which the received pings are either the same or more (compared to the experiment with intersatelliteLinks enable). Finally, Sensor's[5] communication is once again not achieved without intersatelliteLinks.

a) **Smart Santander-Other Test-beds / With & Without intersatelliteLinks / With Ships / .CSV startTimes / 900satellites**

The difference of the present scenario (compared to "a" and "b") is the increased amount of used satellites in the constellation (900 satellites). We run two experiments with and without intersatelliteLinks.

*Table 35: Simulation's characteristics of the "ReWire Case Study" (c) scenarios*

| numOfSats | 900 |
|---|---|
| Planes | 15 |
| satPerPlane | 60 |
| enableInterSatelliteLinks | True / False |

1. *With interSatelliteLinks*

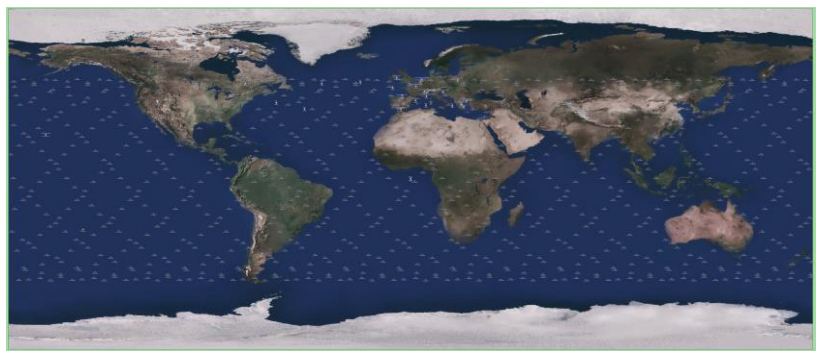

*Figure 52: Full constellation snapshot of the "ReWire Case Study" (c.1) scenario*

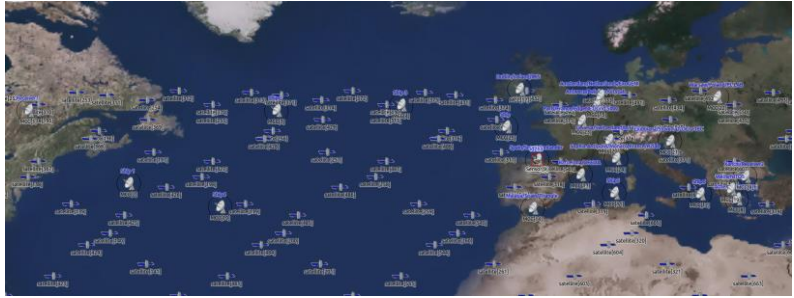

*Figure 53: Simulation snapshots of the "ReWire Case Study" (c.1) scenario*

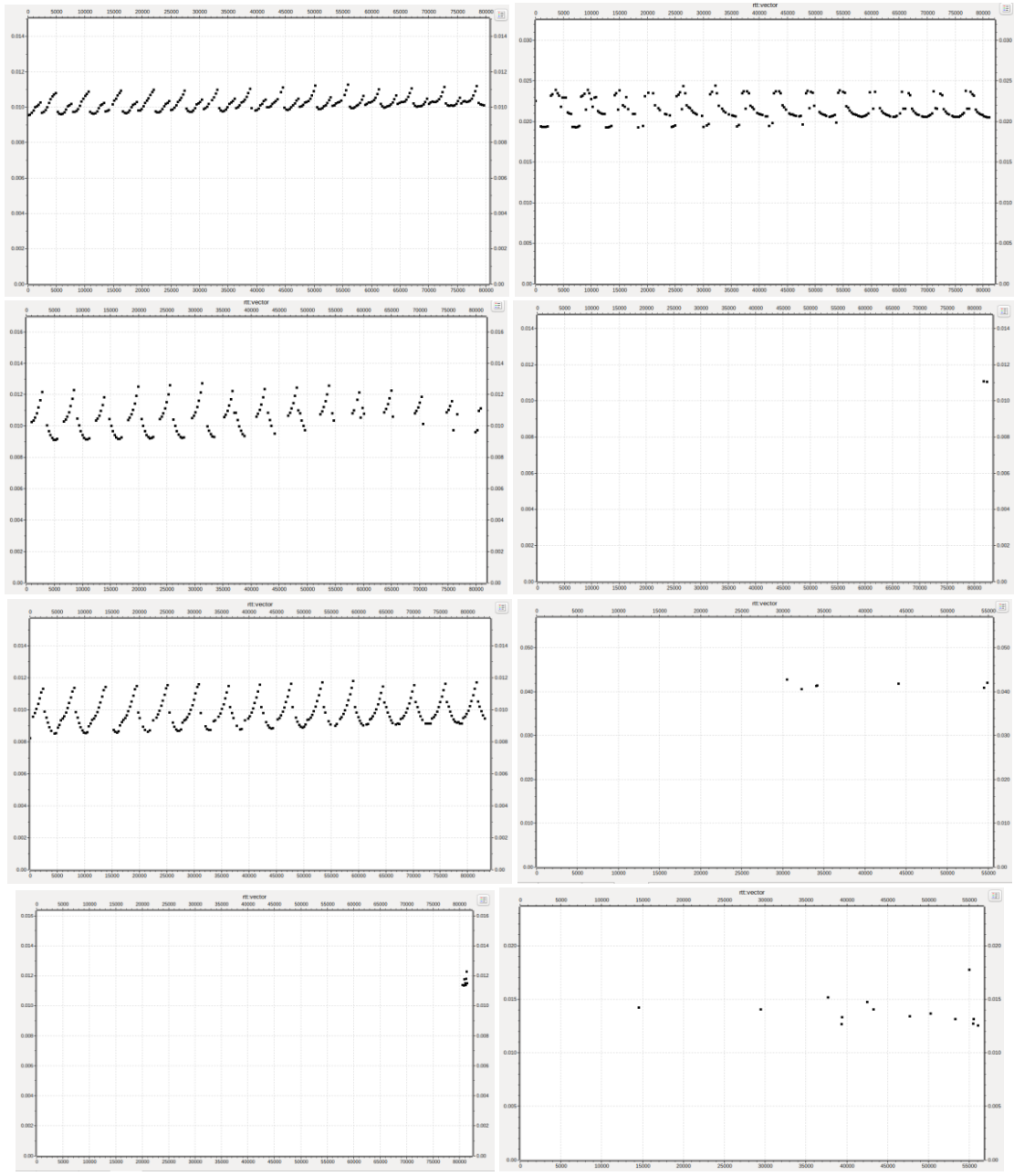

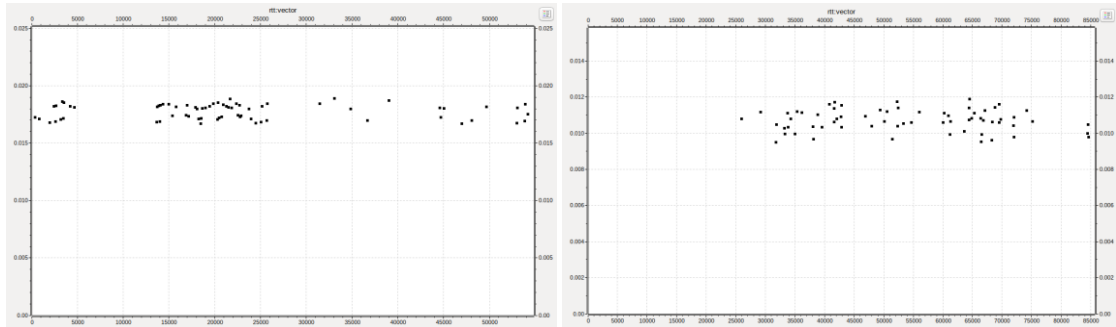

*Figure 54: rtt:vector (rtt/time) for the sensors 0-9 of the "ReWire Case Study" (c.1) scenario*

*Table 36: Statistics / MCC of the "ReWire Case Study" (c.1) scenario*

|  | Pings sent | Pings received | Range rtt (ms) | Mean rtt (ms) | Ping loss (%) |
|---|---|---|---|---|---|
| **Sensor[0]** | 266 | 261 | 9,58-11,21 | 10,08 | 1,88 |
| **Sensor[1]** | 270 | 225 | 19,3-24,43 | 21,56 | 16,67 |
| **Sensor[2]** | 269 | 154 | 9,11-12,72 | 10,89 | 42,75 |
| **Sensor[3]** | 128 | 2 | 11,05-11,08 | 11,65 | 98,43 |
| **Sensor[4]** | 277 | 248 | 8,22-11,8 | 9,76 | 10,47 |
| **Sensor[5]** | 13 | 7 | 40,63-42,79 | 41,49 | 46,15 |
| **Sensor[6]** | 19 | 8 | 11,36-12,27 | 11,85 | 57,89 |
| **Sensor[7]** | 15 | 14 | 12,68-17,75 | 13,65 | 6,67 |
| **Sensor[8]** | 115 | 68 | 16,71-18,9 | 17,6 | 40,87 |
| **Sensor[9]** | 102 | 66 | 9,49-11,89 | 10,74 | 35,29 |

**Results:** In the 900 satellites constellation with intersatelliteLinks, the smallest mean RTT is observed in Sensor[4] at 9,76ms, while Sensor[5] presents a mean RTT of 41,49ms. We detect the largest ping loss (98,43%) in Sensor[3] with 2 out of the total 128 transmitted pings being received. Finally, we notice the lowest ping loss (1,88%) in Sensor[0] with 261 out of 266 pings received.

### 2. *Without interSatelliteLinks*

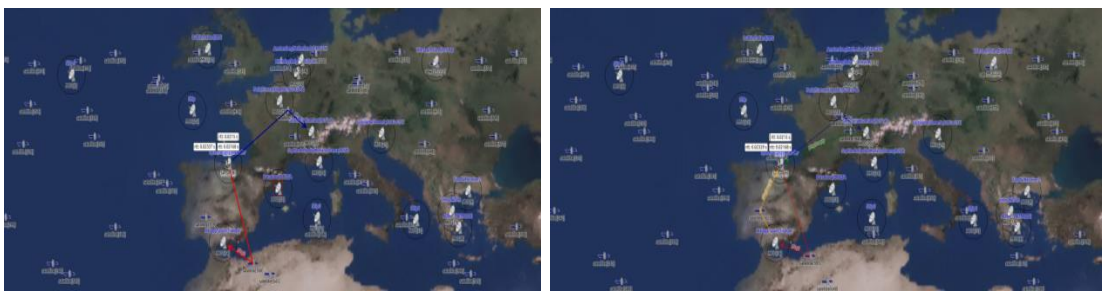

*Figure 55: Simulation snapshots of the "ReWire Case Study" (c.2) scenario*

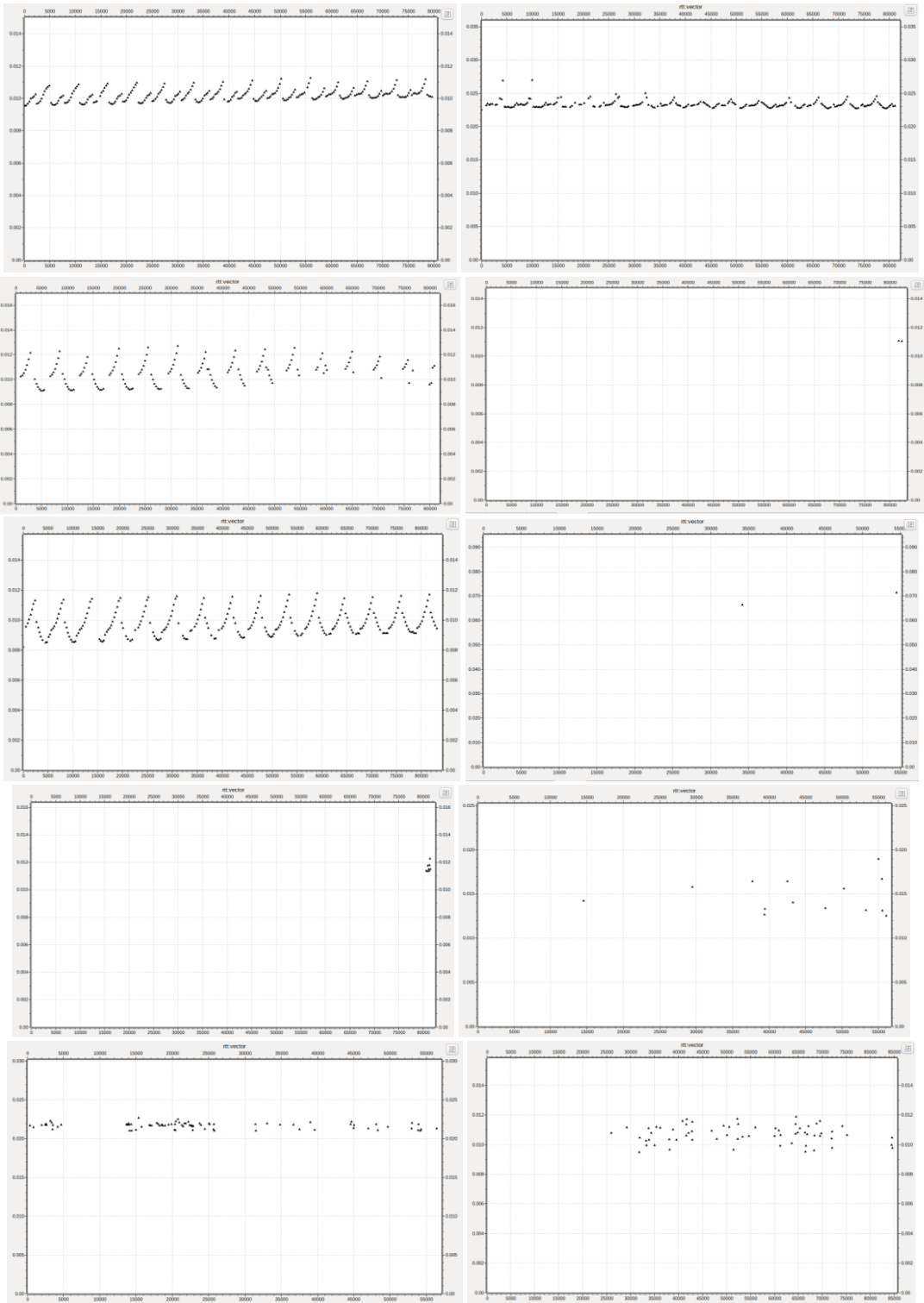

*Figure 56: rtt:vector (rtt/time) for the sensors 0-9 of the "ReWire Case Study" (c.2) scenario*

*Table 37: Statistics / MCC of the "ReWire Case Study" (c.2) scenario*

|  | Pings sent | Pings received | Range rtt (ms) | Mean rtt (ms) | Ping loss (%) |
|---|---|---|---|---|---|
| **Sensor[0]** | 266 | 263 | 9,58-11,27 | 10,14 | 1,12 |
| **Sensor[1]** | 270 | 226 | 22,54-27 | 23,47 | 16,30 |
| **Sensor[2]** | 269 | 157 | 9,11-12,72 | 10,34 | 41,64 |
| **Sensor[3]** | 128 | 2 | 11,05-11,08 | 11,07 | 98,44 |
| **Sensor[4]** | 277 | 249 | 8,22-11,81 | 9,68 | 10,11 |
| **Sensor[5]** | 13 | 2 | 66,5-71,45 | 69,2 | 84,61 |
| **Sensor[6]** | 19 | 8 | 11,36-12,27 | 11,64 | 57,89 |
| **Sensor[7]** | 15 | 14 | 12,56-18,98 | 14,8 | 6,67 |
| **Sensor[8]** | 115 | 83 | 21,04-22,7 | 21,7 | 27,83 |
| **Sensor[9]** | 102 | 66 | 9,49-11,89 | 10,81 | 35,29 |

**Results:** With intersatelliteLinks disable in the 900 satellite constellation, the minimum mean RTT is observed in Sensor[4] at 9,68ms (as with intersatelliteLinks on), while Sensor[5], even with high ping loss, achieves communication for the first time without intersatelliteLinks. Finally, the largest ping loss (98,44%) is presented in Sensor[3] while, we notice the lowest ping loss in Sensor[0] with 263 out of 266 pings received and 1,12% of loss.

Comparing the 900 satellite constellation experiments, the former with intersatelliteLinks enabled presents lower average RTTs and more compressed ranges of RTT with at least the same or smaller minimum and maximum values. While both of the range and mean RTTs are higher in the experiment without intersatelliteLinks, we observe the received pings to be either the same or more in all Sensors.

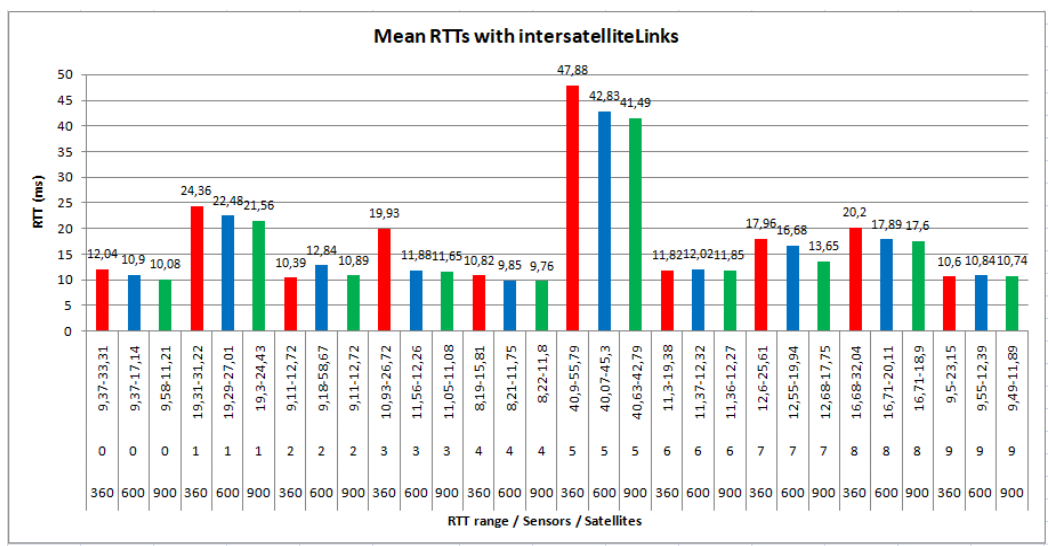

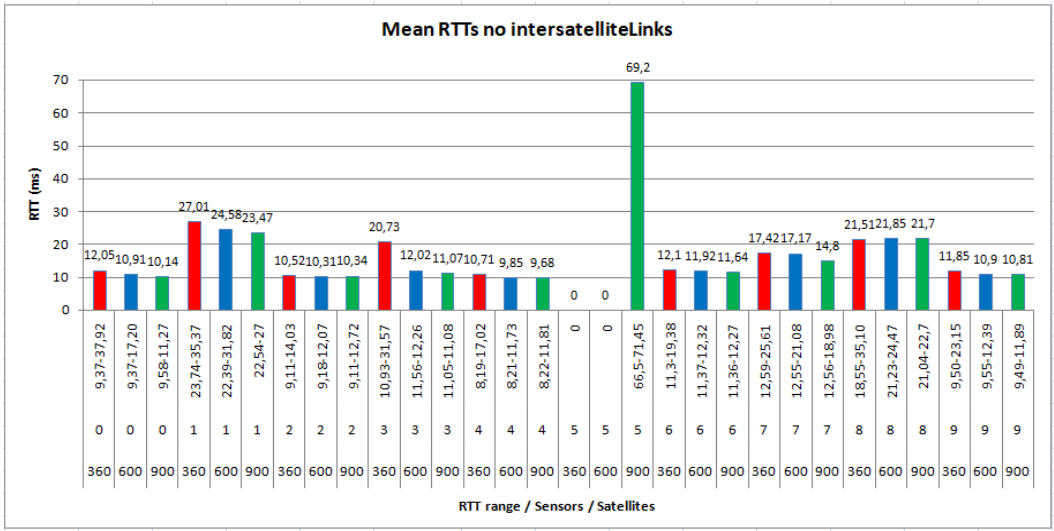

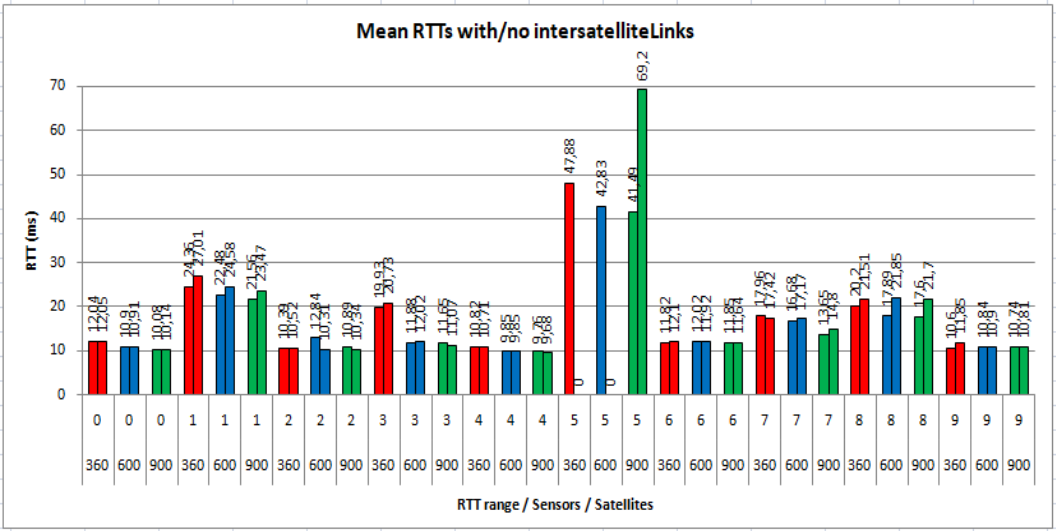

*Figure 57: Mean RTTs & range RTTs with & without intersatelliteLinks per sensor and satellites of the "ReWire Case Study" scenarios*

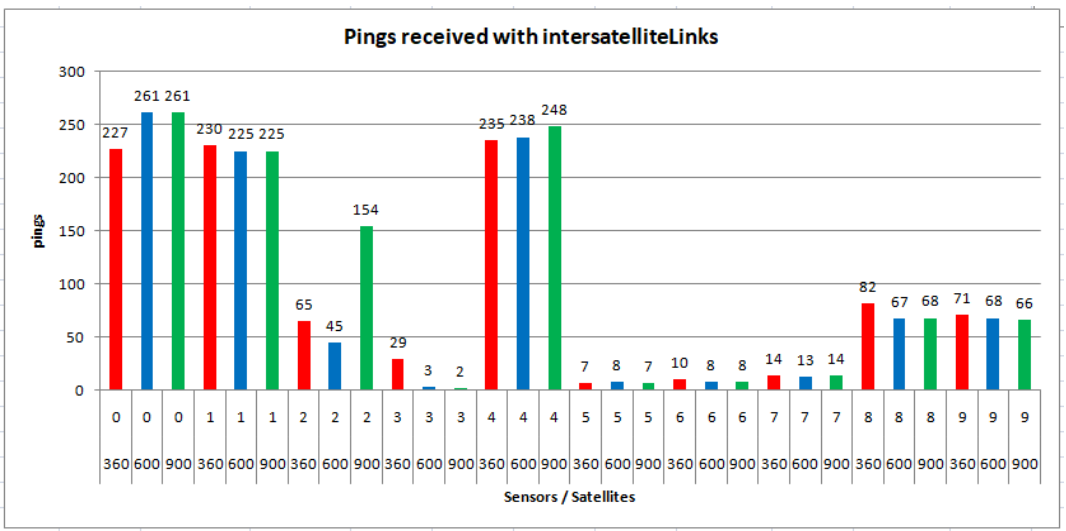

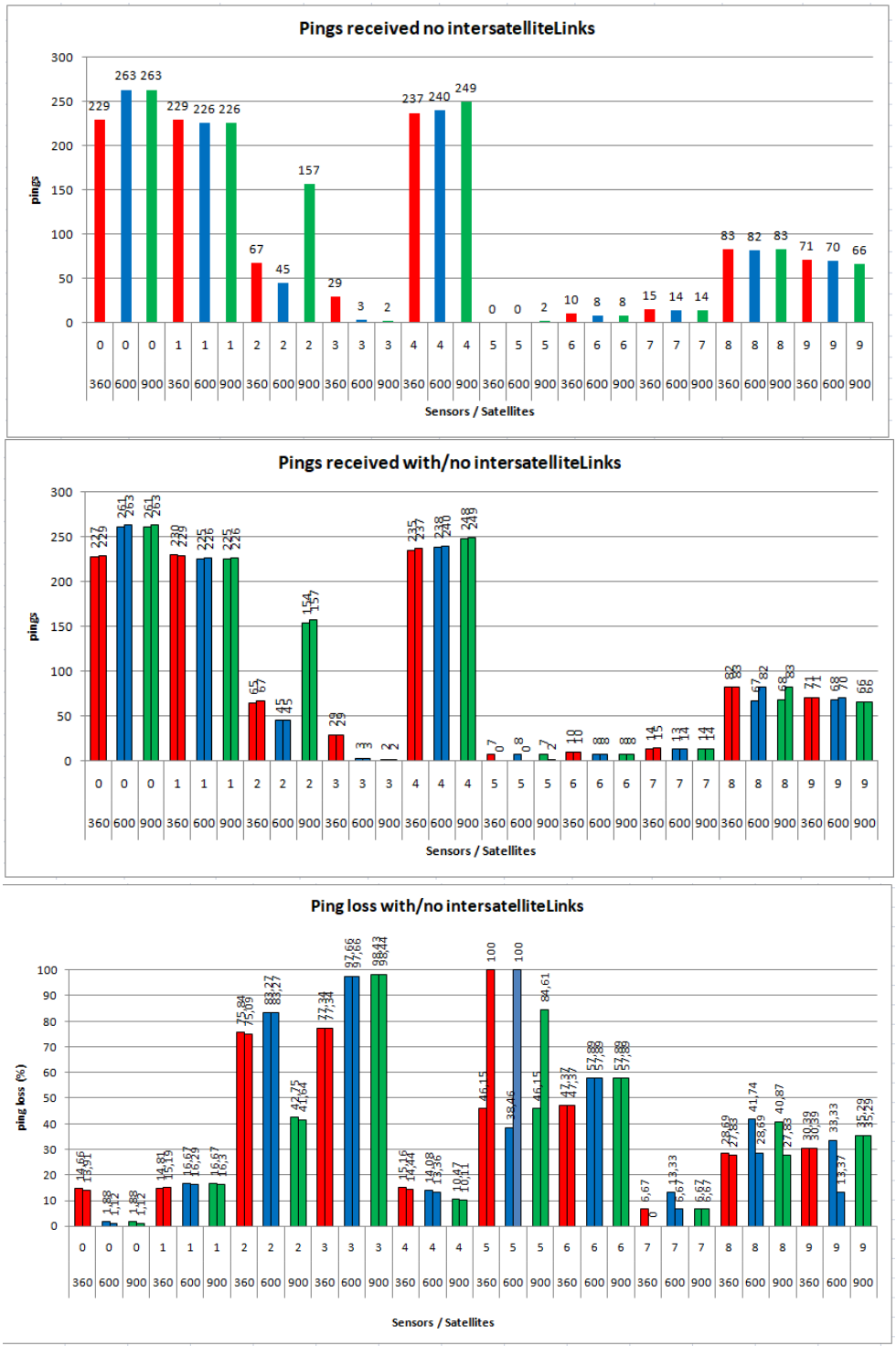

*Figure 58: Pings received and ping loss with & without intersatelliteLinks per sensor and satellites of the "ReWire Case Study" scenarios*

*Figure 57* presents the mean & range RTTs and *Figure 58* the received pings & ping loss, per sensor and satellite constellation. As each Sensor communicates with different destination nodes, the comparison among Sensors is pointless and therefore we compare each Sensor through our different scenarios. In all experiments of the Case Study, the range RTT is more compressed when intersatelliteLinks are enabled, due to the less frequent satellite<->MCC connections. For the same reason we generally observe a higher mean RTT with intersatelliteLinks disable. We also notice similar values of the mean RTT in a few Sensors with and without the intersatelliteLinks parameter. The most likely explanation is the location of the destination nodes, meaning that the connection between Sender and Receiver is achieved through one satellite, utilizing practically no intersatelliteLinks although they are enabled. Moreover Sensor[5] is a special case, communicating with its destination (located in USA) primarily with intersatelliteLinks enabled and only without them through a constellation of 900 satellites. Regarding the received pings, we detect some inconsistencies of the results. While some of them follow the expected pattern, i.e. with the increase of planes and satellites in the constellation the received pings are also increased (e.g., Sensor 0,4), others result in an inversely proportional relationship (e.g., Sensor 1,3) and some "follow their own rules". Specifically for the latter case, Sensor[2] receives more pings with a 360 satellite constellation compared to the 600 satellites but less pings in contrast to the 900 satellite constellation. We justify the peculiar results both by a possible occurrence of overlapping pings, but primarily by the topology of the satellite constellation (location of orbits and satellites) in comparison with the position of the ground MCCs.

# E. Conclusions and Future work

Regarding the effect of intersatelliteLinks, in all experiments with active intersatelliteLinks we observe more compressed values of the range RTT, i.e. lower minimum and maximum RTT, compared to those without the intersatelliteLink parameter. The observations are drawn by the presented *Figures* (rtt:vector and rtt:histogram) and *Tables* through the scenarios. In addition, mean RTTs also present lower values in cases of communication with intersatelliteLinks, as connecting satellites of the same plane result in shorter propagation delays (in contrast to connections of the pattern MCC<->satellite<->MCC). As mentioned, intersatelliteLinks also assist -in fact they are the only alternative in our cases- in scenarios where no MCC exists between a distant Sender and a Receiver (*Section D.I.b*). Nonetheless, we observed cases in which the existence of intersatelliteLinks resulted in higher ping losses. Specifficaly, in the ReWire Case Study (*Section D.III*) we discovered that Sensors 0,4,8 (see *Figure* 58) presented lower ping losses without the intersatelliteLink parameter in all of the examined satellite constellations. For instance we observe Sensor[4] in which scenario of the 360 satellite constellation with intersatelliteLinks results in 15,16% ping loss, while without them the ping loss drops to 14,44%. In addition, with 600 satellites and intersatelliteLinks enabled the loss is valued at 14,07%, higher than without intersatelliteLinks (13,36%). Finally, the 900 satellites constellation with intersatelliteLinks enable results in 10,47% of loss compared to the 10,11% in which intersatelliteLinks are disable. Concluding, the advantages of intersatelliteLinks are easily understood, but we should emphasize that they also significantly increase the cost and complexity of satellites, parameters we didn't consider or analyze in the present work.

The effect of the number of satellites in a constellation with regard to the produced RTT and ping loss was also investigated. Although the topology of each experiment plays the most important role, i.e. the location of the Sender and Receiver (or every MCC) on the map compared to the satellite constellation (planes, satellites, orbits etc.), in the majority of our experiments we detect an inversely proportional relationship. Thus, the increase of the number of satellites usually results in the reduction of the generated RTT and ping loss, which seems reasonable since larger constellations of satellites may utilize more effectively the available routes and paths between Sender & Receiver. Moreover, in experiments where the intersatelliteLink parameter was enabled as well, we noticed a further reduction of the RTT and ping loss. We observed an exception with Sensor[3] of the ReWire Case Study (see *Figure* 58) in which the 360 satellites constellation achieves better communication regarding the ping loss (29 received pings) compared to the constellations of 600 and 900 satellites (2 and 3 received pings respectively) regardless of the existence of intersatelliteLinks.

Furthermore, in our experiments we use "Ships" as relays which operate in the context of the other ground MCCs, establishing or enhancing the communication between Sender & Receiver. Their utilization should be investigated in details regarding the technical specifications and applicability of such a proposal.

Among the examined scenarios, we also explored a case with multiple neighboring Senders (see *Section D.II*). Due to their nearby distance, they contact the same satellites for their communication, resulting in issues when simultaneous transmissions occur (MCCs with identical sendIntervals). We captured this occasion (*Section D.II.a*) in which pings exceed the satellites' buffer queue and congestion appears, with consequent results the chaotic ping losses, the significant increase of RTT or even the loss of connection.

In addition, according to the extensive literature (in the *Related Work Section*), we argue that OMNeT++ is a suitable tool for the simulation of network topologies, the implementation and evaluation of networking protocols and novel paradigms (such as the ICN, DTN & SDN). Its applicability in space works is also considerable, with the combination of frameworks (e.g., OS3 & MiXiM) and platforms (e.g., Matlab & Simulink) providing flexible and adaptable experimental tools in various circumstances.

As for the future work, many parameters can be alternated and evaluated such as different constellation orbits, number of planes and satellites, angles of elevation and satellite altitudes. Also, additional mechanisms could be implemented aiming at the optimization of some outcome. Such examples might include algorithms for routing, queuing and traffic management, congestion control or even the store and forward technique, and the implementation & evaluation of protocols such as the TCP and NDP for satellite infrastructures.

Considering the extension of satellite or in general space usage, many novel research areas can be addressed. Ships, drones, UAVs and every "smart" device with the appropriate characteristics can assist in a various occasions based on the mission's specification requirements. A great example is the utilization of unique micro/nano satellites or drones that are launched for a specific purpose and their influence on communication relief or even security by acting as relays in smart-city / IoT scenarios. For instance, they can aid in cases of jammed, compromised or attacked communication infrastructures, while equipped with specific and secure instructions, they can become the connection bridge and temporarily fix a potential problem (e.g., a security breach on the communication infrastructure).

Concluding with the ReWire project, the challenging and dynamic network conditions of Smart-Cities require the deployment of novel solutions to deal with the scalability, efficiency, reliability and adaptability issues. Such an interpretation is detailed in [2][3] in which authors experiment with the novel Fed4FIRE+ test-beds and the ReWire platform. Consequently, each node in their topology deploys a multi-protocol solution based on most appropriate protocol strategy. Furthermore, we argue that space infrastructure is a supportive alternative, even a viable solution for networking communications in the context of smart-cities and IoT scenarios. Specifically, space infrastructures (e.g., satellites, drones, MANETs, VANETs or FANETs), combined with novel technologies (such as the DTN) can further assist in various circumstances.